\documentclass[a4paper,12pt]{article}
\usepackage{amssymb}
\usepackage{fancybox}
\usepackage{epsfig}
\usepackage{amstex}

\makeatletter    
\long\def\@makecaption#1#2{%
  \vskip\abovecaptionskip
  \sbox\@tempboxa{{\sf #1}: #2}
  \ifdim \wd\@tempboxa >\hsize
     {\sf #1}: #2\par                   
  \else
    \hbox to\hsize{\hfil\box\@tempboxa\hfil}%
  \fi
   \vskip\belowcaptionskip}
\makeatother     

\ifx\href\unDefined\def\href#1#2{#2}\fi
\title{Probability and Measurement Uncertainty in
Physics\\ - a Bayesian Primer\footnote[]{Notes based on
lectures given to graduate students in Rome (May 1995)
and summer students at DESY (September 1995). \\
E-mail: dagostini@@vaxrom.roma1.infn.it\\
Postscript file:
\href{http://zow00.desy.de:8000/zeus_papers/ZEUS_PAPERS/DESY-95-242.ps}
     {http://zow00.desy.de:8000/zeus\_papers/ZEUS\_PAPERS/DESY-95-242.ps}} -}
 \author{Giulio D'Agostini \\
 Universit\`a ``La Sapienza'' and INFN, Roma,
 Italy. }
\date{                 }

\begin{document}
 \maketitle
\begin{abstract}
Bayesian statistics is based on the subjective definition
of probability as {\it ``degree of belief''} and on Bayes' theorem,
the basic tool for assigning probabilities to hypotheses combining
{\it a priori} judgements and experimental information. This
was the original
point of view
of Bayes, Bernoulli, Gauss, Laplace, etc.
and contrasts with
later ``conventional'' (pseudo-)definitions of probabilities,
which implicitly presuppose the concept of probability.
These notes show that
the Bayesian approach is the natural one
for data analysis in the most general sense,
and for assigning uncertainties
to the results of physical measurements
- while at the same time
resolving philosophical aspects of the problems.
The approach, although
little known and usually misunderstood
among the High Energy Physics community,
has become the standard way of reasoning in
several  fields of research
and has recently been adopted by the
international metrology organizations
in their recommendations for
assessing  measurement uncertainty.

These notes describe
a general model for treating
uncertainties originating from random
and systematic errors
 in a consistent way and
 include  examples of
 applications of the model in High Energy Physics, e.g.
``confidence intervals'' in different contexts, upper/lower
limits, treatment of ``systematic errors'',
hypothesis tests and unfolding.
\end{abstract}

\mbox{}
\vspace{-20cm}
\begin{flushleft}
\tt DESY 95-242 \hfill ISSN 0418-9833\\ 
 Roma1 N.1070 \\
hep-ph/9512295\\
 December 1995
\end{flushleft}

\newpage
\tableofcontents
\newpage
\begin{flushright}
{\sl ``The only relevant thing is uncertainty - the extent of our} \\
{\sl knowledge and ignorance. The actual fact of whether or not}\\
{\sl the events considered are in some sense  {\it determined}, or}\\
{\sl known by other people, and so on, is of no consequence''.}\\
{\sl (Bruno de Finetti)} \\
\end{flushright}
\vspace{0.3 cm}

\section{Introduction}
The purpose of a measurement is
to determine the value of a physical quantity.
One often speaks of the {\it true value}, an idealized concept
achieved by an infinitely precise
and accurate measurement, i.e. immune from errors. In practice the result
 of a measurement is expressed in terms of the best
 estimate of the true value and of a related uncertainty.
Traditionally the various
contributions to the overall uncertainty are classified
in terms of {\it ``statistical''} and {\it ``systematic''}
uncertainties, expressions
which reflect the sources of the experimental
errors (the quote marks indicate that a different
way of classifying uncertainties
will be adopted in this paper).

``Statistical'' uncertainties
arise from  variations in the results of repeated observations
under (apparently) identical conditions. They vanish
if the number of observations becomes very large
(``the uncertainty is dominated by systematics'', is the typical
expression used in this case) and can be treated - in most
of cases, but with some exceptions of great relevance
in High Energy Physics - using conventional statistics based on the
frequency-based definition of probability.

On the other hand, it is  not possible to treat
 ``systematic'' uncertainties coherently in the
frequentistic framework. Several ad hoc prescriptions for how
to combine ``statistical'' and ``systematic'' uncertainties
can be found in text books and in the literature:
{\sl ``add them linearly''};
{\sl ``add them linearly if $\ldots$, else add them
      quadratically''};
{\sl ``don't add them at all''}, and so on (see, e.g.,
part 3 of \cite{DIN}). The ``fashion'' at the moment is to
add them quadratically if they are considered
independent, or to build a covariance matrix of ``statistical''
and ``systematic'' uncertainties
to treat general cases.
These procedures are not justified by  conventional
statistical theory, but they are accepted
because of the pragmatic good sense of physicists.
 For example, an experimentalist may be
reluctant to add twenty or more
contributions linearly to evaluated the uncertainty
of a complicated measurement, or decides
 to treat
the correlated ``systematic'' uncertainties
 ``statistically'', in both cases
 unaware of, or simply not caring about, about violating
frequentistic principles.

The only way to deal with these  and related
problems in a consistent way
is to abandon the frequentistic interpretation
of probability introduced at the beginning of this century,
and to recover the intuitive concept of probability
as {\it degree of belief}. Stated differently, one needs to associate
the idea of probability to the lack of  knowledge,
rather than to  the outcome of repeated experiments.
This has been recognized also by the International Organization
for Standardization (ISO) which assumes the subjective definition
of probability
in its
{\it ``Guide to the expression of uncertainty in measurement''}\cite{ISO}.

These notes are organized as follow:
\begin{itemize}
\item
sections 1-5 give a general introduction
to subjective probability;
\item
sections 6-7 summarize some concepts and formulae concerning
random variables, needed for many applications;
\item
section 8 introduces the problem of measurement uncertainty
and deals with the terminology.
\item
sections 9-10 present the analysis model;
\item
sections 11-13 show several physical  applications of the model;
\item
section 14 deals with the approximate methods needed
when the general solution becomes complicated; in this context
the ISO recommendations
will be presented and discussed;

\item
section 15 deals with uncertainty propagation. It
is particularly short because, in this scheme, there
is no difference between the treatment of ``systematic'' uncertainties
and indirect measurements; the section simply
refers the results of sections
11-14;
\item
section 16 is dedicated to a detailed discussion about
the covariance matrix of correlated data and the trouble
it may cause;
\item
section 17 was added as an
example of a more complicate inference (multidimensional unfolding)
than those treated in sections 11-15.
\end{itemize}
\newpage

\section{Probability}
\subsection{What is probability?}
The standard answers to this question are
\begin{enumerate}
\item
``the ratio of the number of favorable cases to the
number of all cases'';
\item
``the ratio of the times the event occurs in a test series
to the total number of trials in the series''.
\end{enumerate}
It is very easy to show that neither of these
statements can define the concept of probability:
\begin{itemize}
\item
Definition (1) lacks
the clause ``if all the cases are
\underline{equally probable}''. This has
been done here intentionally, because people often forget it.
The fact that the definition of probability makes use of the term
``probability'' is clearly embarrassing. Often in text books the
clause is replaced by ``if all the cases are
equally possible'', ignoring that in this
context ``possible''
is just a synonym of ``probable''. There is no way out.
This statement does not
 define probability but
gives, at most, a useful rule for evaluating it -
assuming we
know what probability is, i.e. of what we are talking about.
The fact that this definition is labelled
``classical'' or ``Laplace'' simply shows that some
 authors are not
aware of what the ``classicals'' (Bayes, Gauss, Laplace, Bernoulli, etc)
thought about this matter. We shall call this ``definition''
{\it combinatorial}.
\item
definition (2) is also incomplete, since it lacks the condition
that the number of trials must be very large (``it goes to infinity'').
But this is a minor point. The crucial point is that the
statement merely defines  the relative {\it frequency} with
 which an event
(a ``phenomenon'')
occurred in the past. To use frequency as a measurement of
probability we have to assume that the phenomenon
occurred in the past, and will occur in the future,
\underline{with the same probability}. But who can tell if this hypothesis
is correct? Nobody: \underline{we}
 have to guess in every single case. Notice that, while in the
 first ``definition'' the assumption of equal probability
 was explicitly stated, the analogous clause is often
 missing from the second one. We shall call this ``definition''
 {\it frequentistic}.
\end{itemize}
We have to conclude that if we want to make use of these
statements
to assign a numerical value to probability, in those cases
in which \underline{we judge} that the clauses are satisfied, we need
a better definition of probability.

\subsection{Subjective definition of probability}
\begin{figure}[t]
\centering\epsfig{file=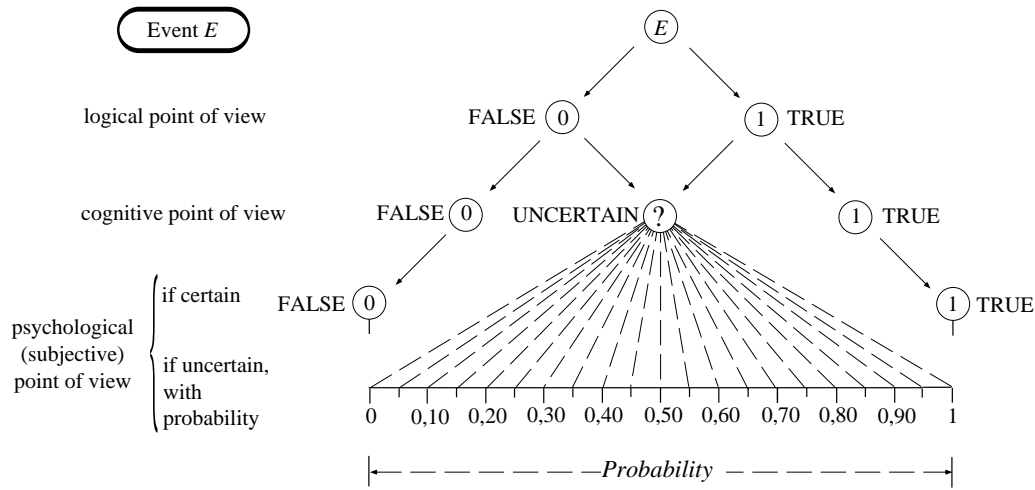,width=\linewidth,clip=}
\caption{\sf Certain and uncertain events.}
\label{fig:probability}
\end{figure}
So, {\it ``what is probability?''}
Consulting
a good dictionary
helps.
Webster's states, for example, that ``{\it probability}
 is the quality, state, or degree of being probable'',
 and then that {\it probable} means ``supported by evidence strong
 enough to make it likely though not certain to be true''.
The concept of probable arises in reasoning when the concept
of {\it certain} is not applicable. When it is impossible to
state firmly if an {\it event} (we use this word as a synonym
for {\it any possible statement}, or {\it proposition},
 relative to past, present or future)
is {\it true} or {\it false}, we just say that this is
{\it possible}, {\it probable}. Different events may have different
{\it levels} of probability, depending whether we think that they are more
likely to be true or false (see Fig. ~\ref{fig:probability}).
The concept of probability
is then simply
\begin{quote}
{\it a measure of the \underline{degree of belief}
that an event will}\footnote{The use of the future tense
does not imply that this definition can only be
applied for future events. ``Will occur'' simply
means that the statement ``will be proven to be true'',
even if it refers to the past. Think for example of
``the probability that it was raining in Rome on the day
of the battle of Waterloo''.} {\it occur}.
\end{quote}
This is the kind of definition that one finds in
Bayesian books\cite{Jeffreys,Winkler,Definetti3,Press,Bernardo}
and the formulation cited here is that given in
the ISO {\it ``Guide to Expression of
Uncertainty in Measurement''}\cite{ISO},
of which we will talk later.

At first sight this definition does not seem to be superior to
the combinatorial or the frequentistic ones.
At least they give some practical
rules to calculate ``something''. Defining
probability as {\it ``degree of belief''} seems too vague to
be of any utility. We need, then, some explanation of its
meaning; a tool to evaluate it - and
we will look at this tool (Bayes' theorem)
later. We will end this section with some
explanatory remarks on the definition, but
first let us discuss the
 \underline{advantages} of this definition:
\begin{itemize}
\item
it is natural, very general and it can be applied to any thinkable
event, independently of the feasibility of making an
inventory of all (equally) possible and favorable cases, or
of repeating the experiment under conditions
of equal probability;
\item
it avoids the linguistic schizophrenia of having to
distinguish  ``scientific'' probability from
 ``non scientific'' probability used in
everyday reasoning (though a meteorologist
might feel offended to hear that
evaluating the probability of rain tomorrow
is ``not scientific'');
\item
as far as measurements are concerned, it allows
us to talk about the probability of the {\it true value} of a physical
quantity. In the frequentistic
frame it is only possible to talk about the probability
of the {\it outcome} of an experiment, as the true value is
considered to be a constant. This approach is
so unnatural that most physicists speak of
``$95\,\%$ probability that the mass of the Top quark is
between $\ldots$'',
\underline{although} they believe that the correct definition of probability
is the limit of the frequency;
\item
it is possible to make a very general theory of uncertainty
which can take into account any source of statistical and
systematic error, independently of their distribution.
\end{itemize}

To get a better understanding of the subjective definition of
probability let us take a look at \underline{odds in betting}.
The higher the
degree of belief
that an event will occur, the higher
the amount of money $A$ that someone (``a rational better'')
is ready to pay in order to receive a sum of money $B$ if the event
occurs. Clearly the bet must be acceptable (``coherent''
is the correct adjective), i.e. the amount of money $A$
must be smaller or equal to $B$
and not negative (who would accept such a bet?).
The cases of $A=0$ and $A=B$ mean that the events are considered
to be false or true, respectively,
and obviously it is not worth betting on certainties.
They are just limit cases, and in fact they can be
treated with standard logic.
It seems reasonable\footnote{This is not always
true in real life. There are also other
practical problems related to  betting which have been
treated in the
literature. Other variations of the
definition have also been proposed, like the one
based on the {\it penalization rule}. A discussion of the
problem goes beyond the purpose of these notes.}
that the amount of money $A$ that one is willing to pay
grows linearly
with the degree of belief.
It follows that  if someone thinks that
the probability of the event $E$ is $p$, then  he
will bet $A=pB$ to get
$B$ if the event occurs, and to lose $pB$
if it does not. It is easy to
demonstrate that the condition of ``coherence''
implies that $0\le p\le 1$.

What has  gambling to do with physics? The
definition of probability through
betting odds has to be considered {\it operational}, although there is no
need to make a bet (with whom?) each time one
presents a result. It has the important role of forcing
one to make an
{\it honest} assessment of the value of probability that
one believes. One could replace  money with  other forms
of gratification or penalization, like the increase or
the loss of scientific reputation. Moreover, the
fact that this operational procedure is not to
be taken literally should not be surprising. Many
physical quantities are defined in a similar way.
Think, for example, of the text book definition of
the electric field, and try to use it
 to measure $\vec{E}$
 in the proximity of an electron.
A nice example comes from the definition of a poisonous chemical
compound: it {\it would be lethal \underline{if} ingested}.
Clearly  it is preferable to keep this operational definition
at a hypothetical level, even though it is the
best definition of the concept.

\subsection{Rules of probability}
\begin{figure}
\centering\epsfig{file=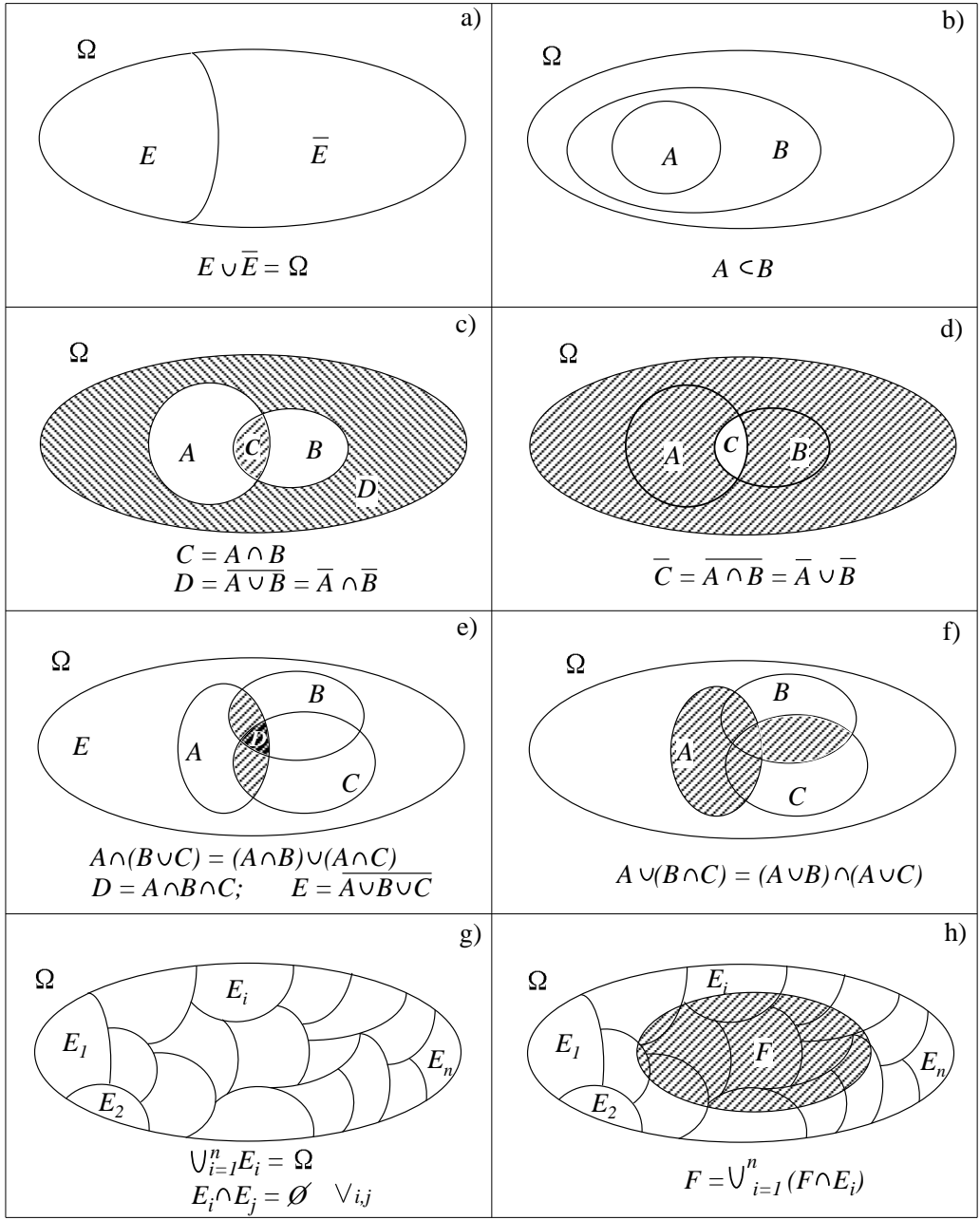,width=\linewidth,clip=}
\caption{\sf Venn diagrams and set properties.}
\label{fig:Venn}
\end{figure}
The subjective definition of probability, together with the condition
of {\it coherence}, requires that $0 \le p \le 1$. This is one of the rules
 which probability has to obey. It is possible, in fact, to demonstrate
that  coherence yields to the standard rules of probability,
generally known as {\it axioms}. At this point
it is worth
clarifying the relationship between the axiomatic approach
and the others:
\begin{itemize}
\item
combinatorial and frequentistic ``definitions''
give
useful rules for evaluating probability, although
they do not, as it is often claimed,
define the concept;
\item
in the axiomatic approach one refrains
 from defining what the probability
is and how to evaluate it: probability
is just any real number which satisfies the axioms.
It is easy to demonstrate that the probabilities
evaluated using the combinatorial and the frequentistic
prescriptions do in fact satisfy the axioms;
\item
the subjective approach to probability, together with the
coherence requirement,
\underline{defines} what probability is and provides
the rules which its evaluation must obey; these rules
turn out to
be the same as the axioms.
\end{itemize}

Since everybody is familiar with the axioms and with the analogy
$events\Leftrightarrow sets$ (see Tab. ~\ref{tab:eventi_insiemi}
and Fig. ~\ref{fig:Venn})
let us remind ourselves of the {\it rules of probability} in this form:
{ \small
\begin{table}[tb]
\begin{center}
\begin{tabular}{|l|l|c|} \hline
\multicolumn{1}{|c|}{Events} & \multicolumn{2}{|c|}{sets} \\ \hline
       &    & symbol \\ \hline
event & set & $E$ \\
certain event & sample space & $\Omega$ \\
impossible event & empty set & $\emptyset$ \\
implication & inclusion & $E_1\subseteq E_2$\\
             & (subset) & \\
opposite event & complementary set & $\overline{E}$
                     \hspace{0.4cm}($E\cup \overline{E} = \Omega$) \\
(complementary) &        & \\
logical product (``AND'')  & intersection  & $E_1 \cap E_2$ \\
logical sum (``OR'')     & union        & $E_1 \cup E_2$ \\
incompatible events&  disjoint sets & $E_1 \cap E_2 = \emptyset$\\
complete class               & finite partition&
  $\left\{ \begin{array}{l}
  E_i \cap E_j  = \emptyset \ (i\ne j)\\
  \cup_i E_i  = \Omega
  \end{array}\right.$ \\ \hline
  \end{tabular}
\end{center}
\caption{\sf Events versus sets.}
\label{tab:eventi_insiemi}
\end{table}
} 
\begin{description}
\item[Axiom 1] $0 \leq P(E) \leq 1$;
\item[Axiom 2] $  P(\Omega) = 1$ (a certain event has probability 1);
\item[Axiom 3]
       $ P(E_1 \cup E_2) = P(E_1)+P(E_2)$, if
                            $ E_1 \cap E_2 = \emptyset$
\end{description}
{}From the basic rules  the following properties can be derived:
\begin{description}
\item[1:]
 $P(E) = 1 - P(\overline E) $;
\item[2:]
     $P(\emptyset) = 0$;
\item[3:]
    if $A\subseteq B$ then $P(A) \leq P(B) $;
\item[4:]
     $P(A\cup B) = P(A) + P(B) - P(A\cap B)$\,.
\end{description}
We also anticipate here a fifth property which will be discussed
in section ~\ref{ss:conditional}:
\begin{description}
\item[5:] $  P(A\cap B) = P(A|B)\cdot P(B) = P(A)\cdot P(B|A)\,.$
\end{description}

\subsection{Subjective probability and ``objective''%
description of the physical world}
The subjective definition of probability seems
to contradict
the aim of physicists to describe the laws of Physics
in the most
objective way (whatever this means $\ldots$).
This is one of the reasons why many regard
the subjective definition of probability
with suspicion (but
probably the main reason is because
we have been taught at University that
``probability is frequency''). The main philosophical
difference between this concept of probability and an
objective definition that
{\it ``we would have liked''} (but which does not exist in reality)
is the fact that $P(E)$ is not an intrinsic characteristic
of the event $E$, but  depends on the {\it status of information}
available to whoever evaluates $P(E)$.
The ideal concept of ``objective''
probability is recovered when everybody has the ``same'' status of
information. But even in this case it would be better to speak of
{\it intersubjective} probability. The best way to  convince
ourselves  about
this aspect of probability is to try to ask practical
questions and to evaluate the probability in specific cases,
instead of seeking refuge in abstract questions. I find, in fact,
that, paraphrasing a famous statement about Time,
``Probability is objective as long as I am not asked to evaluate it''.
Some examples:
\begin{description}
\item[Example 1:]
``What is the probability
that a molecule of nitrogen at atmospheric pressure and room
temperature has a velocity between 400 and 500 m/s?''. The answer
appears easy: ``take the Maxwell distribution formula from a text
book, calculate an integral and get a number. Now
let us change the question:
{\it ``I give you a vessel containing nitrogen
 and a detector}
{\it capable of
measuring the speed of a single molecule and you
set up the apparatus. Now, what is the probability that the
\underline{first}
molecule that hits the detector has a velocity between
 400 and 500 m/s?''}. Anybody who has minimal experience (direct
 or indirect) of experiments would hesitate before answering.
 He would study the problem carefully and  perform
 preliminary measurements and checks.
 Finally he would {\it probably}
 give not just a single number, but a range of possible numbers
 compatible with the formulation of the problem. Then
 he starts the experiment and eventually, after 10 measurements,
 he may form
 a different  opinion about the outcome of the eleventh measurement.
\item[Example 2:]
``What is the probability that the gravitational constant $G_N$
has a value between $6.6709\cdot 10^{-11}$ and $6.6743 \cdot 10^{-11}$
$\mbox{m}^3\mbox{kg}^{-1}\mbox{s}^{-2}$?''. Last year you
could have  looked at the latest
 issue of the Particle Data Book\cite{PDG}
and answered that the probability was $95\,\%$. Since then - as you
probably know - three new measurements of  $G_N$ have been
 performed\cite{gn}
 and  we now have \underline{four} numbers which do not agree
with each other (see Tab. \ref{tab:Gn}).
The probability of the true value of $G_N$
being in that range is currently dramatically decreased.
{ \small
\begin{table}
\begin{center}
\begin{tabular}{cccc} \hline
Institut & $G_N$ $\left(10^{-11}
\frac{\mbox{m}^3}{\mbox{kg}\cdot\mbox{s}^{2}}\right)$ &
$\frac{\sigma(G_N)}{G_N}$  (ppm) &
$\frac{G_N-G_N^C}{G_N^C}$ ($10^{-3}$)\\ \hline
CODATA 1986 (``$G_N^C$'') & $6.6726\pm 0.0009$ & 128 & -- \\
PTB  (Germany) 1994  & $6.7154\pm 0.0006$ & 83 & $+6.41\pm 0.16$ \\
MSL (New Zealand) 1994& $6.6656\pm 0.0006$ & 95 & $-1.05\pm 0.16$ \\
Uni-Wuppertal        & $6.6685\pm 0.0007$ & 105& $-0.61\pm 0.17$ \\
  (Germany) 1995     &                    &    &                  \\ \hline
\end{tabular}
\end{center}
\caption{\sf Measurement of $G_N$ (see text).}
\label{tab:Gn}
\end{table}
} 

\item[Example 3:]
``What is the probability that the mass of the Top
quark, or that of any of the supersymmetric particles, is below
20 or $50\,\mbox{GeV}/c^2$?''. Currently it looks
as if it must be zero. Ten years ago
many experiments were intensively looking
for these particles in those energy ranges.
Because so
many people where searching for them, with
enormous human and capital investment, it means that,
\underline{at that time},
the probability was considered rather \underline{high}, $\ldots$
 high enough for fake signals
to be reported as strong evidence for them\footnote{We will talk
later about the influence of {\it a priori} beliefs on the
outcome of an experimental investigation.}.
\end{description}
The above examples show how the evaluation of probability
is \underline{conditioned} by some {\it a priori} (``theoretical'')
prejudices and by some facts (``experimental data''). ``Absolute''
probability makes no sense. Even the classical example
of probability $1/2$ for each of the results in tossing a coin
is only acceptable if: the coin is regular;
it does not remain vertical (not impossible
when playing on the beach);
it does not fall into a manhole; etc.

 The subjective point of view  is expressed
 in a provocative way
by de Finetti's\cite{Definetti3}
\begin{quote}
\begin{center}
``PROBABILITY DOES NOT EXIST''.
\end{center}
\end{quote}
\section{Conditional probability and Bayes' theorem}
\subsection{Dependence of the probability from the
status of information}\label{ss:conditional}
If the status of information changes, the evaluation of
the probability also has to be modified. For example
most people would agree that the probability
of a car being stolen depends on the model, age and parking site.
To take an example from physics, the probability that
in a HERA detector a charged particle
of $1\, \mbox{GeV}$
gives a certain number of ADC counts due
to the energy loss in a gas detector can be evaluated
in a very general way - using High Energy Physics jargon - making a
(huge) Monte Carlo simulation which takes into account all
possible reactions (weighted with their cross sections),
all possible backgrounds, changing all physical and detector
parameters within {\it reasonable} ranges, and also taking into
account the trigger efficiency. The probability
changes if one knows that the particle is a $K^+$: instead of very
complicated Monte Carlo one can just run a single particle
generator. But then it changes further if one also knows the
exact gas mixture,
pressure, $\ldots$, up to the latest
determination of the pedestal and the temperature of the ADC module.

\subsection{Conditional probability}
Although everybody knows the formula of conditional probability,
it is  useful to derive it here. The notation is $P(E|H)$,
to be read ``probability of $E$ given $H$'', where $H$ stands for
{\it hypothesis}.
This means: the probability that $E$ will occur if one
already knows that  $H$ has occurred\footnote{$P(E|H)$ should not be
confused with $P(E\cap H)$, ``the probability that both
events occur''. For example $P(E\cap H)$ can be very small, but
nevertheless $P(E|H)$ very high: think of the limit case
$$ P(H)\equiv P(H\cap H) \le P(H|H) = 1 \,:$$
``$H$ given $H$'' is a certain event no matter how small $P(H)$ is,
even if $P(H)=0$ (in the sense of Section ~\ref{sec:cont_var}).}.
\newpage
The event $E|H$ can have three values:
\begin{description}
\item[TRUE:] if $E$ is TRUE \underline{and} $H$ is TRUE;
\item[FALSE:] if $E$ is FALSE \underline{and} $H$ is TRUE;
\item[UNDETERMINED:] if $H$ is FALSE; in this case we are merely
                     uninterested as to what happens to $E$. In terms
                     of betting, the bet is
                     invalidated and none loses or gains.
\end{description}
Then $P(E)$ can be written $P(E|\Omega)$,
to state explicitly that it is the probability of
$E$ whatever happens to the rest of the world
($\Omega$ means all possible events). We realize immediately
that this condition is really too vague and nobody would
bet a cent on a such a statement. The reason for usually
writing $P(E)$
is that many conditions
are implicitly - and reasonably - assumed in most
circumstances. In the classical problems of coins and dice, for example,
 one \underline{assumes} that they are regular. In the example
of the energy loss,
it was implicit -``obvious''-  that the
High Voltage was on (at which voltage?)
and that HERA was running (under which condition?).
But one has to take care:  many riddles are
based on the fact that one tries to find a solution which is
valid under more strict conditions than those explicitly stated
in the question (e.g. many people make bad business deals signing
contracts in which what ``was obvious''
was not explicitly stated).

In order to derive the formula of conditional probability
let us assume for a moment that it is reasonable to
talk about
``absolute probability'' $P(E)=P(E|\Omega)$,
and let us rewrite
\begin{eqnarray}
P(E) \equiv P(E|\Omega)
&\underset{\bf a}{=}& P(E\cap\Omega) \nonumber \\
&\underset{\bf b}{=} &
   P\left(E\cap (H \cup \overline{H})\right) \nonumber \\
  & \underset{\bf c}{=}&
    P\left((E\cap H) \cup (E\cap\overline{H})\right) \nonumber \\
  & \underset{\bf d}{=}& P(E\cap H) + P(E\cap\overline{H})\,,
\label{eq:cond0}
\end{eqnarray}
where the result has been achieved through the following steps:
\begin{description}
\item[(a):] $E$ implies $\Omega$ (i.e. $E\subseteq \Omega$)
and hence $E\cap\Omega=E$;
\item[(b):] the complementary events $H$ and $\overline{H}$
            make a {\it finite partition} of $\Omega$,
            i.e. $H \cup \overline{H} = \Omega$;
\item[(c):] distributive property;
\item[(d):] axiom 3.
\end{description}
The final result of (\ref{eq:cond0}) is very simple:
$P(E)$ is equal to the probability that $E$ occurs and $H$ also
occurs, plus the probability that $E$ occurs but $H$ does not
occur. To obtain $P(E|H)$ we just get rid of the subset of
$E$ which does not contain $H$ (i.e. $E\cap\overline{H}$)
and renormalize the probability
dividing by $P(H)$, assumed to be different from zero. This guarantees
that if $E=H$ then $P(H|H)=1$.
The expression of the conditional probability is finally
\begin{equation}
P(E|H) = \frac{P(E\cap H)}{P(H)}\hspace{1.0cm}(P(H)\ne 0)\,.
\label{eq:cond1}
\end{equation}
In the most general (and realistic)
case, where both $E$ and $H$ are conditioned by the occurrence of
a third event  $H_\circ$, the formula becomes
\begin{equation}
P(E|H, H_\circ) =
\frac{P\left(E\cap (H| H_\circ)\right) }
{P(H|H_\circ)}\hspace{1.0cm}(P(H|H_\circ)\ne 0)\,.
\label{eq:cond2}
\end{equation}
Usually we shall make use of (\ref{eq:cond1})
(which means $H_\circ=\Omega$) assuming that $\Omega$ has been
properly chosen.
We should also remember that (\ref{eq:cond1}) can be resolved
with respect to $P(E\cap H)$, obtaining the well known
\begin{equation}
P(E\cap H) = P(E|H)P(H)\,,
\label{eq:pcomp1}
\end{equation}
and by symmetry
\begin{equation}
P(E\cap H) = P(H|E)P(E)\,.
\label{eq:pcomp2}
\end{equation}
Two events are called {\it independent} if
\begin{equation}
P(E\cap H) = P(E)P(H)\,.
\end{equation}
This is equivalent to saying that $P(E|H) = P(E)$ and $P(H|E)=P(H)$,
i.e. the knowledge that one event has occurred does not change the
probability of the other. If $P(E|H) \ne P(E)$ then the events
$E$ and $H$ are {\it correlated}. In particular:
\begin{itemize}
\item
if $P(E|H) > P(E)$ then $E$ and $H$ are {\it positively} correlated;
\item
if $P(E|H) < P(E)$ then $E$ and $H$ are {\it negatively} correlated;
\end{itemize}
\subsection{Bayes' theorem}
Let us think of all the possible, mutually exclusive,
 hypotheses $H_i$ which could condition
the event $E$. The  problem here is the inverse of the previous one:
what is the probability of $H_i$ under the hypothesis that $E$
has occurred? For example,
``what is the probability that a charged
particle which went in a certain direction and has lost
between 100 and $120\,\mbox{keV}$
in the detector,
is a $\mu$, a $\pi$, a $K$, or a $p$?" Our event $E$
is ``energy loss between 100 and $120\,\mbox{keV}$'',
and $H_i$
are the four ``particle hypotheses''.
This example sketches the basic problem for
any kind of measurement: having observed an {\it effect},
to assess the probability of each of the {\it causes } which
could have produced it. This intellectual
process is called {\it inference}, and it will be discussed
after section ~\ref{sec:inference}.

In order to calculate $P(H_i|E)$ let us rewrite the
joint probability $P(H_i\cap E)$, making use of
(\ref{eq:pcomp1}-\ref{eq:pcomp2}),
 in two different ways:
\begin{equation}
P(H_i|E)P(E) = P(E|H_i)P(H_i)\,,
\end{equation}
obtaining
\begin{equation}
\boxed{
P(H_i|E) = \frac{P(E|H_i)P(H_i)}{P(E)}\,,
}
\label{eq:bayes1}
\end{equation}
or
\begin{equation}
\boxed{
\frac{P(H_i|E)}{P(H_i)} = \frac{P(E|H_i)}{P(E)}\,.
}
\label{eq:bayes1a}
\end{equation}
Since the hypotheses $H_i$ are mutually exclusive
 (i.e. $H_i\cap H_j=\emptyset$, $\forall\, i,j$) and exhaustive
 (i.e. $\bigcup_i H_i = \Omega$),
 $E$ can be written as $E\cup H_i$, the union  of
  $E$ with each of the hypotheses $H_i$. It follows that
\begin{eqnarray}
P(E) \equiv P(E \cap \Omega) &=& P\left(E \cap\bigcup_i H_i\right)
= P\left(\bigcup_i (E \cap  H_i)\right) \nonumber \\
&=& \sum_i P(E\cap H_i) \nonumber \\
&=& \sum_i P(E|H_i)P(H_i)\,,
\end{eqnarray}
where we have made use of (\ref{eq:pcomp1})
again in the last step.
It is then possible to rewrite (\ref{eq:bayes1})
as
\begin{equation}
\boxed{
P(H_i|E) = \frac{P(E|H_i)P(H_i)}{\sum_j P(E|H_j)P(H_j)}\,.
}
\label{eq:bayes2}
\end{equation}
This is the standard form by which  {\it Bayes' theorem}
is known. (\ref{eq:bayes1})
and (\ref{eq:bayes1a}) are also different ways
of writing it. As the denominator of
(\ref{eq:bayes2}) is nothing but a normalization
factor (such that $\sum_i P(H_i|E)=1$), the formula
(\ref{eq:bayes2}) can be
written as
\begin{equation}
\boxed{
P(H_i|E) \propto P(E|H_i)P(H_i) \,.
}
\label{eq:bayes3}
\end{equation}
Factorizing $P(H_i)$ in (\ref{eq:bayes2}), and explicitly writing
the fact that all the events were already
conditioned by $H_\circ$, we can rewrite the formula
as
\begin{equation}
\boxed{
P(H_i|E, H_\circ) = \alpha P(H_i|H_\circ)\,,
}
\label{eq:bayes4}
\end{equation}
with
\begin{equation}
\alpha=\frac{P(E|H_i,H_\circ)}
              {\sum_i P(E|H_i, H_\circ)P(H_i|H_\circ)}\,.
\label{eq:bayes5}
\end{equation}
These five ways of rewriting the same formula simply reflect
the importance that we shall give to this simple theorem.
They  stress different aspects of the same concept:
\begin{itemize}
\item
(\ref{eq:bayes2}) is the standard way of writing it (although some
prefer (\ref{eq:bayes1}));
\item
(\ref{eq:bayes1a}) indicates  that $P(H_i)$ is altered
by the condition $E$ with the same ratio with which
$P(E)$ is altered by the condition $H_i$;
\item
(\ref{eq:bayes3})
is the simplest and the most intuitive way to
formulate the theorem: ''the probability of $H_i$ given $E$ is
proportional to the {\it initial} probability of $H_i$ times
the probability of $E$ given $H_i$'';
\item
(\ref{eq:bayes4}-\ref{eq:bayes5})
show explicitly how
the probability of a certain hypothesis is updated when the
{\it status of information} changes:
 \begin{description}
 \item[\fbox{ $P(H_i|H_\circ)$}] (also indicated as $P_\circ(H_i)$) is
 the {\it initial}, or {\it a priori}, probability (or simply
 {\it ``prior''}) of $H_i$, i.e. the probability of this hypothesis
 with the status of information available
 \underline{before} the
 knowledge that $E$ has occurred;
 \item[\fbox{$P(H_i|E, H_\circ)$}] (or simply $P(H_i|E)$) is the
 {\it final}, or {\it ``a posteriori''}, probability of $H_i$
 \underline{after} the new information;
 \item[\fbox{ $P(E|H_i, H_\circ)$}] (or simply $P(E|H_i)$) is
 called {\it likelihood}.
 \end{description}
\end{itemize}
To better understand the terms ``initial'', ``final'' and
``likelihood'', let us formulate the problem in a way closer
to the physicist's mentality, referring to {\it causes} and
{\it effects}: the causes can be all the physical sources
which may produce a certain {\it observable} (the effect). The
likelihoods are - as the word says - the
likelihoods that the effect follows from each of the causes.
Using our example of the $dE/dx$ measurement again, the
causes are all the possible charged particles which can
pass through the detector; the effect is the amount of observed
ionization;
the likelihoods are the probabilities that each of the particles
give that amount of ionization.
Notice that in this example we have fixed all
the other sources of influence: physics process,
HERA running conditions, gas mixture, High Voltage,
track direction, etc.. This is our $H_\circ$.
The problem immediately gets rather complicated (all real cases,
apart from tossing coins and dice, are complicated!).
The real inference would be of the kind
\begin{equation}
P(H_i|E,H_\circ) \propto P(E|H_i, H_\circ)
                                   P(H_i|H_\circ)P(H_\circ),.
\end{equation}
For each status of $H_\circ$ (the set of all the possible values
of the influence parameters) one gets a different result
for the final probability\footnote{The symbol $\propto$
could be misunderstood if one forgets that the proportionality
factor depends on all likelihoods and priors (see (\ref{eq:bayes4})).
This means that, for a given hypothesis $H_i$,
as the status of information $E$ changes,
$P(H_i|E,H_\circ)$ may change
if $P(E|H_i, H_\circ)$ and
$P(H_i|H_\circ)$ remain constant,
if some of the other likelihoods get modified by the new information.}.
So, instead of getting a single number
for the final probability we have a distribution of values. This spread
will result in a large uncertainty of $P(H_i|E)$. This is what
every physicist knows: if the calibration constants of the detector
and the physics process \underline{are not under control},
the \underline{``systematic errors''} are large and the result is
of poor quality.
\subsection{Conventional use of Bayes' theorem}
Bayes' theorem follows
 directly from the rules of probability,
and it can be used in any kind of approach. Let us take an
example:
\begin{description}
\item[Problem 1:]
A particle detector has a $\mu$ identification efficiency of $95\,\%$,
and a probability of identifying a $\pi$ as a $\mu$ of $2\,\%$. If a
particle is identified as a $\mu$, then
 a trigger is issued. Knowing that
the particle beam is a mixture of $90\,\%$ $\pi$ and $10\,\%$ $\mu$,
what is the probability that a trigger is really fired by a $\mu$?
What is the signal-to-noise ($S/N$) ratio?
\item[Solution:]
The two hypotheses (causes) which could condition the event (effect)
$T$ (=``trigger fired'') are ``$\mu$'' and ``$\pi$''. They are incompatible
(clearly) and exhaustive (90\,\%+10\,\%=100\,\%). Then:
\begin{eqnarray}
P(\mu|T) & = & \frac{P(T|\mu)P_\circ(\mu)}
                     {P(T|\mu)P_\circ(\mu) + P(T|\pi)P_\circ(\pi)} \\
         & = & \frac{0.95\times 0.1}{0.95\times 0.1 + 0.02\times 0.9}=0.84\,,
         \nonumber
\label{eq:pi_mu}
\end{eqnarray}
and $P(\pi|T)=0.16$.

The signal to noise ratio is $P(\mu|T)/P(\pi|T)=5.3$. It is interesting
to rewrite the general expression of the signal to noise ratio
if the effect  $E$ is observed as
\begin{equation}
S/N = \frac{P(S|E)}{P(N|E)}=\frac{P(E|S)}{P(E|N)}\cdot
\frac{P_\circ(S)}{P_\circ(N)}\,.
\end{equation}
This formula explicitly shows that when there are
 {\it noisy conditions}
$$P_\circ(S) \ll P_\circ(N)$$ the experiment must be {\it very selective}
$$P(E|S) \gg P(E|N)$$ in order to have a decent $S/N$ ratio.\\
(How does the $S/N$ change if the particle has to be identified by
two independent detectors in order to give the trigger?
Try it yourself, the answer is $S/N=251$.)
\item[Problem 2:]
Three boxes contain two rings each, but in one of
them they are both gold, in the second both silver,
and in the third one of each type. You have the choice of
randomly  extracting
a ring from one of the boxes, the content of which
is unknown to you. You look at the selected ring,
and you then have the possibility of extracting a second ring,
again from
any of the three boxes. Let us assume
the first ring you extract is a gold one.
Is it then preferable to extract the second one
from the same or from a different box?
\item[Solution:] Choosing the same box you have a $2/3$ probability
                 of getting a second gold ring.
                 (Try to apply the theorem,
                 or  help yourself with intuition.)
\end{description}
The difference between the two problems, from the conventional
statistics point of view, is that the first
is only meaningful
in the frequentistic approach, the second only in the
combinatorial one. They are, however, both acceptable
from the Bayesian point of view. This is simply
because in this framework there is no
restriction on the definition of probability.
In many and important cases of life and science,
 neither of the two
conventional definitions are applicable.

\subsection{Bayesian statistics: learning by experience}
The advantage of the Bayesian approach
(leaving aside the ``little philosophical detail''
of trying to define what probability is) is that one
may talk about the probability of
\underline{any} kind of event, as already
 emphasized.
Moreover, the procedure of updating the probability
with increasing information is very similar
to that followed by the
mental processes of rational people. Let us consider
a few examples of ``Bayesian use'' of Bayes' theorem:
\begin{description}
\item[Example 1:] Imagine some persons
listening to a common friend
having a phone conversation with an unknown person $X_i$,
and who
are trying to guess who $X_i$ is. Depending on the knowledge
they have about the friend, on the  language spoken,
on the tone of voice, on the subject of conversation, etc.,
they will attribute some probability to several
possible persons. As the conversation goes on they begin
to consider some possible candidates for $X_i$, discarding others,
and eventually fluctuating between two possibilities,
until
the status of information $I$ is such that they are
{\it practically sure} of the identity of $X_i$. This experience
has happened to must of us, and it is not difficult to
recognize the Bayesian scheme:
\begin{equation}
P(X_i|I,I_\circ) \propto P(I|X_i,I_\circ)P(X_i|I_\circ)\,.
\label{eq:telefonata}
\end{equation}
We have put the initial status of information
$I_\circ$ explicitly in (\ref{eq:telefonata})
to remind us that likelihoods and initial probabilities
depend on it. If we know nothing about the person, the final
probabilities will be very {\it vague}, i.e.
for many persons $X_i$ the probability  will
be different from zero,  without necessarily
favoring any particular person.
\item[Example 2:] A person $X$ meets an old friend $F$ in a pub.
$F$ proposes
that the drinks should be payed for by
whichever
of the two extracts
the card of lower value
from a pack
(according to some rule which is of no
interest to us). $X$ accepts and $F$
wins. This situation happens again in the following days
and it is always $X$ who has to pay.
What is the probability that $F$ has become a cheat, as
the number of consecutive wins $n$ increases?

The two hypotheses are: {\it cheat} ($C$) and {\it honest} ($H$).
$P_\circ(C)$ is low because $F$ is an ``old friend'',
but certainly not zero (you know $\ldots$): let us \underline{assume}
$5\,\%$. To make the problem simpler let us make the approximation
that a cheat always wins (not very clever$\ldots$):
$P(W_n|C)=1)$. The probability of winning if he is honest is, instead,
given by the rules of probability {\it assuming} that
the chance
of winning at each trial is $1/2$ ("why not?", we shall
come back to this point later): $P(W_n|H)=2^{-n}$. The result
\begin{eqnarray}
P(C|W_n) & = & \frac{P(W_n|C)\cdot P_\circ(C)}
                    {P(W_n|C)\cdot P_\circ(C) + P(W_n|H)\cdot P_\circ(H)}
\nonumber \\
 & = &  \frac{1\cdot P_\circ(C)}
             {1\cdot P_\circ(C) + 2^{-n} \cdot P_\circ(H)}
\label{eq:baro}
\end{eqnarray}
is shown in the following table:
\begin{center}
\vspace{0.5 cm}
\begin{tabular}{|c|c|c|}\hline
$n$ & $ P(C|W_n)$ &  $P(H|W_n)$ \\
& (\%) & (\%) \\ \hline
0  &  5.0 & 95.0 \\
1  &  9.5  & 90.5\\
2  &  17.4 & 82.6\\
3  &  29.4 & 70.6\\
4  &  45.7 & 54.3\\
5  &  62.7 & 37.3\\
6  &  77.1 & 22.9\\
$\ldots$  & $\ldots$ & $\ldots$ \\ \hline
\end{tabular}
\vspace{0.5 cm}
\end{center}
Naturally, as $F$ continues to win the suspicion
of $X$ increases. It is important to make two remarks:
\begin{itemize}
 \item
 the answer is always probabilistic. $X$ can never reach
  absolute certainty that $F$ is a cheat,
 unless he catches $F$ cheating, or $F$
  confesses to having cheated. This is coherent
 with the fact that we are dealing with random events
 and with the fact that any sequence of outcomes has the
 same probability (although there is only one possibility over
 $2^n$ in which $F$ is \underline{always} luckier). Making \underline{use}
 of $P(C|W_n)$, $X$ can take a \underline{decision} about the
 next action to take:
 \begin{itemize}
 \item
 \underline{continue} the game, with
 probability $P(C|W_n)$
 of \underline{losing}, with certainty, the next time too;
 \item
 \underline{refuse} to play further, with probability $P(H|W_n)$
 of \underline{offending} the innocent friend.
 \end{itemize}
\item
If $P_\circ(C)=0$ the final probability will
always remain zero: if $X$ fully trusts $F$,
then he has just to
record the occurrence of a rare event when $n$ becomes large.
\end{itemize}

To better follow the process of updating the probability
when new experimental data become available,
according to the Bayesian scheme
\begin{quote}
  {\it ``the final probability of the
  present  inference is the initial probability
  of the next one''}\,,
\end{quote}
let us call $P(C|W_{n-1})$ the probability assigned
after the previous win. The iterative application
of the Bayes formula yields:
\begin{eqnarray}
P(C|W_n) &=& \frac{P(W|C)\cdot P(C|W_{n-1})}
                    {P(W|C)\cdot P(C|W_{n-1}) +
                    P(W|H)\cdot P(H|W_{n-1})} \nonumber \\
 & = &  \frac{1\cdot P(C|W_{n-1})}
         {1\cdot P(C|W_{n-1}) + \frac{1}{2} \cdot P(H|W_{n-1})}\,,
\end{eqnarray}
where $P(W|C)=1$ and $P(W|H)=1/2$ are the probabilities of
\underline{each} win.
The interesting result is that
\underline{exactly} the same values of $P(C|W_n)$ of (\ref{eq:baro})
are obtained (try to believe it!).
\end{description}

 It is also instructive to see the dependence of the final
 probability on the initial probabilities, for a given
 number of wins $n$:
\begin{center}
\vspace{0.5 cm}
\begin{tabular}{|c|c|c|c|c|}\hline
             & \multicolumn{4}{|c|}{$ P(C|W_n)$} \\
$P_\circ(C)$ & \multicolumn{4}{|c|}{ $(\%)$} \\ \hline
             & $n=5$  & $n=10$  &$n=15$  & $n=20$  \\ \hline
 $1\,\%$     & 24     &   91    &  99.7  &  99.99  \\
 $5\,\%$     & 63     &   98    &  99.94 &  99.998 \\
 $50\,\%$    & 97     &   99.90 &  99.997&  99.9999 \\ \hline
\end{tabular}
\vspace{0.5 cm}
\end{center}
As the number of experimental observations increases the conclusions
no longer depend, practically,
on the initial assumptions. This is a crucial
point in the Bayesian scheme and it will be discussed in more detail
later.

\section{Hypothesis test (discrete case)}\label{sec:hyp_test_discr}
Although in conventional statistics books this
argument is usually dealt with in one of the
later chapters, in the Bayesian
approach this is so natural that it is in fact
the first application, as we have seen in the
above examples. We summarize here the procedure:
\begin{itemize}
\item
{\it probabilities are attributed to the different
hypotheses} using  initial probabilities and
experimental data (via the likelihood);
\item
the person who makes the inference
- or the ``user'' -  will take a decision
of which he is \underline{fully responsible}.
\end{itemize}
If one needs to compare
two hypotheses, as in the example of the signal to noise
calculation, the ratio of the final probabilities
can be taken as a quantitative result of the test.
Let us rewrite the $S/N$ formula in the most general case:
\begin{equation}
\frac{P(H_1|E,H_\circ)}{P(H_2|E,H_\circ)}
= \frac{P(E|H_1, H_\circ)}{P(E|H_2,H_\circ)} \cdot
  \frac{ P(H_1|H_\circ)}{P(H_2|H_\circ)}\,,
\label{eq:bayes_factor}
\end{equation}
where again we have reminded ourselves
of the existence of $H_\circ$.
The ratio depends on the
product of two terms: the ratio of the priors
and the ratio of the likelihoods. When there is absolutely
no reason for  choosing between the two hypotheses the
prior ratio is 1 and the decision depends only on the
other term, called {\it the Bayes factor}.
If one firmly believes in either hypothesis,
 the Bayes
factor is of minor importance, unless it is zero or infinite
(i.e. one and only one of the likelihoods is vanishing).
Perhaps this is disappointing for those who expected
objective certainties from a probability theory, but
\underline{this} is in the nature of things.

\section{Choice of the initial probabilities
(discrete case)}\label{sec:choice1}
\subsection{General criteria}
\begin{figure}
\centering\epsfig{file=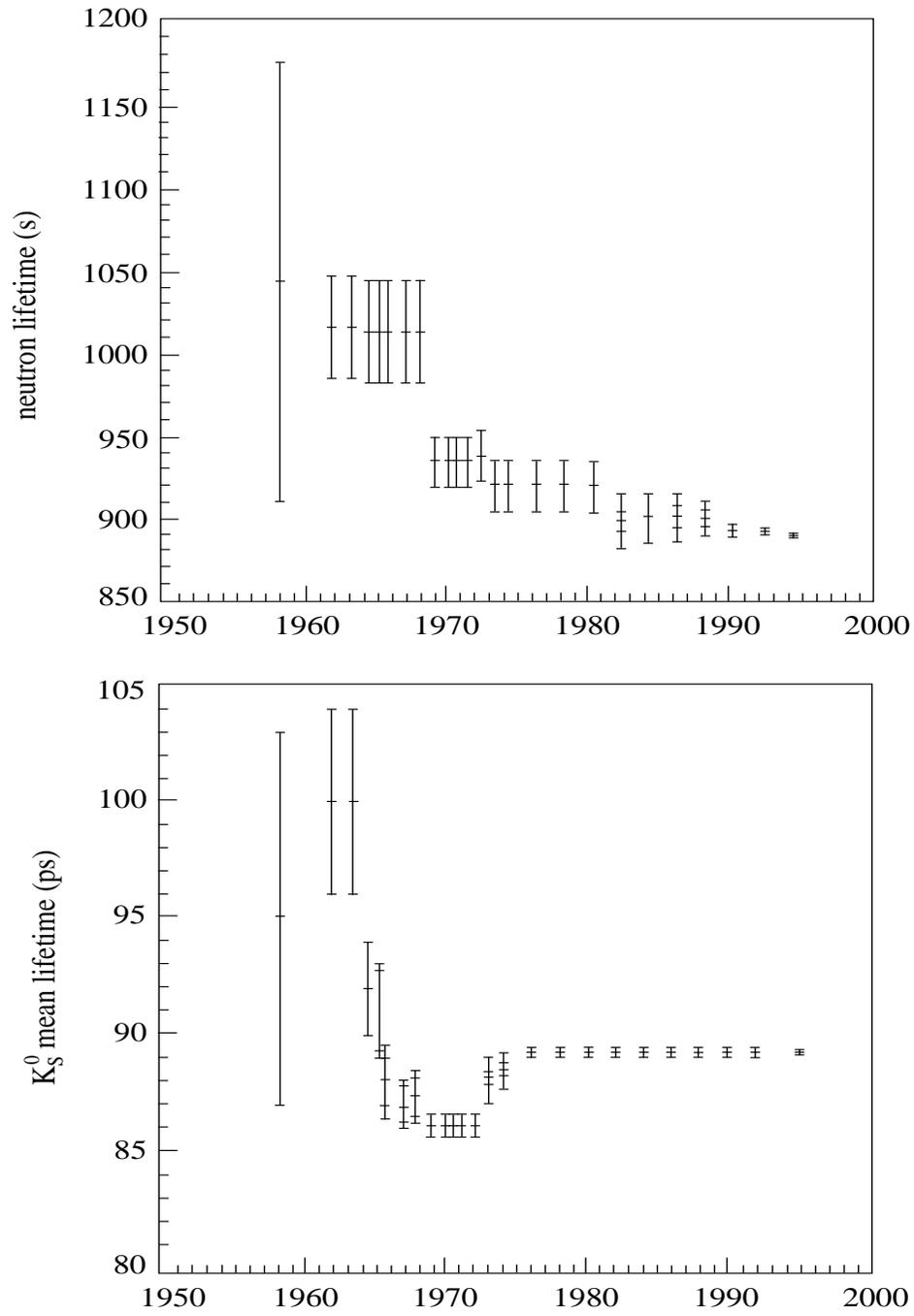,width=12.5cm,height=18cm,clip=}
\caption{\sf Results on two physical quantities as a function
of the publication date.}
\label{fig:sistematiche}
\end{figure}

\begin{figure}
\centering\epsfig{file=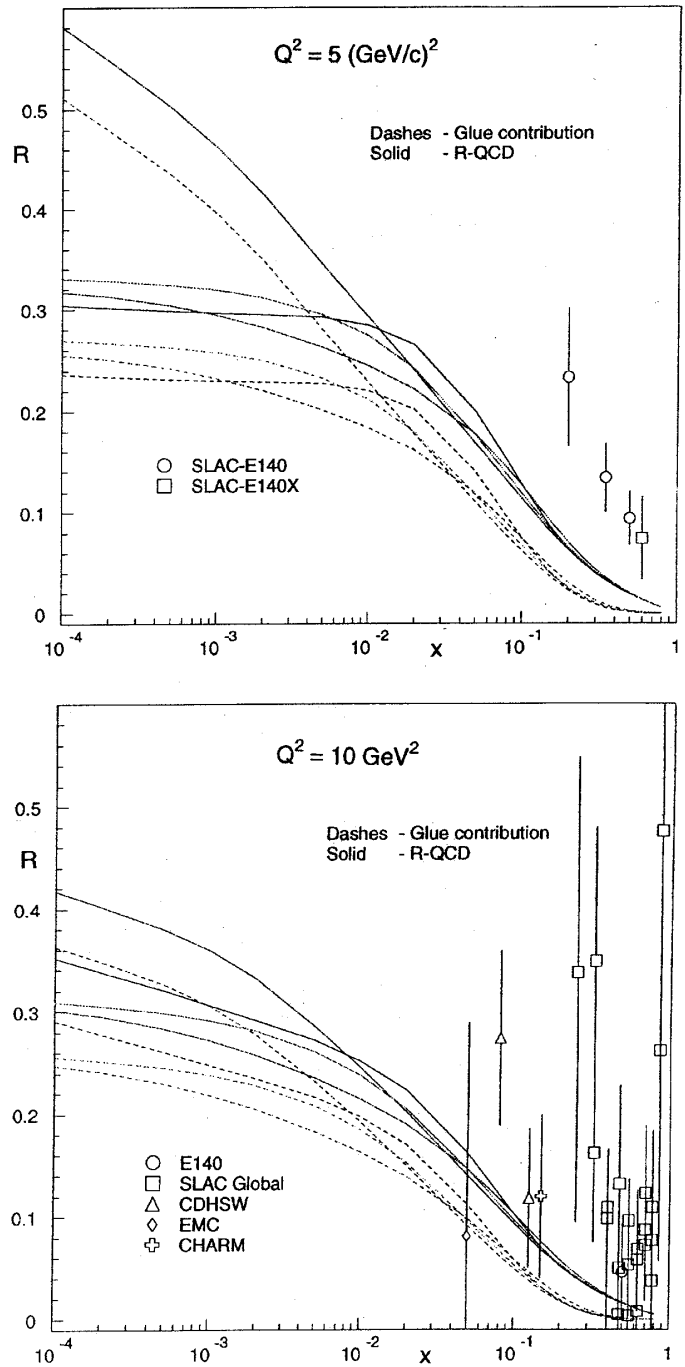,height=19cm,clip=}
\caption{\sf $R=\sigma_L/\sigma_T$ as a function of the Deep Inelastic
Scattering variable $x$ as measured by experiments and as predicted by
QCD.}
\label{fig:RDIS}
\end{figure}

The dependence of
Bayesian inferences on initial probability is
pointed to by  opponents as
the fatal flaw in the theory.
But this criticism is less severe than one might think
at first sight. In fact:
\begin{itemize}
\item
It is impossible to construct a theory
of uncertainty which is not affected by this
``illness''. Those methods which are advertised as being
``objective'' tend in reality  to hide the hypotheses on
which they are grounded.
 A typical example is
the maximum likelihood method, of which we
will talk later.
\item
as the amount of information increases
the dependence on initial prejudices diminishes;
\item
when the amount of information is  very limited,
or completely lacking, there is nothing to be ashamed of if
the inference is dominated by  {\it a priori} assumptions;
\end{itemize}
The fact that
conclusions drawn from an experimental result
(and sometimes even the ``result'' itself!)
\underline{often} depend
on prejudices about the phenomenon under study is well known
to all experienced physicists. Some examples:
\begin{itemize}
\item
when doing quick checks on  a device, a single
measurement is usually performed if the value is
 ``what it should be'', but if it is not then
many measurements tend to be made;
\item
results are sometimes influenced by
 previous results or by theoretical
predictions. See for example Fig. ~\ref{fig:sistematiche} taken from
the Particle Data Book\cite{PDG}.
The interesting  book {\it ``How experiments end''}\cite{end}
discusses, among others, the issue
of \underline{when}
experimentalists are ``happy with the result'' and stop
``correcting for the systematics'';
\item
it can happen that slight deviations from the background
are interpreted as a signal
(e.g. as for the first claim of discovery of
the Top quark in spring '94),
while larger ``signals'' are viewed with suspicion if they
are unwanted by the physics ``establishment''\footnote{A case,
concerning
the search for electron compositeness in $e^+e^-$ collisions,
is discussed in \cite{comp}.};
\item
experiments are planned and financed according to the
prejudices of the moment\footnote{For a recent delightful report,
see \cite{wroblewski}.};
\end{itemize}
These comments are not intended to justify unscrupulous behaviour
or sloppy analysis. They are intended, instead, to remind us
- if need be - that scientific research is ruled by
subjectivity much more than
outsiders imagine. The transition from subjectivity
to ``objectivity'' begins when there
is a large consensus among the most influential people about
how to  interpret  the results\footnote{{\sl ``A theory
needs to be confirmed by experiments. But it is
also true that an experimental result needs to be} {\it confirmed
by a theory''}. This sentence
expresses clearly - though paradoxically -
the idea that it is difficult to accept a result which is
not rationally justified. An example of results ``not confirmed by
the theory'' are the $R$ measurements in Deep Inelastic Scattering
shown in Fig.~\ref{fig:RDIS}. Given the  conflict
in this  situation,
physicists tend to believe more in QCD and use the
``low-x'' extrapolations (\underline{of what?})
to correct the data for the
unknown values of $R$.}.

In this context, the subjective approach to  statistical
inference at least teaches us that every assumption must be
\underline{stated clearly}
and all available
information which could influence conclusions
must be weighed
with the maximum \underline{attempt at objectivity}.

What are the rules for choosing the ``right''
initial probabilities?
As one can imagine, this is an open and
debated question
among scientists and  philosophers.
My personal point of view is that
one should avoid pedantic discussion of the matter,
because the idea of universally true  priors
reminds me terribly  of the famous ``angels' sex'' debates.

If I had to give recommendations, they would be:
\begin{itemize}
\item
the {\it a priori} probability should be chosen in the same
spirit as the rational person  who places a bet,
seeking to minimize the risk
of losing;
\item
general principles - like those that we will discuss in a while -
may help, but since it may be difficult to apply
elegant theoretical ideas
in all practical situations,
in many circumstances the {\it guess} of the ``expert''
can be relied on for guidance.
\item
avoid using as prior the results of other experiments
dealing with the same open problem, otherwise correlations
between the results would prevent all comparison between the experiments
and thus the detection of any
systematic errors. I find that this point is
generally overlooked by statisticians.
\end{itemize}

\subsection{Insufficient Reason and Maximum Entropy}
The first and most famous criterion for choosing
 initial probabilities is the simple
{\it Principle of Insufficient Reason}
(or {\it Indifference Principle}): if there is no reason
to prefer one hypothesis over alternatives, simply attribute
the same probability to all of them. This
 was stated as a principle
by Laplace\footnote{It may help in understanding
  Laplace's approach
if we consider that he called the theory of probability ``good sense
turned into calculation''.}
in contrast to
Leibnitz' famous {\it Principle of Sufficient Reason}, which, in simple
words, states that "nothing happens without a reason".
The indifference principle applied to coin and die tossing,
to card games or to other simple and symmetric
problems, yields to the well known rule of probability
evaluation that we have called combinatorial.
Since it is impossible not to agree with this point of
view, in the cases that \underline{one judges} that it does apply,
the combinatorial ``definition'' of probability is
recovered in the Bayesian approach if the
word ``definition'' is simply replaced by ``evaluation rule''.
We have in fact already used this reasoning
in previous examples.

A modern and more sophisticated version of the Indifference Principle
is the Maximum Entropy Principle. The information entropy
function of $n$
mutually exclusive events, to each of which a probability $p_i$
is assigned, is defined as
\begin{equation}
H(p_1, p_2,\ldots p_n) = - K\sum_{i=1}^np_i\ln{p_i},
\end{equation}
with $K$ a positive constant. The principle states that
``in making inferences on the basis of partial
information we must use that probability distribution which
has the maximum entropy subject to whatever is known\cite{Jaynes}''.
Notice that, in this case, ``entropy'' is synonymous with
``uncertainty''\cite{Jaynes}.
One can show that, in the case of \underline{absolute}
ignorance about the events $E_i$, the maximization of the
information uncertainty, with the constraint that $\sum_{i=1}^np_i=1$,
yields the classical
$p_i=1/n$ (any other result would have been worrying$\ldots$).

Although this principle is sometimes used in combination
with the Bayes' formula for inferences
(also applied to measurement uncertainty, see
\cite{Weise2}), it will not be
used for applications
in these notes.

\section{Random variables}\label{sec:variables}
In the discussion which follows I will assume that the reader is
familiar with random variables, distributions, probability
density functions, expected values, as well as with the
most frequently used distributions. This section is
only intended as a summary of concepts and as a presentation of the
notation used in the subsequent sections.

\subsection{Discrete variables}
Stated  simply, to define a {\it random variable} $X$
means to find a rule which allows a real number
to be related univocally
(but not biunivocally)
to an event ($E$), chosen from those events which
constitute a finite partition of $\Omega$ (i.e. the events
must be exhaustive and mutually exclusive).
One could write this expression
$X(E)$. If the number of possible events is finite then the
random variable is discrete, i.e. it can assume only a
finite number of values.
Since the chosen set of events are mutually exclusive,
the probability of $X=x$ is the sum of the probabilities of all
the events for which $X(E_i)=x$. Note that we shall indicate
the variable
with $X$
and
its numerical realization
with $x$,
and  that, differently from
 other notations, the symbol $x$
(in place of $n$ or $k$) is also used for discrete variables.

After this short introduction, here is a list of
definitions, properties and notations:
\begin{description}
\item[Probability function:]
\begin{equation}
 f(x)=P(X=x)\,.
\end{equation}
It has the following properties:
\begin{eqnarray}
1) & &  0 \leq f(x_i) \leq 1\,;\\
2) & & P(X = x_i\,\cup\, X = x_j)= f(x_i)+f(x_j)\,;\\
3) & & \sum_i f(x_i) = 1\,.
\end{eqnarray}
\item[Cumulative distribution function:]
\begin{equation}
F(x_k) \equiv P(X\leq x_k) =  \sum_{x_i\leq x_k} f(x_i)
\, .
\end{equation}
Properties:
\begin{eqnarray}
1) &  &  F(-\infty) = 0\\
2) &  &  F(+\infty) = 1\\
3) &  & F(x_i) - F(x_{i-1}) = f(x_i)\\
4) &  & \lim_{\epsilon \rightarrow o} F(x+\epsilon) = F(x)
        \hspace{1.0 cm} (right\ side\   continuity)\,.
\end{eqnarray}
\item[Expected value (mean):]
\begin{equation}
\mu \equiv E[X] = \sum_i x_i f(x_i)\,.\label{eq:media}
\end{equation}
In general, given a function $g(X)$ of $X$:
\begin{equation}
E[g(X)] = \sum_i g(x_i) f(x_i)\,.
\end{equation}
$E[\cdot]$ is a linear operator:
\begin{equation}
E[a X+b] =  a E[X] + b \,.
\end{equation}
\item[Variance and standard deviation:]
Variance:
\begin{equation}
\sigma ^2 \equiv Var(X) = E[(X-\mu)^2] =  E[X^2] - \mu ^2 \,.
\label{eq:varianza}
\end{equation}
Standard deviation:
\begin{equation}
 \sigma = +\sqrt{\sigma^2}\,.
\end{equation}
Transformation properties:
\begin{eqnarray}
Var(a X+b) & = & a^2 Var(X)\,;\\
\sigma(aX+b) & = & |a|\sigma(X)\,.
\end{eqnarray}
\item[Binomial distribution:]
$X\sim {\cal B}_{n,p}$ (hereafter ``$\sim$'' stands for ``follows'');
${\cal B}_{n,p}$ stands for {\it binomial} with parameters
$n$ (integer) and $p$ (real):
\begin{equation}
f(x|{\cal B}_{n,p}) =
\frac{n!}{(n-x)!x!} p^x (1-p)^{n-x} \, ,
\hspace{1.0 cm}
\left\{ \begin{array}{l}   n = 1, 2, \ldots,  \infty \\
                           0 \le p \le 1 \\
                           x = 0, 1, \ldots, n \end{array}\right.\,.
\label{eq:binomial}
\end{equation}
Expected value, standard deviation and {\it variation coefficient}:
\begin{eqnarray}
\mu &  = &  np \\
\sigma   & = & \sqrt{np (1-p)}\\
v \equiv \frac{\sigma}{\mu} &=&
\frac{\sqrt{np(1-p)}}{n p} \propto \frac{1}{\sqrt{n}}\, .
\end{eqnarray}
$1-p$ is often indicated by $q$.
\item[Poisson distribution:]
$X\sim {\cal P}_\lambda$:
\begin{equation}
f(x|{\cal P}_\lambda)=\frac{\lambda^x}{x!} e^{-\lambda}
\hspace{1.0 cm}
\left\{ \begin{array}{l}   0 < \lambda < \infty \\
                           x = 0, 1, \ldots,  \infty\\
                           \end{array} \right.\,.
\end{equation}
($x$ is integer, $\lambda$ is real.)\\
Expected value, standard deviation and variation coefficient:
\begin{eqnarray}
\mu  & = &  \lambda \\
\sigma  & = &   \sqrt{\lambda} \\
v &=& \frac{1}{\sqrt{\lambda}}
\end{eqnarray}
\item[Binomial $\rightarrow$ Poisson:]
\begin{equation}
{\cal B}_{n,p} @>>{\begin{array}{l} n\rightarrow ``\infty'' \\
                   p\rightarrow ``0''       \\
                   (\lambda = np) \end{array}}>{\cal P}_\lambda\,.
\end{equation}
\end{description}

\subsection{Continuous variables: probability and
density  function}\label{sec:cont_var}
Moving from discrete to continuous variables there are the
usual problems with infinite possibilities,
similar to those found in
Zeno's  ``Achilles and the tortoise'' paradox.
In both cases
the answer is given by infinitesimal
calculus. But some comments are needed:
\begin{itemize}
\item
 the probability of
each of the realizations of $X$ is zero ($P(X=x)=0$),
but this does \underline{not} mean that each value
is \underline{impossible}, otherwise it would be impossible
 to get \underline{any} result;
\item
although all values $x$ have zero probability, one usually
assigns different
degrees of belief to them, quantified by the
{\bf probability density function} $f(x)$. Writing
$f(x_1) > f(x_2)$,
for example,
indicates that our degree of belief in
$x_1$ is greater than that in $x_2$.
\item
The probability that  a random variable
lies inside a finite interval, for example
$P(a\leq X \leq b)$,
is instead finite.
 If the distance between $a$ and $b$ becomes
infinitesimal, then  the probability becomes infinitesimal too.
If all the values of $X$ have the same degree of belief
(and not only equal numerical
probability $P(x)=0$) the infinitesimal
probability is simply proportional to the infinitesimal
interval $dP=kdx$. In the general case the ratio between
two infinitesimal probabilities around two different points
will be equal to the ratio of the degrees of belief in the
points (this argument implies the continuity of $f(x)$
on either side of the values). It follows that $dP=f(x)dx$
and then
\begin{equation}
P(a \leq X \leq b)= \int_{a}^{b}f(x)dx\,;
\end{equation}
\item
$f(x)$ has a dimension inverse to that of the random variable.
\end{itemize}
After this short introduction, here is a list of
definitions, properties and notations:
\begin{description}
\item[Cumulative distribution function:]
\begin{equation}
F(x) = P(X\leq x) = \int_{-\infty}^{x}f(x^\prime)dx^\prime\,,
\end{equation}
or
\begin{equation}
f(x) = \frac{dF(x)}{dx}
\end{equation}
\item[Properties of $f(x)$ and $F(x)$:]\
\begin{itemize}
\item
$f(x) \geq 0\,\,$;
\item
$\int_{-\infty}^{+\infty} f(x)dx=1\,\,$;
\item
$0\leq F(x)\leq 1$;
\item
$P(a\leq X \leq b) = \int_a^bf(x)dx = \int_{-\infty}^bf(x)dx
  -\int_{-\infty}^af(x)dx\\
  \hspace{2.64cm} = F(b)-F(a)$;
\item
if $x_2 > x_1$ then
$F(x_2) \ge F(x_1)$\,.
\item
$ \lim_{x\rightarrow -\infty} F(x) = 0$\,;\\
$ \lim_{x\rightarrow +\infty} F(x) = 1$\,;
\end{itemize}
\item[Expected value:]
\begin{eqnarray}
 E[X] & = & \int_{-\infty}^{+\infty}x f(x)dx\\
 E[g(X)] & = & \int_{-\infty}^{+\infty}g(x) f(x)dx.
\end{eqnarray}
\item[Uniform distribution:]\!\footnote{
The symbols of the following distributions
have the parameters within parentheses to indicate that
the variables are continuous.}
$X \sim {\cal K}(a,b)$:
\begin{eqnarray}
 f(x|{\cal K}(a,b)) & = & \frac{1}{b-a}
 \hspace{0.6cm}(a\le x \le b)\\
  F(x|{\cal K}(a,b)) & = &
\frac{x-a}{b-a}\, .
\end{eqnarray}
Expected value and standard deviation:
\begin{eqnarray}
\mu&=& \frac{a+b}{2} \\
\sigma &=&\frac{b-a}{\sqrt{12}}\,.
\end{eqnarray}
\item[Normal (gaussian) distribution:] $X\sim {\cal N}(\mu,\sigma)$:
\begin{equation}
f(x|{\cal N}(\mu,\sigma))
=\frac{1}{\sqrt{2\pi}\sigma}e^{-\frac{(x-\mu)^2}{2\sigma^2}}
\hspace{1.0 cm}
\left\{ \begin{array}{l}
                           -\infty < \mu < +\infty\\
                           0 < \sigma < \infty\\
                           -\infty < x < +\infty
         \end{array}\right.\,,
\end{equation}
where $\mu$ and $\sigma$ (both real) are the expected value and standard
deviation, respectively.
\item[Standard normal distribution:]
the particular normal distribution of mean 0 and standard
deviation 1, usually indicated by $Z$:
\begin{equation}
Z\sim {\cal N}(0,1)\,.
\end{equation}
\item[Exponential distribution:]
$T \sim {\cal E}(\tau)$:
\begin{eqnarray}
f(t|{\cal E}(\tau)) & = &
 \frac{1}{\tau} e^{-t/\tau} \hspace{1.3 cm}
  \left\{ \begin{array}{c}
   0 \le \tau < \infty \\
   0 \le t < \infty
  \end{array}\right. \\
 F(t|{\cal E}(\tau)) & = & 1-e^{-t/\tau}
\end{eqnarray}
we use of the symbol $t$ instead of $x$  because this distribution
will be applied to the {\it time domain}.\\
{\it Survival probability}:
\begin{equation}
P(T>t) = 1-  F(t|{\cal E}(\tau)) = e^{-t/\tau}
\end{equation}
Expected value and standard deviation:
\begin{eqnarray}
\mu & = & \tau\\
\sigma & = & \tau.
\end{eqnarray}
The real parameter $\tau$ has the physical meaning of {\it lifetime}.
\item[Poisson $\leftrightarrow$ Exponential:]
If $X$ (= ``number of counts during the time $\Delta t$'') is
Poisson distributed then $T$ (=``interval of time to be waited -
starting \underline{from any instant} - before the first count
is recorded'') is exponentially distributed:
\begin{eqnarray}
  X \sim  f(x|{\cal P}_\lambda)
  & \Longleftrightarrow &
  T  \sim  f(x|{\cal E}(\tau)) \\
 &(\tau = \frac{\Delta T}{\lambda})&
\end{eqnarray}
\end{description}
\subsection{Distribution of several random variables}
We  only consider the case of two continuous variables ($X$ and $Y$).
The extension to more variables is straightforward.
The infinitesimal element of probability is
$dF(x,y) = f(x,y)dxdy$, and the probability
density function
\begin{equation}
f(x,y) = \frac{\partial^2F(x,y)}{\partial x\partial y}\,.
\end{equation}
The probability of finding the variable inside a certain
area $A$ is
\begin{equation}
\iint\limits_A f(x,y)dx dy\,.
\end{equation}
\begin{description}
\item[Marginal distributions:]
\begin{eqnarray}
f_X(x) & = & \int_{-\infty}^{+\infty}f(x,y)dy \\
f_Y(y) & = & \int_{-\infty}^{+\infty}f(x,y)dx \,.
\end{eqnarray}
The subscripts $X$ and $Y$ indicate that
$f_X(x)$ and $f_Y(y)$
 are only function of
$X$ and $Y$, respectively (to avoid  fooling around with different
symbols to indicate the generic function).
\item[Conditional distributions:]
\begin{eqnarray}
 f_X(x|y) & = & \frac{f(x,y)}{f_Y(y)} = \frac{f(x,y)}{\int f(x,y)dx} \\
 f_Y(y|x) & = & \frac{f(x,y)}{f_X(x)} \\
 f(x,y)   & = &  f_X(x|y)f_Y(y) \\
          & = &  f_Y(y|x)f_Y(x)\,.
\end{eqnarray}
\item[Independent random variables]
\begin{equation}
f(x,y) = f_X(x) f_Y(y)
\end{equation}
(it implies $f(x|y)=f_X(x)$ and $f(y|x)=f_Y(y)$\,.)
\item[Bayes' theorem for continuous random variables]
\begin{equation}
\boxed{
f(h|e) = \frac{f(e|h)f_h(h)}
                   {\int f(e|h)f_h(h)dh}\, .
}
\label{eq:bayes_cont}
\end{equation}
\item[Expected value:]
\begin{eqnarray}
\mu_X=E[X] & = & \int\!\!\int_{-\infty}^{+\infty}
                 \!x f(x,y)dxdy \\
           & = & \int_{-\infty}^{+\infty}\!x f_X(x) dx\,,
\end{eqnarray}
and analogously for $Y$. In general
\begin{equation}
E[g(X,Y)] = \int\!\!\int_{-\infty}^{+\infty}
            \!g(x,y) f(x,y) dxdy\,.
\end{equation}
\item[Variance:]
\begin{equation}
\sigma_X^2=E[X^2]-E^2[X]\,,
\end{equation}
and analogously for $Y$.
\item[Covariance:]
\begin{eqnarray}
 Cov(X,Y) & = & E\left[\left(X-E[X]\right) \left(Y-E[Y]\right)\right]\\
& = & E[X Y]-E[X] E[Y]\,.
\end{eqnarray}
If $Y$ and $Y$ are independent, then  $ E[XY]=E[X] E[Y]$
and hence $Cov(X,Y) =0$ (the opposite is true only if $X$,
$Y\sim {\cal N}(\cdot)$).
\item[Correlation coefficient:]
\begin{eqnarray}
\rho(X,Y)&=&\frac{Cov(X,Y)}{\sqrt{Var(X) Var(Y)}}\\
         &=& \frac{Cov(X,Y)}{\sigma_X \sigma_Y}\, .
\end{eqnarray}
$$( -1 \le \rho \le 1)$$
\item[Linear combinations of random variables:]\ \\
If $Y=\sum_i c_iX_i$, with $c_i$ real, then:
\begin{eqnarray}
\mu_Y=E[Y] & = & \sum_ic_i E[X_i] = \sum_ic_i\mu_i
\label{eq:linc1} \\
\sigma_Y^2=Var(Y)& = &
 \sum_i c_i^2Var(X_i) + 2\sum_{i< j}c_ic_jCov(X_i,X_j)
\label{eq:linc2} \\
 & = &
 \sum_i c_i^2Var(X_i) + \sum_{i\ne j}c_ic_jCov(X_i,X_j)
\label{eq:linc3} \\
 & = & \sum_i c_i^2\sigma_i^2
       +\sum_{i\ne j}\rho_{ij}c_ic_j\sigma_i\sigma_j
\label{eq:linc4} \\
 & = & \sum_{ij}\rho_{ij}c_ic_j\sigma_i\sigma_j
\label{eq:linc5} \\
 & = & \sum_{ij}c_ic_j\sigma_{ij}
\label{eq:linc6} \,.
\end{eqnarray}
$\sigma^2_Y$ has been written in the different ways, with
 increasing levels of compactness, that can be found
 in the literature. In particular, (\ref{eq:linc6}) uses the convention
 $\sigma_{ii}=\sigma^2_i$ and the fact that,
 by definition,  $\rho_{ii}=1$.
\item[Bivariate normal distribution:] joint probability density function
of $X$ and $Y$ with correlation coefficient $\rho$
(see Fig ~\ref{fig:bivar}):
\begin{figure}
\centering\epsfig{file=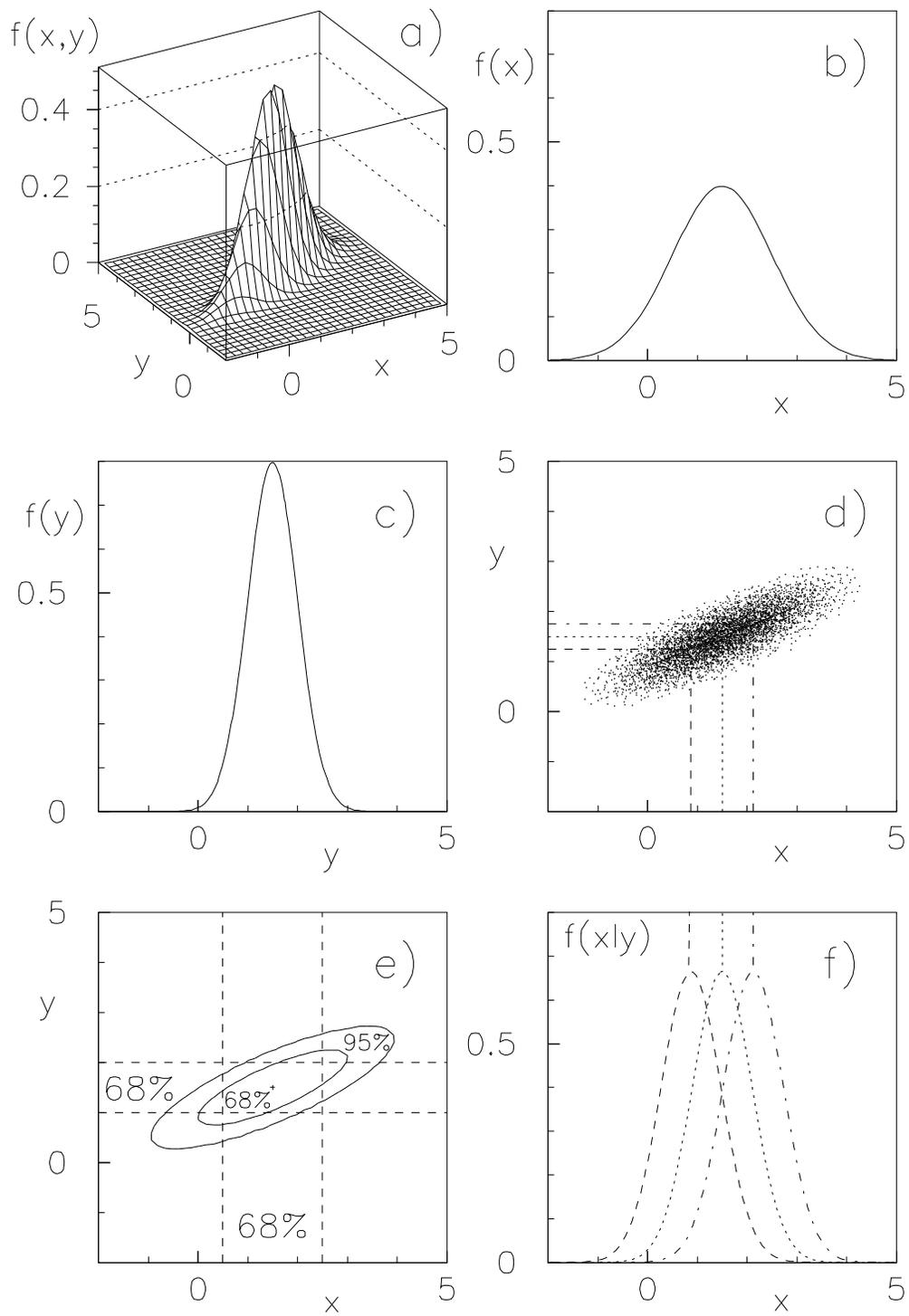,width=\linewidth,clip=}
\caption{\sf Example of bivariate normal distribution.}
\label{fig:bivar}
\end{figure}
\begin{eqnarray}
f(x,y) &=&
\frac{1}{2\pi\sigma_x\sigma_y\sqrt{1-\rho^2}}\cdot  \\
&&  \exp{\left\{
                -\frac{1}{2(1-\rho^2)}
                 \left[  \frac{(x-\mu_x)^2}{\sigma_x^2}
     - 2\rho\frac{(x-\mu_x)(y-\mu_y)}{\sigma_x\sigma_y}
        + \frac{(y-\mu_y)^2}{\sigma_y^2}
                 \right]
         \right\}}\,. \nonumber
\label{eq:bivar}
\end{eqnarray}
Marginal distributions:
\begin{eqnarray}
X &\sim & {\cal N}(\mu_x,\sigma_x) \\
Y &\sim & {\cal N}(\mu_y,\sigma_y) \,.
\end{eqnarray}
Conditional distribution:
\begin{equation}
f(y|x_\circ) = \frac{1}{\sqrt{2\pi}\sigma_y\sqrt{1-\rho^2}}
\exp{\left[
-\frac{\left(y-\left[\mu_y+\rho\frac{\sigma_y}{\sigma_x}
\left(x_\circ-\mu_x\right)\right]
\right)^2}
{2\sigma_y^2(1-\rho^2)}
\right]}\,,
\label{eq:y_cond}
\end{equation}
i.e.
\begin{equation}
Y_{|x_\circ}\sim {\cal N}\left( \mu_y+\rho\frac{\sigma_y}{\sigma_x}
                  \left(x_\circ-\mu_x\right),\,
                 \sigma_y\sqrt{1-\rho^2}\right):
\label{eq:y_cond1}
\end{equation}
the condition $X=x_\circ$ squeezes the standard deviation and shifts
 the mean of $Y$.
\end{description}
\section{Central limit theorem}\label{sec:clim}
\subsection{Terms and role}
\begin{figure}
\centering\epsfig{file=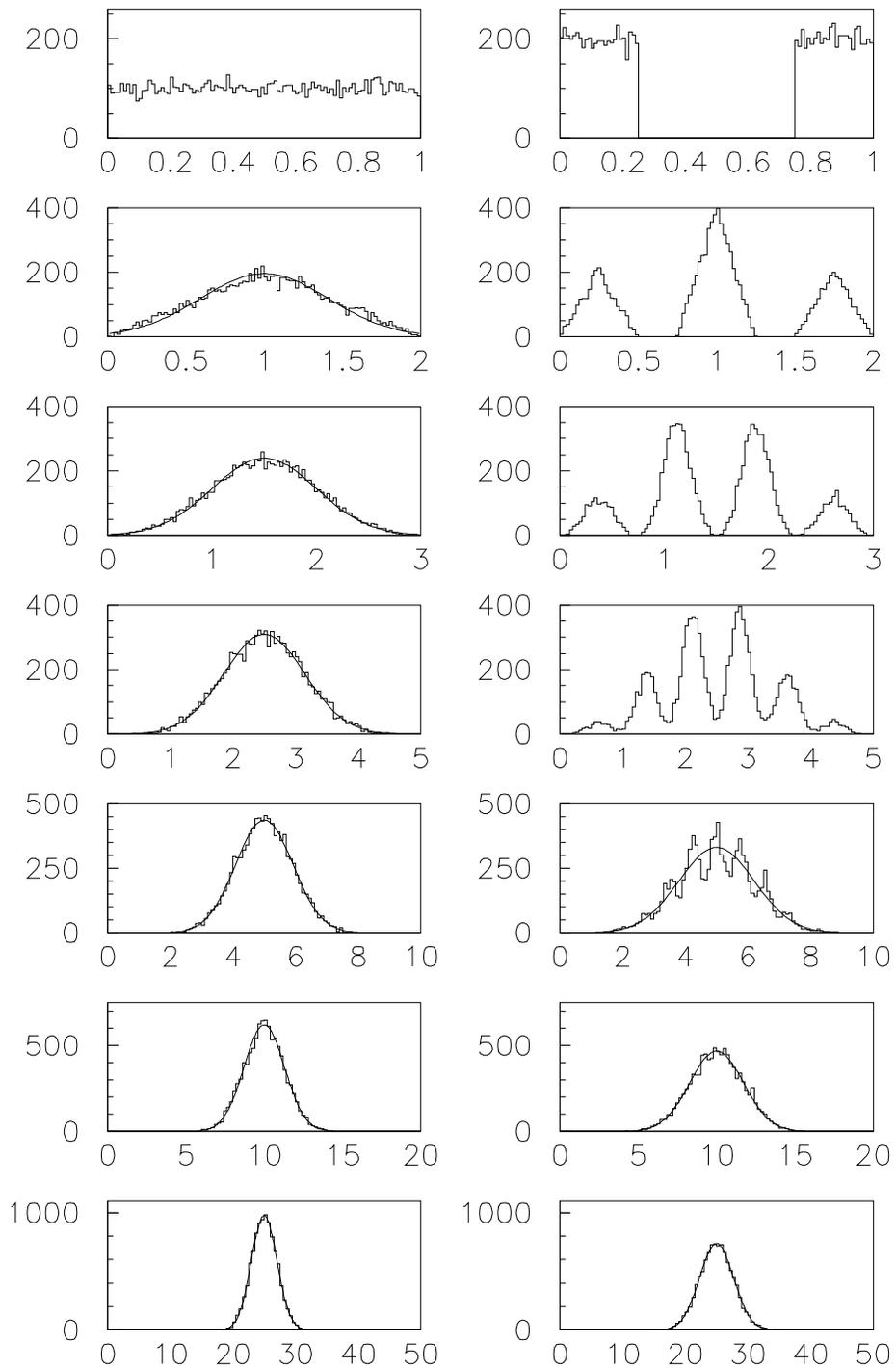,width=12.5cm,clip=}
\caption{\sf Central limit theorem at work: the sum of $n$
variables, for two different distribution, is shown. The values
of $n$ (top-down) are: 1,2,3,5,10,20,50.}
\label{fig:cen_lim}
\end{figure}
The well known central limit theorem plays
a crucial role in statistics
and justifies the enormous importance
that the normal distribution has in many practical applications
 (this is the reason why it appears on 10 DM notes).

We have reminded ourselves in (\ref{eq:linc1}-\ref{eq:linc2})
of the expression of the mean and variance of a linear combination
of random variables
 $$Y=\sum_{i=1}^n X_i$$
in the most general case, which includes
correlated variables ($\rho_{ij}\ne0$). In the case of
independent variables the variance is
given by the simpler, and better known,
expression
\begin{equation}
\sigma_Y^2= \sum_{i=1}^n c_i^2\sigma_i^2 \hspace{1.0cm}
(\rho_{ij}=0,\ i\ne j) \,.
\end{equation}
This is a very general statement, valid for any
number and  kind of variables
(with the obvious clause that all $\sigma_i$ must be finite) but
it does not give any information about the probability distribution
of $Y$. Even if all $X_i$ follow the same distributions $f(x)$,
$f(y)$ is different from $f(x)$, with some exceptions,
one of these being the normal.

The central limit theorem states that
the distribution of a linear combination
$Y$ will be {\it approximately normal} if the variables $X_i$
are independent and $\sigma_Y^2$ is much larger than any
single component $c_i^2\sigma_i^2$ from a non-normally distributed
$X_i$. The last condition is just to guarantee that there is
no single random variable which dominates the fluctuations.
The accuracy of the approximation improves as the number of
variables $n$ increases (the theorem says ``when $n\rightarrow\infty$''):
\begin{equation}
n\rightarrow\infty  \Longrightarrow Y \sim
{\cal N}\left(\sum_{i=1}^n c_i X_i,
\left(\sum_{i=1}^n c_i^2\sigma_i^2\right)^{\frac{1}{2}}\right)
\end{equation}
 The proof of the theorem
can be found in standard text books.
For practical purposes, and if one is not very interested
in the detailed behavior of the tails, $n$ equal to 2 or 3
may already  gives a satisfactory approximation, especially
if the $X_i$ exhibits a gaussian-like shape. Look for example
at Fig. ~\ref{fig:cen_lim}, where samples of 10000 events have
been simulated starting from a uniform distribution and from a
crazy square wave distribution. The latter, depicting
a kind of ``worst practical case'', shows that, already
for $n=20$ the distribution of the sum is practically normal.
In the case of the uniform distribution $n=3$ already
gives an acceptable approximation as far as probability intervals of
one or two standard deviations
from the mean value are concerned. The figure also shows
that, starting from a triangular distribution (obtained
in the example from the sum of 2 uniform distributed variables),
$n=2$ is already sufficient (the sum of 2 triangular distributed
variables is equivalent to the sum of 4
uniform distributed variables).

\subsection{Distribution of a sample average}\label{ss:media}
As first application of the theorem let us  remind ourselves that
a sample average $\overline{X}_n$of $n$ \underline{independent} variables
\begin{eqnarray}
\overline{X}_n &=& \sum_{i=1}^n\frac{1}{n}X_i,
\end{eqnarray}
is normally distributed, since it is a  linear combination
of $n$ variables $X_i$, with $c_i=1/n$. Then:
\begin{eqnarray}
\overline{X}_n & \sim & {\cal N}(\mu_{\overline{X}_n},
                                  \sigma_{\overline{X}_n}) \\
\mu_{\overline{X}_n}  &=& \sum_{i=1}^n\frac{1}{n}\mu = \mu \\
\sigma^2_{\overline{X}_n}& = & \sum_{i=1}^n
\left(\frac{1}{n}\right)^2\sigma^2 = \frac{\sigma^2}{n} \\
\sigma_{\overline{X}_n}& = & \frac{\sigma}{\sqrt{n}}\,.
\end{eqnarray}
This result, we repeat, is independent of the distribution
of $X$ and is already {\it approximately valid} for small values of $n$.
\subsection{Normal approximation  of the binomial and
of  the Poisson distribution}
Another important application of the theorem is that the binomial
and the Poisson distribution can be approximated, for ``large numbers'',
by a  normal distribution. This is a general result, valid for
all distributions which have the {\it reproductive property
under the sum}. Distributions of this kind are the binomial,
the Poisson and the $\chi^2$. Let us go into more detail:
\begin{description}
\item[\fbox{${\cal B}_{n,p} \rightarrow {\cal N}
\left(np, \sqrt{np(1-p)}\right)$}]
The reproductive property of the binomial states that if $X_1$,
$X_2$, $\ldots$, $X_m$ are $m$ independent variables,
each following a binomial distribution of parameter $n_i$ and $p$,
then their sum $Y=\sum_iX_i$ also follows a binomial distribution
with parameters $n=\sum_i n_i$ and $p$. It is easy to be convinced
of this property  without
any mathematics: just think of what happens if one tosses bunches
of  three, of five and of ten coins, and then one considers
the global result:
a binomial with a large $n$ can then always
be seen as a sum of many binomials with smaller $n_i$. The
application of the central limit theorem is straightforward,
apart from deciding when the convergence is acceptable:
the parameters on which one has to judge
are in this case $\mu=np$ and the
complementary quantity $\mu^c=n(1-p)=n-\mu$. If they are
\underline{both} $\gtrsim 10$ then the approximation starts to
be reasonable.
\item[\fbox{${\cal P}_{\lambda} \rightarrow
      {\cal N}\left(\lambda, \sqrt{\lambda}\right)$}]
The same argument holds for the Poisson distribution.
In this case the approximation starts to be reasonable
when $\mu=\lambda \gtrsim 10$.
\end{description}
\subsection{Normal distribution of measurement errors}
The central limit theorem is also important to {\it justify}
why in  many cases the distribution followed by
the measured values around their average is approximately normal.
 Often, in fact, the  random experimental error $e$,
which causes the fluctuations of the measured values
around the unknown true value of the physical quantity, can be seen
as an \underline{incoherent} sum of smaller contributions
\begin{equation}
e = \sum_i e_i\,,
\end{equation}
each contribution
having a distribution which satisfies
the conditions of the central limit theorem.
\subsection{Caution}
After this commercial in favour of the miraculous properties
of the central limit theorem, two remarks of caution:
\begin{itemize}
\item
sometimes the conditions of the theorem are not satisfied:
  \begin{itemize}
  \item
  a single component dominates the fluctuation of the
  sum:
  a typical case is the well known Landau distribution;
  also systematic errors may have the same effect on the global error;
  \item
  the condition of independence is lost if systematic
  errors affect a set of measurements, or
  if there is coherent noise;
  \end{itemize}
\item
 the
\underline{tails} of the distributions \underline{do exist}
and they are not always gaussian! Moreover,
realizations of a random variable several standard deviations
 away from the mean are \underline{possible}. And they show up
without notice!
\end{itemize}
\section{Measurement errors and measurement
uncertainty}
One might assume that the concepts of error and uncertainty
are well enough known to be not worth discussing.
Nevertheless a
few comments are needed
(although for more details
to the DIN\cite{DIN} and ISO\cite{ISO,ISOD} recommendations
should be consulted):
\begin{itemize}
\item
the first concerns the terminology. In fact the words
{\it error} and {\it uncertainty} are
currently used almost  as synonyms:
\begin{itemize}
\item
``error'' to mean  both error
and uncertainty (but nobody says ``Heisenberg
Error Principle'');
\item
``uncertainty'' only for the uncertainty.
\end{itemize}
``Usually'' we understand
 what each other is talking about, but a more precise
use of these nouns would really help. This is strongly
called for
by the DIN\cite{DIN} and ISO\cite{ISO,ISOD} recommendations.
They state in fact that
 \begin{itemize}
 \item
 \underline{error} is {\it ``the result of a measurement minus a
 true value of the measurand''}: it follows that
 the \underline{error} is usually
 \underline{unkown};
 \item
 \underline{uncertainty} is a ``{\it parameter, associated with the result
 of a measurement, that characterizes the dispersion of the values that could
 reasonably be attributed to the measurand}'';
 \end{itemize}
\item
Within  the High Energy Physics community
there is an established
 practice for reporting the final uncertainty of a measurement in the form
of \underline{standard deviation}.
This is also recommended by these norms.
However this should be done
at each step of the analysis, instead of estimating
 ``maximum error bounds'' and using
  them as standard deviation in the
 ``error propagation'';
\item
the process of measurement is a complex one  and it is difficult
to disentangle the different contributions which cause the total
error. In particular,
the active role of the experimentalist
is sometimes overlooked.
For this reason it is
often incorrect to quote the (``nominal'') uncertainty due to the
instrument as if it were \underline{the} uncertainty
of the measurement.
\end{itemize}
\section{Statistical Inference}\label{sec:inference}
\subsection{Bayesian inference}
\label{ss:bayes_inf}
In the Bayesian framework the inference
is performed calculating
the final distribution of the random variable
associated to the true
values of the physical quantities
from all available information.
 Let us call
$\underline{x}=\{x_1, x_2, \ldots, x_n\} $
the {\it n-tuple} (``vector'') of observables,
$\underline{\mu}=\{\mu_1, \mu_2, \ldots, \mu_n\}$ the n-tuple
of the true
values of the physical quantities of interest,
 and $\underline{h}=\{h_1, h_2, \ldots, h_n\}$
the n-tuple of all the
possible realizations of the {\it influence variables} $H_i$.
The term  ``influence variable'' is used here with
an extended meaning, to indicate not only external factors which
could influence the result (temperature, atmospheric pressure,
and so on) but also any possible calibration constant and any
source of systematic errors.
In fact the distinction between $\underline{\mu}$ and
$\underline{h}$ is artificial, since they are all conditional
hypotheses. We separate them simply because at the end we will
``marginalize'' the final joint distribution functions
with respect to $\underline{\mu}$, integrating the joint distribution
with respect to the other hypotheses
considered as influence variables.

The likelihood of the {\it sample} $\underline{x}$  being
produced from $\underline{h}$ and $\underline{\mu}$ and the
initial probability are
$$ f(\underline{x}|\underline{\mu}, \underline{h}, H_\circ)$$
and
\begin{equation}
f_\circ(\underline{\mu}, \underline{h}) =
f(\underline{\mu}, \underline{h}| H_\circ)\,,
\end{equation}
respectively.
$H_\circ$ is intended to remind us, yet again, that
likelihoods and priors
- and hence conclusions - depend
on all explicit and implicit assumptions within the problem,
and in particular on the parametric  functions used to
model  priors and likelihoods.
To simplify the formulae, $H_\circ$
will  no longer be written explicitly.

Using the Bayes formula for multidimensional continuous
distributions (an extension of (~\ref{eq:bayes_cont}))
we obtain the most general formula
of inference
\begin{equation}
f(\underline{\mu}, \underline{h}|\underline{x}) =
\frac{f(\underline{x}|\underline{\mu}, \underline{h})
      f_\circ(\underline{\mu}, \underline{h})}
     {\int
      f(\underline{x}|\underline{\mu}, \underline{h})
       f_\circ(\underline{\mu}, \underline{h})
       d\underline{\mu} d\underline{h}}\,,
\label{eq:ginf0}
\end{equation}
yielding the joint distribution of all conditional variables
$\underline{\mu}$ and $\underline{h}$ which are responsible
for the observed sample $\underline{x}$.
To obtain the final distribution of $\underline{\mu}$
one has to integrate (\ref{eq:ginf0})
over all possible values of $\underline{h}$,
obtaining
\begin{equation}
\boxed{
f(\underline{\mu}|\underline{x}) =
\frac{\int f(\underline{x}|\underline{\mu}, \underline{h})
      f_\circ(\underline{\mu}, \underline{h})d\underline{h}}
     {\int
      f(\underline{x}|\underline{\mu}, \underline{h})
       f_\circ(\underline{\mu}, \underline{h})
       d\underline{\mu} d\underline{h}}\,.
}
\label{eq:ginf1}
\end{equation}
Apart from the technical problem of evaluating the integrals,
if need be
numerically or using Monte Carlo
methods\footnote{This is conceptually what experimentalists
do when they change all the parameters of the Monte Carlo simulation
in order to estimate the ``systematic error''.},
(\ref{eq:ginf1}) represents the most general form
of {\it hypothetical inductive inference}.
The word ``hypothetical''
reminds us of $H_\circ$.

When all the sources of influence are under control,
i.e. they can be assumed to take a precise value,
the initial distribution can be factorized by a
$f_\circ(\underline{\mu})$
and a Dirac $\delta(\underline{h}-\underline{h}_\circ)$,
obtaining the much simpler formula
\begin{eqnarray}
f(\underline{\mu}|\underline{x}) &=&
\frac{\int f(\underline{x}|\underline{\mu}, \underline{h})
      f_\circ(\underline{\mu})
      \delta(\underline{h}-\underline{h}_\circ)d\underline{h}}
     {\int
      f(\underline{x}|\underline{\mu}, \underline{h})
      f_\circ(\underline{\mu})\delta(\underline{h}-\underline{h}_\circ)
      d\underline{\mu} d\underline{h}}
      \nonumber  \\
& = &
\frac{f(\underline{x}|\underline{\mu}, \underline{h}_
\circ)f_\circ(\underline{\mu})}
     {\int f(\underline{x}|\underline{\mu}, \underline{h}_\circ)
           f_\circ(\underline{\mu}) d\underline{\mu}}\,.
\label{eq:ginf2}
\end{eqnarray}
Even if formulae (\ref{eq:ginf1}-\ref{eq:ginf2})
look complicated because of the
multidimensional integration and  of the continuous nature
of $\underline{\mu}$, conceptually they are
 identical to the example
of the $dE/dx$ measurement discussed in Sec. ~\ref{ss:bayes_inf}

The final probability density function provides the
most complete and detailed information about the
unknown quantities, but sometimes (almost always $\ldots$) one
is not interested in
full knowledge of $f(\underline{\mu})$, but just in a
few numbers which summarize at best the position and the width
of the distribution (for example when publishing the result
in a journal in the most compact way).
The most natural quantities for this purpose
are the expected value and the variance, or the standard deviation.
Then the Bayesian best estimate of a physical quantity
is:
\begin{eqnarray}
\widehat{\mu}_i = E[\mu_i] & = &
\int \mu_i f(\underline{\mu}|\underline{x}) d\underline{\mu}
\label{eq:best_mu1} \\
\sigma_{\mu_i}^2\equiv Var(\mu_i) & = & E[\mu_i^2] - E^2[\mu_i] \\
\sigma_{\mu_i} & \equiv & +\sqrt{\sigma_{\mu_i}^2}
\end{eqnarray}

When many true values are inferred
from the same data
the numbers which synthesize the result are not
only the expected values and variances, but also the covariances,
which give \underline{at least} the (linear!)
correlation coefficients between the variables:
\begin{equation}
\rho_{ij}\equiv\rho(\mu_i,\mu_j) = \frac{Cov(\mu_i,\mu_j)}
{\sigma_{\mu_i}\sigma_{\mu_j}}\,.
\end{equation}
In the following sections we will deal in most cases
with only one value to infer:
\begin{equation}
f(\mu|\underline{x}) = \ldots \,,
\end{equation}
\subsection{Bayesian inference and maximum likelihood}
We have already said
that the dependence of the final probabilities
on the initial ones gets weaker as the amount of
experimental information increases. Without going into mathematical
complications (the proof of this statement can be found
for example in\cite{Jeffreys})
 this simply means that, asymptotically,
whatever $f_\circ(\mu)$
one puts in (\ref{eq:ginf2}),
$f(\mu|\underline{x})$ is unaffected. This is ``equivalent'' to
dropping
 $f_\circ(\mu)$ from
(\ref{eq:ginf2}). This results in
\begin{equation}
f(\mu|\underline{x}) \approx
\frac{f(\underline{x}|\mu, \underline{h}_\circ)}
     {\int f(\underline{x}|\mu, \underline{h}_\circ) d\mu}\,.
\end{equation}
Since the denominator of the Bayes formula has the
technical role of properly normalizing the probability
density function,
the result can be written in the simple form
\begin{equation}
f(\mu|\underline{x}) \propto
f(\underline{x}|\mu, \underline{h}_\circ) \equiv
{\cal L}(\underline{x};\mu, \underline{h}_\circ)\,.
\end{equation}
Asymptotically the final probability is just the (normalized)
likelihood! The notation  ${\cal L}$ is that used in the
maximum likelihood literature (note that, not only does  $f$
become ${\cal L}$,
but also ``$|$'' has been replaced by ``;'':
${\cal L}$ has no probabilistic interpretation in conventional statistics.)

If the mean value of $f(\mu|\underline{x})$
coincides with the value for which $f(\mu|\underline{x})$
has a maximum, we obtain the
maximum likelihood method. This does not mean that the
Bayesian methods are ``blessed'' because
of this achievement, and that
they can be used only in those cases where they provide the same results.
It is the
other way round, the maximum likelihood method
gets justified \underline{when} all the
the limiting conditions of the approach
($\rightarrow$ insensitivity of the result from the initial
probability $\rightarrow$ large number of events)
are satisfied.

Even if in this asymptotic limit the two approaches yield the same
numerical results, there are  differences in their interpretation:
\begin{itemize}
\item
the likelihood, after proper normalization, has a probabilistic
meaning for Bayesians but not for
 frequentists; so Bayesians can say that the probability
that $\mu$ is in a certain interval is, for example, $68\,\%$,
while this statement is blasphemous for a frequentist (``the
true value is a constant'' from his point of view);
\item
frequentists prefer to choose
$\widehat{\mu}_L$
the value which maximizes the likelihood,
as estimator. For  Bayesians, on the other hand,
the expected value
$\widehat{\mu}_B=E[\mu]$ (also called the {\it prevision})
is more appropriate. This is justified by the fact that
the assumption of the $E[\mu]$ as best estimate of $\mu$
minimizes the risk of a bet (always keep the bet in mind!).
For example, if the final distribution is exponential
with parameter $\tau$ (let us think for a moment of particle
decays) the maximum likelihood method would {\it recommend betting} on
the value $t=0$, whereas the Bayesian
approach suggests the value $t=\tau$. If the terms of the bet
are ``whoever gets \underline{closest} wins'' what is the best strategy?
And then, what is the best strategy if the terms are
``whoever gets the \underline{exact} value wins''?
But now think of the probability of getting the exact value and
of the probability of getting closest?
\end{itemize}
\subsection{The dog, the hunter and the biased Bayesian estimators}
One of the most important tests to judge
how good an estimator is,
is whether or not it is
{\it correct} (not biased).
Maximum likelihood estimators are
usually correct, while Bayesian estimators - analysed within
the maximum likelihood framework - often are not.
This could be considered  a weak point - however the
Bayes estimators are simply
naturally consistent with the status
of information before new data
become available.
In the maximum
likelihood method, on the other hand, it is not clear what
the assumptions are.

Let us take an example which shows the logic of frequentistic
inference and why the use of reasonable prior distributions
yields results which
that frame classifies as distorted.
Imagine meeting a hunting dog in the country. Let us assume we
know that there is a $50\,\%$ probability
of finding  the dog within a radius of 100 meters centered
on the position of the hunter (this is our likelihood).
Where is the hunter? He is with $50\,\%$ probability
within a radius of 100 meters around the position of the dog,
with equal probability in all directions. ``Obvious''.
This is exactly the
logic scheme used in the frequentistic approach to
 build confidence regions from the estimator (the dog in this
 example). This however assumes that the hunter can be anywhere
 in the country. But now let us change the status of information:
 ``the dog is by a river''; ``the dog has collected a duck and
 runs in a certain direction''; ``the dog is sleeping'';
 ``the dog is in a field surrounded by a fence through which he
 can pass without problems, but the hunter cannot''. Given
 any new condition the conclusion changes.
 Some of the new conditions change our likelihood, but
 some others only influence the initial distribution.
 For example, the case of the dog in an enclosure
 inaccessible to the hunter is exactly the problem encountered
 when measuring a quantity close to the edge of its physical region,
 which is quite common in frontier research.
\section{Choice of the initial probability density function}
\subsection{Difference with respect to the discrete case}
\begin{figure}
\centering\epsfig{file=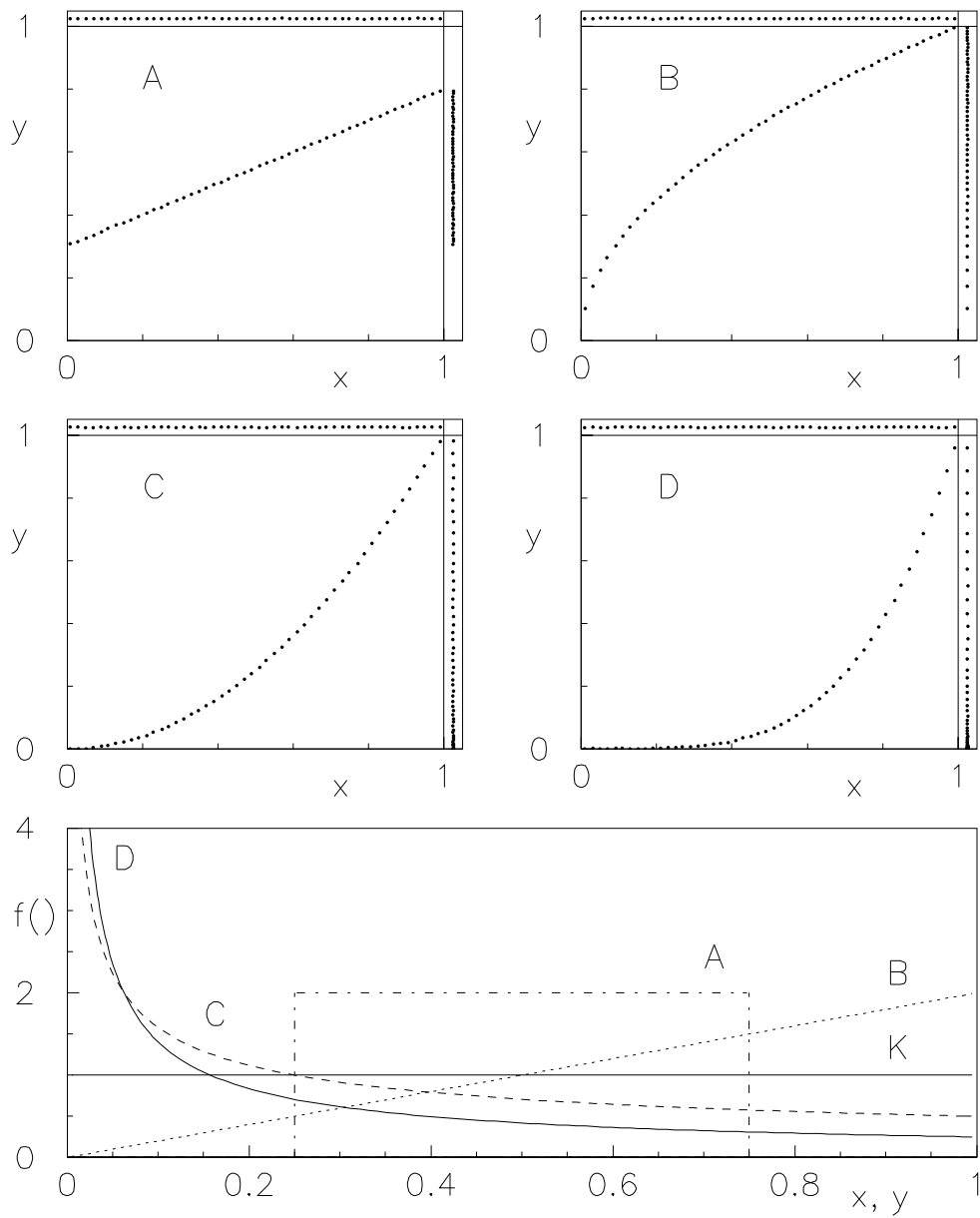,width=\linewidth,clip=}
\caption{\sf Examples of variable changes.}
\label{fig:var_tr}
\end{figure}
The title of this section is similar to that of Sec. ~\ref{sec:choice1}, but
the problem and the conclusions will be different. There we said that
the Indifference Principle (or, in its refined modern version, the
Maximum Entropy Principle) was a good choice. Here there are problems
with  {\it infinities}
 and with the fact that it is possible to map an infinite
number of points contained in a finite region onto an infinite
number of points contained in a larger or smaller
finite region. This changes the probability density
function. If, moreover, the transformation from one
 set of variables to the
other is not linear (see, e.g. Fig. ~\ref{fig:var_tr})
what is uniform in one variable ($X$)
 is not uniform in another variable (e.g. $Y=X^2$). This problem
 does not exist in the
 case of discrete variables, since if $X=x_i$ has a probability
 $f(x_i)$ then $Y=x_i^2$ has the same probability.
 A different way of stating the problem is that the
 {\it Jacobian} of the
 transformation squeezes or stretches the metrics, changing the
 probability density function.

We will not enter into the open
discussion about the optimal choice
of the distribution. Essentially we shall use the uniform distribution,
being careful to employ the variable which ``seems'' most appropriate
for the problem, but \underline{You} may disagree
- surely with good reason - if You have a different
kind of experiment in mind.

The same problem is also present, but well hidden,
 in the maximum likelihood method.
For example, it is possible to demonstrate
that, in the case of normally distributed likelihoods,
a uniform distribution of the mean $\mu$ is implicitly assumed
(see section \ref{sec:normal_results}).
 There is nothing wrong with this, but one should be aware
 of it.
\subsection{Bertrand paradox and angels' sex}
A good example to help understand the problems outlined
in the previous section
is the so-called Bertrand paradox:
\begin{description}
\item[Problem:]
Given a circle of radius $R$ and a chord  drawn randomly on it,
  what is the probability that the length $L$ of the chord
  is  smaller than
  $R$?
\item[Solution 1:] Choose ``randomly'' two points on the circumference
and draw a chord between them: $\Rightarrow P(L<R)=1/3=0.33$\,.
\item[Solution 2:] Choose a straight line passing through
 the center
of the circle; then draw a second line, orthogonal to the first,
and which intersects it inside the circle at a
``random'' distance from the center:
$\Rightarrow P(L<R)=1-\sqrt{3}/2 = 0.13$\,.
\item[Solution 3:] Choose ``randomly'' a point inside the circle and
draw a straight line orthogonal to the radius
that passes through
the chosen point $\Rightarrow P(L<R)=1/4 = 0.25$;
\item[Your solution:] $\ldots$ $\ldots$ $\ldots$?
\item[Question:] What is the origin of the paradox?
\item[Answer:] The problem does not specify how to ``randomly''
choose the chord. The three solutions take a
\underline{uniform} distribution:
along the circumference; along the the radius; inside
the area. What is uniform in one variable is not uniform in the others!
\item[Question:] Which is the \underline{right} solution?
\end{description}
In principle you may imagine an infinite number of different solutions.
{}From a physicist's viewpoint
any attempt to answer this question is a waste of time.
The reason why the paradox
 has been compared to the Byzantine discussions
about the sex of angels is that there are indeed people arguing
about it. For example, there is a school of thought which
insists that  Solution 2 is the \underline{right} one.

In fact this kind of paradox, together with abuse of the Indifference
Principle for problems like ``what is the probability that the
sun will rise tomorrow morning'' threw a  shadow over
Bayesian methods at the end of last century. The maximum likelihood
method, which does not make explicit use of prior distributions,
was then seen as a valid solution to the problem. But
in reality
 the ambiguity of the proper metrics on which
the initial distribution is uniform has an equivalent
on the arbitrariness of the variable used in the likelihood function.
In the end, what was criticized
when it was stated explicitly in the Bayes formula \underline{is}
accepted passively when it is hidden in the maximum
likelihood method.

\section{Normally distributed observables}\label{sec:normal_results}
\subsection{Final distribution, prevision and credibility intervals of
the true value}\label{sec:normal_results1}
The first application of the Bayesian inference will be that
of a normally distributed quantity. Let us take
a data sample $\underline{q}$ of $n_1$ measurements, of which
we calculate the average $\overline{q}_{n_1}$. In our formalism
$\overline{q}_{n_1}$ is a realization of the random variable
$\overline{Q}_{n_1}$. Let us assume \underline{we know} the
standard deviation $\sigma$ of the variable $Q$, either
because $n_1$ is very large and it can be estimated
accurately from the sample or because it was known {\it a priori}
(we are not going to discuss in these notes the case
of small samples and unknown variance).
The property of the average (see ~\ref{ss:media})
tells us that the
likelihood $f(\overline{Q}_{n_1}|\mu,\sigma)$  is gaussian:
\begin{equation}
\overline{Q}_{n_1} \sim {\cal N}(\mu, \sigma/\sqrt{n_1}).
\label{eq:lik_q}
\end{equation}
To simplify the following notation, let us call $x_1$
this average and $\sigma_1$ the standard  deviation of the average:
\begin{eqnarray}
x_1 & = &\overline{q}_{n_1}\\
\sigma_1 & = & \sigma/\sqrt{n_1}\,,
\end{eqnarray}

We then apply (\ref{eq:ginf2}) and get
\begin{equation}
f(\mu|x_1,
{\cal N}(\cdot,\sigma_1)) =
                \frac{\frac{1}{\sqrt{2\pi}\sigma_1}
                     e^{-\frac{(x_1-\mu)^2}{2\sigma_1^2}}f_\circ(\mu)}
                     { \int \frac{1}{\sqrt{2\pi}\sigma_1}
                     e^{-\frac{(x_1-\mu)^2}{2\sigma_1^2}}f_\circ(\mu)d\mu}\, .
\label{eq:invg}
\end{equation}
At this point we have to make a choice for
$f_\circ(\mu)$. A reasonable choice
is to take, as a first guess,
 a uniform distribution defined over a ``large''
interval which includes $x_1$. It is not really important
how large the interval is,
for a few $\sigma_1$'s away
from $x_1$ the integrand at the denominator
tends to zero because of the gaussian function. What is important
is that a  constant $f_\circ(\mu)$ can be simplified
in (\ref{eq:invg}) obtaining
\begin{equation}
f(\mu|x_1,
{\cal N}(\cdot,\sigma_1)) =
                \frac{\frac{1}{\sqrt{2\pi}\sigma_1}
                     e^{-\frac{(x_1-\mu)^2}{2\sigma_1^2}}}
                     { \int_{-\infty}^{\infty}
                      \frac{1}{\sqrt{2\pi}\sigma_1}
                     e^{-\frac{(x_1-\mu)^2}{2\sigma_1^2}}d\mu}\, .
\label{eq:invg2}
\end{equation}
The integral in  the denominator is equal to unity, since
integrating with
respect to $\mu_1$ is equivalent to integrating with respect to $x_1$.
The final result is then
\begin{equation}
f(\mu) =
f(\mu|x_1,{\cal N}(\cdot,\sigma_1)) =
                 \frac{1}{\sqrt{2\pi}\sigma_1}
                 e^{-\frac{(\mu-x_1)^2}{2\sigma_1^2}}\, :
\label{eq:invg3}
\end{equation}
\begin{itemize}
\item
the true value is normally distributed around $x_1$;
\item
its best estimate ({\it prevision}) is $E[\mu]=x_1$;
\item
its variance is $\sigma_{\mu}=\sigma_1$;
\item
the ``confidence intervals'', or {\it credibility intervals},
in which there is a certain probability of finding the
true value are easily calculable:
\begin{center}
\vspace{0.5 cm}
\begin{tabular}{|c|rcl|} \hline
Probability level  & \multicolumn{3}{|c|}{credibility interval} \\
(confidence level) & \multicolumn{3}{|c|}{(confidence interval)} \\
 $(\%)$   & \multicolumn{3}{|c|}{}  \\ \hline
68.3  & \hspace{0.4cm}$x_1$  & $\pm$  & $ \sigma_1$ \\
90.0  & $x_1$& $\pm$& $ 1.65\sigma_1$ \\
95.0 & $  x_1$& $\pm$& $ 1.96\sigma_1$ \\
99.0 & $ x_1$& $\pm$& $ 2.58\sigma_1$ \\
99.73 & $ x_1$& $\pm$& $ 3\sigma_1$ \\ \hline
\end{tabular}
\vspace{0.5 cm}
\end{center}
\end{itemize}

\subsection{Combination of several measurements}
Let us imagine  making a second set of measurements of the physical
quantity, which \underline{we assume} unchanged from the previous
set of measurements. How will our knowledge of $\mu$ change after
this new information? Let us call $x_2  = \overline{q}_{n_2}$
and $\sigma_2  =  \sigma^\prime/\sqrt{n_2}$ the new average and standard
deviation of the average
($\sigma^\prime$ may be different from $\sigma$ of the sample of
numerosity $n_1$).
Applying
Bayes' theorem
a second time
we now have to use {\it as initial distribution
the final probability of the previous inference}:
\begin{equation}
f(\mu|x_1,\sigma_1, x_2, \sigma_2, {\cal N}) =
                \frac{\frac{1}{\sqrt{2\pi}\sigma_2}
e^{-\frac{(x_2-\mu)^2}{2\sigma_2^2}}f(\mu|x_1,{\cal N}(\cdot,\sigma_1))}
{ \int \frac{1}{\sqrt{2\pi}\sigma_2}
e^{-\frac{(x_2-\mu)^2}{2\sigma_2^2}}f(\mu|x_1,{\cal N}(\cdot,\sigma_1))d\mu}\,
{}.
\label{eq:recg}
\end{equation}
The integral is not as simple as the previous one, but still
feasible analytically. The final result is
\begin{equation}
f(\mu|x_1,\sigma_1, x_2, \sigma_2, {\cal N}) =
       \frac{1}{\sqrt{2\pi}\sigma_A}
       e^{-\frac{(\mu-x_A)^2}{2\sigma_A^2}}\, ,
\label{eq:waver}
\end{equation}
where
\begin{eqnarray}
x_A      & = & \frac{x_1/\sigma_1^2 + x_2/\sigma_2^2}
                 {1/\sigma_1^2 + 1/\sigma_2^2}\, ,
                 \label{eq:waver1} \\
\frac{1}{\sigma_A^2} & = & \frac{1}{\sigma_1^2} + \frac{1}{\sigma_2^2}\, .
\label{eq:waver2}
\end{eqnarray}
One recognizes the famous formula of the weighted
average with the inverse of the variances, usually obtained
from maximum likelihood.
Some remarks:
\begin{itemize}
\item
Bayes' theorem updates the knowledge about $\mu$
in an automatic and natural way;
\item
if $\sigma_1 \gg \sigma_2$ (and $x_1$ is not ``too far'' from
$x_2$) the final result is only determined by the second
sample of measurements.
This suggests that an alternative {\it vague} {\it a priori} distribution
can be, instead of the uniform, a gaussian with a
{\it large enough} variance
and a {\it reasonable} mean;
\item
the combination of the samples requires a subjective judgement
that the two samples are really coming from the same true
value $\mu$. We will not discuss this point in these notes, but
a hint on how to proceed is: take the inference on the
difference of two measurements, $D$, as explained at the end of
Section ~\ref{sec:offset} and judge yourself  if $D=0$ is
consistent with the probability density function of $D$.
\end{itemize}
\subsection{Measurements close to the edge
of the physical region}\label{sec:neutrino}
A case which has essentially no solution
in the maximum likelihood approach is when a measurement is performed
at the edge of the physical region and the measured
value comes out very close to it, or even on the
\underline{unphysical} region.
Let us take a numeric example:
\begin{description}
\item[Problem:]
An experiment is planned to measure the
(electron) neutrino mass. The
simulations show that the mass resolution is $3.3\,\mbox{eV}/c^2$,
largely independent of the mass value, and that the measured
mass is normally distributed around
the true mass\footnote{In reality, often
what is normally distributed is $m^2$ instead of $m$.
Holding this hypothesis the terms of the problem change
and a new solution should be worked out, following the
trace indicated in this example.}.
The mass value which results from the elaboration,\footnote{
We consider detector and analysis machinery as a black box,
no matter how complicated it was, and treat the numerical
outcome as a result of a direct measurement\cite{DIN}.}
and corrected for all
known systematic effects, is $x=-5.41\,\mbox{eV}/c^2$. What have we learned
about the neutrino mass?
\item[Solution:]
Our {\it a priori} value of the mass is that it is \underline{positive}
and not too large (otherwise it would already have been measured
in other experiments). One
can take any vague distribution which assigns a probability
density function between 0 and 20 or 30 $\mbox{eV}/c^2$.
In fact, if  an experiment having a resolution of $\sigma=3.3\,\mbox{eV}/c^2$
 has been planned and financed by rational people, with
 the {\it hope} of finding evidence of non negligible mass
 it means that the mass was thought to be in that range.
 If there is no reason to prefer one of the values in that interval
 a uniform distribution can be used, for example
 \begin{equation}
 f_{\circ K}(m)=k=1/30\hspace{1.0cm} (0\le m \le 30)\,.
 \end{equation}

  Otherwise, if one thinks
there is a greater chance of the mass having
small rather than high values,
a prior which reflects
 such an assumption could be chosen,
 for example a half normal with $\sigma_\circ=10\,eV$
 \begin{equation}
 f_{\circ N}(m) =\frac{2}{\sqrt{2\pi}\sigma_\circ}
                 \exp{\left[-\frac{m^2}{2\sigma_\circ^2}\right]}
                 \hspace{1.0cm} (m \ge 0)\,,
 \end{equation}
 or a triangular distribution
 \begin{equation}
 f_{\circ T}(m) = \frac{1}{450}(30-x) \hspace{.6cm} (0\le m \le 30)\,.
 \end{equation}

Let us consider for simplicity the uniform distribution
\begin{eqnarray}
f(m|x, f_{\circ K})
&=&  \frac{
          \frac{1}{\sqrt{2\pi}\sigma}
          \exp{\left[-\frac{(m-x)^2}{2\sigma^2}\right]} k
          }
          {\int_0^{30}
          \frac{1}{\sqrt{2\pi}\sigma}
          \exp{\left[-\frac{(m-x)^2}{2\sigma^2}\right]}
          k d\mu} \\
 &= &
 \frac{19.8}{\sqrt{2\pi}\sigma}\exp{\left[-\frac{(m-x)^2}{2\sigma^2}\right]}
 \hspace{0.7 cm}(0 \le m \le 30)\,.
\end{eqnarray}
The  value which has the highest degree of belief is $m=0$,
but $f(m)$ is non vanishing up to $30\,\mbox{eV}/c^2$ (even if very small).
We can define an interval, starting from $m=0$,
in which we believe that $m$ should have a certain
probability. For example
this level of probability can be $95\,\%$. One has to find the value
$m_\circ$ for which the cumulative function $F(m_\circ)$
equals 0.95.
This value of $m$ is called the {\it upper limit} (or {\it upper bound}).
The result is
\begin{equation}
m < 3.9\, \mbox{eV}/c^2\hspace{0.5 cm} at\ 0.95\,\%\ C.L. \,.
\end{equation}

If we had assumed the other initial distributions the
limit would have been in both cases
\begin{equation}
m < 3.7\, \mbox{eV}/c^2\hspace{0.5 cm} at\ 0.95\,\%\ C.L.\,,
\end{equation}
practically the same (especially if compared with the experimental
resolution of $3.3\, \mbox{eV}/c^2$).
\item[Comment:] Let us assume an {\it a priori} function
sharply peaked at zero and see what happens. For example it could be
of the kind
\begin{equation}
f_{\circ S}(m)\propto \frac{1}{m}\,.
\end{equation}
To avoid singularities in the integral,
let us take a power of $m$ a bit greater
than $-1$, for example $-0.99$, and let us limit its domain
to 30, getting
\begin{equation}
f_{\circ S}(m) = \frac{0.01\cdot 30^{0.01}}{m^{0.99}}\,.
\end{equation}
The upper limit becomes
\begin{equation}
m < 0.006\, \mbox{eV}/c^2\hspace{0.5 cm} at\ 0.95\,\%\ C.L.\,.
\end{equation}
Any experienced physicist would find this result \underline{ridiculous}.
The upper limit is less then $0.2\,\%$ of the experimental resolution;
like expecting to resolve objects having dimensions smaller  than
a micron with a design ruler!
Notice instead that in the previous examples the limit was always of the
\underline{order of magnitude} of the experimental resolution
$\sigma$.
As $f_{\circ S}(m)$ becomes more and more peaked at zero (power
of $x\rightarrow 1$) the limit gets smaller and smaller. This means
that, asymptotically, the degree of belief that $m=0$ is so high
that whatever you measure you will conclude that $m=0$: you could use
the measurement to calibrate the apparatus!
This means that this choice of initial distribution was unreasonable.
\end{description}
\section{Counting experiments}
\subsection{Binomially distributed quantities}\label{ss:binom}
\begin{figure}
\centering\epsfig{file=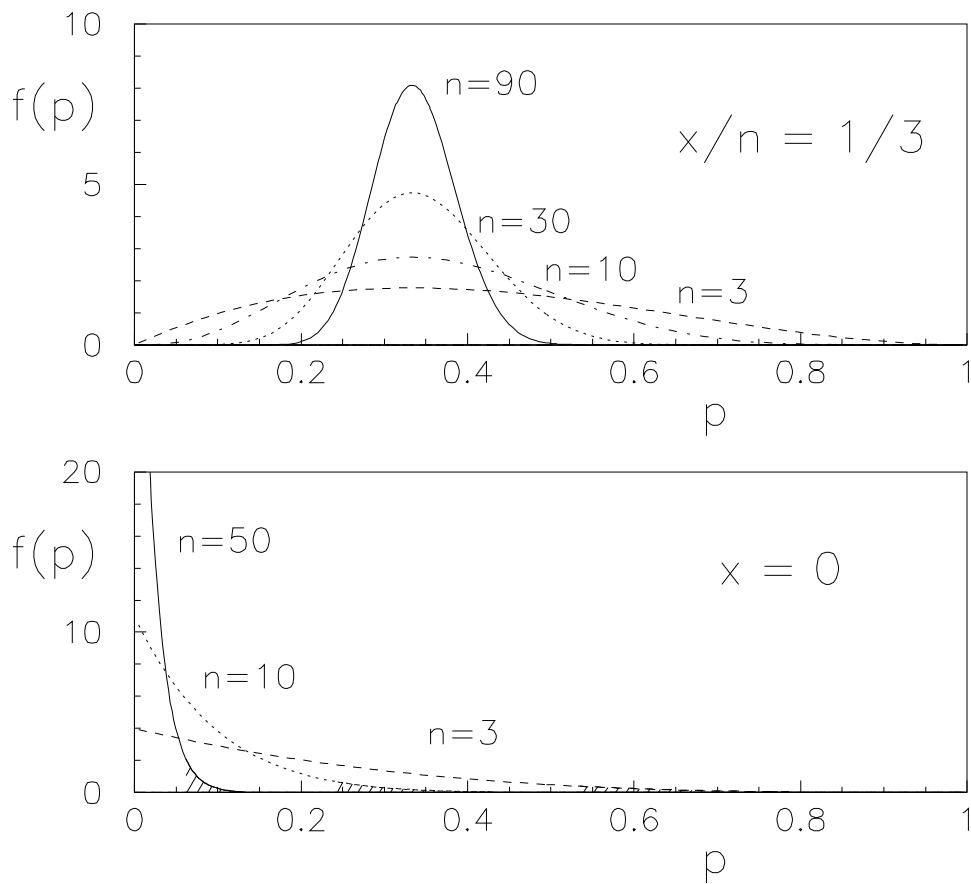,clip=,width=\linewidth}
\caption{\sf Probability density function of the binomial parameter
$p$, having observed $x$ successes in $n$ trials.}
\label{fig:beta}
\end{figure}
Let us assume we have performed $n$ trials and obtained $x$
favorable events. What is the probability of the next event?
This situation happens frequently when measuring efficiencies,
branching ratios,  etc. Stated more generally,
one tries to infer the ``constant
and unknown probability''\footnote{
This concept, which  is very close to the physicist's mentality,
is not correct from the probabilistic - \underline{cognitive} -
point of view. According to the Bayesian scheme, in fact,
the probability changes with the new observations. The final
inference of $p$, however, does not depend on the particular sequence
yielding $x$ successes over $n$ trials. This can be seen in the next table
where $f_n(p)$ is given as a function of the number of trials $n$,
for the three sequences which give 2 successes (indicated by ``1'')
 in three trials
(the use of (\ref{eq:inv_binom}) is anticipated):
\begin{center}
\begin{tabular}{|r|ccc}
\multicolumn{1}{c}{}  & \multicolumn{3}{c}{Sequence} \\
n & 011 & 101 & 110 \\ \hline
0 & 1 & 1 & 1 \\
1 & $2(1-p)$  & $2p$ & $2p$ \\
2 & $6p(1-p)$ & $6p(1-p)$ & $3p^2$ \\
3 & $12p^2(1-p)$ & $12p^2(1-p)$ & $12p^2(1-p)$
\end{tabular}
\end{center}
This important result, related to the concept of
{\it interchangeability},
``allows'' a physicist who is
reluctant to give up the concept
``unknown constant probability'', to see the problem from his
point of view,
ensuring that the same numerical result is obtained.}
of an event occurring.

Where we can assume that the probability is constant
and the observed number of favorable events are binomially
distributed, the unknown quantity to be measured is the parameter
$p$ of the binomial. Using Bayes' theorem we get
\begin{eqnarray}
f(p|x,n,{\cal B}) & = &  \frac{
f(x|{\cal B}_{n,p})f_\circ(p)
                    }{
\int_0^1 f(x|{\cal B}_{n,p})f_\circ(p)dp
                    }\nonumber \\
& = &  \frac{
\frac{n!}{(n-x)!x!}p^x(1-p)^{n-x}f_\circ(p)
}{
\int_0^1
\frac{n!}{(n-x)!x!}p^x(1-p)^{n-x}f_\circ(p)dp
} \nonumber \\
& = &  \frac{
 p^x(1-p)^{n-x}
}{
\int_0^1
 p^x(1-p)^{n-x} dp
}\, ,
\end{eqnarray}
where an initial uniform distribution has been assumed.
The final distribution is known to statisticians as $\beta$ distribution
since the integral at the denominator is the special
function called $\beta$, defined also for real values of $x$ and $n$
(technically this is a $\beta$ with parameters
$a=x+1$ and $b=n-x+1$). In our case
these two numbers are integer and the integral becomes
equal to $x!(n-x)!/(n+1)!$. We then get
\begin{equation}
f(p|x,n,{\cal B})
= \frac{(n+1)!}{x!(n-x)!}p^x(1-p)^{n-x}\,.
\label{eq:inv_binom}
\end{equation}
The expected value and the variance of this distribution
are:
\begin{eqnarray}
E[p] &=& \frac{x+1}{n+2}
\label{eq:infbinom1}\\
Var(p) &=& \frac{(x+1)(n-x+1)}{(n+3)(n+2)^2} \\
       &=& \frac{x+1}{n+2}\left(\frac{n}{n+2} \nonumber
          -\frac{x+1}{n+2}\right)\frac{1}{n+3} \\
       &=& E[p]\left(\frac{n}{n+2} - E[p]\right)\frac{1}{n+3}
\label{eq:infbinom2}\,.
\end{eqnarray}
The value of $p$ for which $f(p)$ has the maximum is
instead $p_m=x/n$. The expression $E[p]$
gives the {\it prevision}
 of the probability for the $(n+1)$-th event
occurring and is called the
``recursive Laplace formula'', or ``Laplace's rule of succession''.

When $x$ and $n$ become large, and $0 \ll x \ll n$,
$f(p)$ has the following asymptotic properties:
\begin{eqnarray}
E[p] &\approx & p_m=\frac{x}{n}\,; \\
Var(p) &\approx & \frac{x}{n}\left(1-\frac{x}{n}\right)\frac{1}{n}
                  = \frac{p_m(1-p_m)}{n}\,; \\
\sigma_p & \approx & \sqrt{\frac{p_m(1-p_m)}{n}}:\\
p &\sim & {\cal N}(p_m, \sigma_p)\,.
\end{eqnarray}
Under these conditions the frequentistic ``definition'' of probability
($x/n$)
is recovered.

Let us see two particular situations: when $x=0$ and $x=n$. In these
cases one gives the result as upper or lower limits, respectively.
Let us sketch the solutions:
\begin{itemize}
\item
\underline{x=n}:
\begin{eqnarray}
f(n|{\cal B}_{n,p}) & = & p^n ;\\
f(p|x=n,{\cal B}) & = & \frac{p^n}{\int_0^1p^ndp} = (n+1)\cdot p^n;\\
F(p|x=n,{\cal B}) & = & p^{n+1}\, .
\end{eqnarray}
To get the $95\,\%$ {\it \underline{lower} bound} ({\it limit}):
\begin{eqnarray}
F(p_\circ|x=n,{\cal B}) & = & 0.05\, , \nonumber \\
& & \nonumber \\
p_\circ & = &  \sqrt[n+1]{0.05}\, .
\end{eqnarray}
An increasing number of trials $n$ constrain more and more
$p$ around 1.

\item
\underline{x=0}:
\begin{eqnarray}
f(0|{\cal B}_{n,p}) & = & (1-p)^n ;\\
 f(p|x=0,n,{\cal B}) & = & \frac{(1-p)^n}{\int_0^1(1-p)^ndp}
  = (n+1)\cdot (1-p)^n;\\
  F(p|x=0, n, {\cal B}) & = & 1 - (1-p)^{n+1}\, .
  \end{eqnarray}
To get the $95\,\%$ {\it \underline{upper} bound (limit)}:
\begin{eqnarray}
F(p_\circ|x=0,n,{\cal B}) & = & 0.95; \nonumber \\
& & \nonumber \\
p_\circ & = &  1 - \sqrt[n+1]{0.05}\, .
\end{eqnarray}
\end{itemize}

The following table shows the $95\,\%$ C.L. limits as a function
of $n$.
The Poisson approximation, to be discussed
in the next section, is also shown.
\begin{center}
\vspace{0.5 cm}
\begin{tabular}{|r|c|c|c|}\hline
& \multicolumn{3}{c|}{Probability level =  $95\,\%$} \\ \hline
$n$ & $x= n$ & \multicolumn{2}{c|}{$x=0$} \\ \hline
& binomial  & binomial &  Poisson approx.  \\
& & & (\,$p_\circ=3/n$\,)\\ \hline
3 & $p\ge  0.47$ & $p\le 0.53$ & $p\le 1$ \\
5 & $p\ge  0.61$ & $p\le  0.39$ & $p\le 0.6$ \\
10 & $p\ge  0.76$ & $p\le  0.24$ & $p\le 0.3$ \\
50 & $p\ge  0.94$ & $p\le  0.057$ & $p\le 0.06$ \\
100& $p\ge  0.97$ & $p\le  0.029$ & $p\le 0.03$ \\
1000 &  $p\ge  0.997$ & $p\le 0.003$ & $p\le 0.003$ \\ \hline
\end{tabular}
\vspace{0.5 cm}
\end{center}

To show in this simple case how $f(p)$ is updated by the new information,
let us imagine we have performed two experiments. The results
 are $x_1=n_1$ and $x_2=n_2$, respectively. Obviously the global
information
is equivalent to $x=x_1+x_2$ and $n=n_1+n_2$, with $x=n$.
We then get
\begin{equation}
f(p|x = n,{\cal B}) = (n+1)p^n = (n_1+n_2+1)p^{n_1+n_2}\, .
\end{equation}
A different way of proceeding would have been to calculate the final
distribution from the information $x_1=n_1$
\begin{equation}
f(p|x_1 = n_1,{\cal B})   =  (n_1+1)p^{n_1}\, ,
\end{equation}
and feed it as initial
distribution to the next inference:
\begin{eqnarray}
f(p|x_1 = n_1, x_2=n_2,{\cal B}) & = & \frac{p^{n_2}
f(p|x_1=n_1, {\cal B})}
{\int_{0}^{1}p^{n_2}f(p|x_1=n_1, {\cal B})dp} \\
& = & \frac{p^{n_2}(n_1+1)p^{n_1}}
{\int_{0}^{1}p^{n_2}(n_1+1)p^{n_1}dp} \\
& = & (n_1+n_2+1)p^{n_1+n_2}\, ,
\end{eqnarray}
getting the same result.
\subsection{Poisson distributed quantities}\label{ss:poisson}
\begin{figure}[t]
\centering\epsfig{file=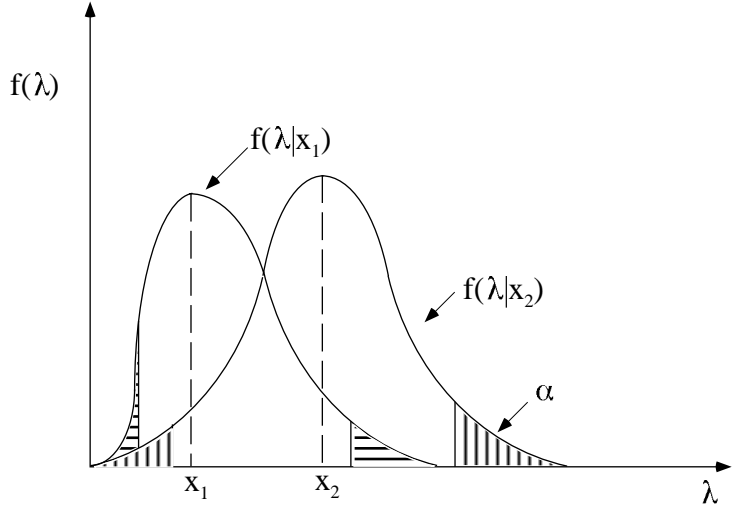,clip=}
\caption{\sf Poisson parameter $\lambda$ inferred from
an observed  number $x$ of counts.}
\label{fig:distr_lambda}
\end{figure}

\begin{figure}
\centering\epsfig{file=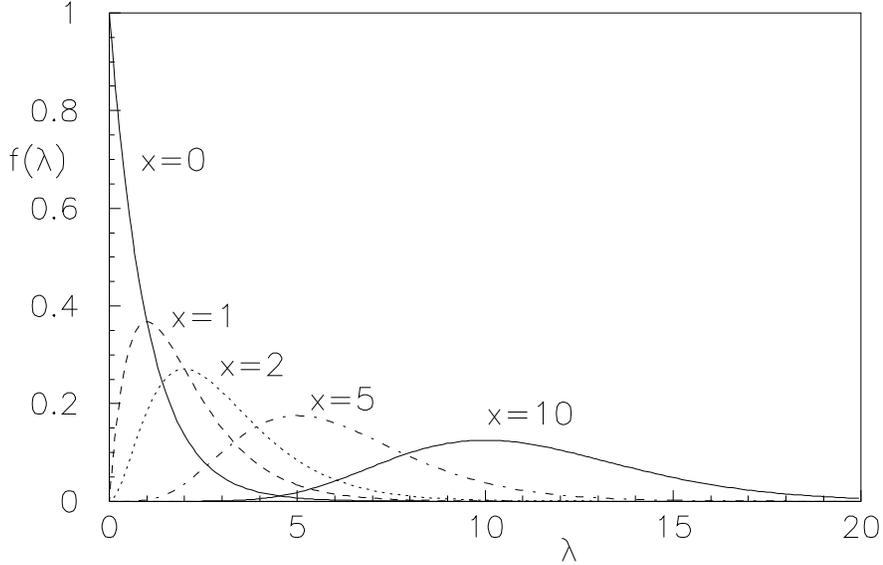,clip=}
\caption{\sf Examples of $f(\lambda|x_i)$.}
\label{fig:invpois}
\end{figure}

As is well known, the typical application of the Poisson
distribution is in counting experiments
such as source activity,
cross sections, etc. The unknown parameter to be
inferred is $\lambda$. Applying  Bayes formula
we get
\begin{eqnarray}
f(\lambda|x,{\cal P}) &=& \frac{\frac{\lambda^xe^{-\lambda}}{x!}
f_\circ(\lambda)}
{\int_0^\infty\frac{\lambda^xe^{-\lambda}}{x!}
f_\circ(\lambda) d\lambda}\, .
\end{eqnarray}
Assuming\footnote{There is a school
of thought according to which the most appropriate
function is $f(\lambda)\propto1/\lambda$.
If \underline{You} think that it is reasonable
for your problem, it may be a good prior.
Claiming that this is ``the Truth''  is one
of the many
claims of the
angels' sex determinations. For didactical purposes
a uniform distribution is more than enough. Some comments
about the $1/\lambda$ prescription will be given
when discussing the particular case $x=0$.}
  $f_\circ(\lambda)$ constant up to a certain
$\lambda_{max}\gg x$ and making the integral by parts we obtain
\begin{eqnarray}
f(\lambda|x,{\cal P}) & = & \frac{\lambda^x e^{-\lambda}}{x!}
\label{eq:inv_poiss1} \\
F(\lambda|x,{\cal P}) & = &
1 - e^{-\lambda}\left(\sum_{n=0}^x \frac{\lambda^n}{n!}\right)\,,
\label{eq:inv_poiss2}
\end{eqnarray}
where the last result has been obtained integrating
(\ref{eq:inv_poiss1}) also
by parts.
Fig. ~\ref{fig:distr_lambda} shows how to build the
credibility intervals, given a certain measured
number of counts $x$.
Fig. ~\ref{fig:invpois} shows some numerical examples.

$f(\lambda)$ has the following properties:
\begin{itemize}
\item
the expected values, variance,  value of maximum
probability are
\begin{eqnarray}
E[\lambda] & = & x+1 \\
Var(\lambda) & = & x+2 \\
\lambda_m &=& x \,;
\end{eqnarray}
the fact that the best estimate of $\lambda$ in the Bayesian sense
is not the intuitive value $x$ but $x+1$ should neither surprise,
nor disappoint us: according to the
initial distribution used ``there are always more possible
values of $\lambda$ on the right side than on the left side of $x$'',
and they pull the distribution to their side; the full information
is always given by  $f(\lambda)$ and the use of the mean is just a
rough approximation; the difference from the ``desired'' intuitive value
$x$ in units of the standard deviation  goes as $1/\sqrt{n+2}$
and becomes immediately negligible;
\item
when $x$ becomes large we get:
\begin{eqnarray}
E[\lambda] &\approx& \lambda_{m} = x \,; \\
Var(\lambda)  &\approx& \lambda_{m} = x \,; \\
\sigma_\lambda  &\approx& \sqrt{x} \,; \label{eq:radice_l}\\
\lambda & \sim & {\cal N}(x, \sqrt{x})\,.
\end{eqnarray}
(\ref{eq:radice_l}) is one of the most familar formulae
used by physicists to assess the uncertainty of a measurement,
although it is sometimes misused.
\end{itemize}
Let us conclude with a special case: $x=0$. As one might imagine,
the inference is highly sensitive
to the initial distribution.
Let us assume that {\it the experiment was planned with
the hope of \underline{observing} something}, i.e. that it could
detect a handful of events within its lifetime. With this hypothesis
one may use any vague prior function not strongly peaked
at zero. We have already come
across a similar case in section
{}~\ref{sec:neutrino},
concerning the upper limit of the neutrino mass. There it was shown
that reasonable hypotheses based on the \underline{positive attitude}
of the experimentalist are almost equivalent
and that they give results consistent with  detector performances.
 Let us use then
the uniform distribution
\begin{eqnarray}
f(\lambda|x=0,{\cal P}) & = & e^{-\lambda} \\
F(\lambda|x=0,{\cal P}) & = & 1-e^{-\lambda} \\
\lambda & < & 3 \ at\ 95\,\%\ C.L. \, .
\end{eqnarray}

\begin{figure}
\centering\epsfig{file=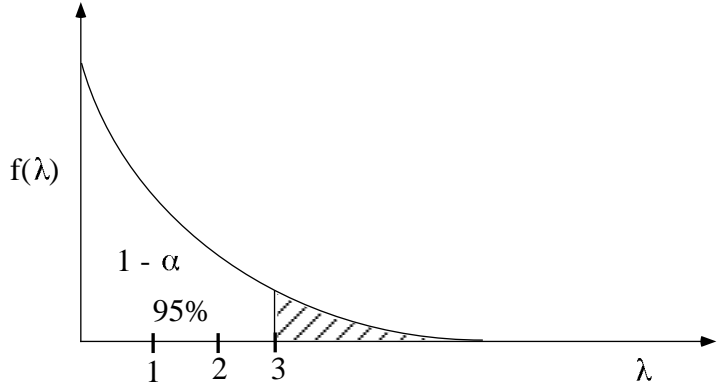,clip=}
\caption{\sf Upper limit to $\lambda$ having observed 0 events.}
\label{fig:lim_lambda}
\end{figure}
\section{Uncertainty due to unknown systematic errors}\label{sec:unknown}
\subsection{Example: uncertainty of the instrument scale
 offset}\label{sec:offset}
In our scheme any quantity of influence of which we don't know the exact
value is a source of systematic error. It will change the final
distribution of $\mu$ and hence its uncertainty.
We have already discussed the most general case in
{}~\ref{ss:bayes_inf}. Let us  make a simple
application making a small variation to the example in section
{}~\ref{sec:normal_results1}: the ``zero'' of the instrument
is not  known exactly, owing to calibration uncertainty.
This can be parametrized assuming that its true value
$Z$ is normally distributed around 0 (i.e. the calibration
was properly done!) with a standard deviation $\sigma_Z$.
Since, most probably, the true value of $\mu$ is independent from
the true value of $Z$, the initial joint probability density
function can be written as the product of the marginal ones:
\begin{equation}
f_\circ(\mu,z)=f_\circ(\mu)f_\circ(z)=
k\frac{1}{\sqrt{2\pi}\sigma_Z}
\exp{\left[-\frac{z^2}{2\sigma_Z^2}\right]}\,.
\end{equation}
Also the likelihood changes with respect to
{}~\ref{eq:lik_q}:

\begin{equation}
f(x_1|\mu,z) = \frac{1}{\sqrt{2\pi}\sigma_1}
\exp{\left[-\frac{(x_1-\mu-z)^2}{2\sigma_1^2}\right]}\,.
\end{equation}
Putting all the pieces together and making use of
(\ref{eq:ginf1}) we finally get
\begin{equation}
f(\mu|x_1, \ldots,f_\circ(z))
=
  \frac{
        \int
        \frac{1}{\sqrt{2\pi}\sigma_1}
        \exp{\left[-\frac{(x_1-\mu-z)^2}{2\sigma_1^2}\right]}
        \frac{1}{\sqrt{2\pi}\sigma_Z}
        \exp{\left[-\frac{z^2}{2\sigma_Z^2}\right]}
        dz
       }
       {
        \int\!\!\int
        \frac{1}{\sqrt{2\pi}\sigma_1}
        \exp{\left[-\frac{(x_1-\mu-z)^2}{2\sigma_1^2}\right]}
        \frac{1}{\sqrt{2\pi}\sigma_Z}
        \exp{\left[-\frac{z^2}{2\sigma_Z^2}\right]}
        d\mu dz
       }\,.
       \nonumber
\end{equation}
Integrating\footnote{It may help to know that
 $$\int_{-\infty}^{+\infty}\exp{\left[bx-\frac{x^2}{a^2}\right]}dx
     = \sqrt{a^2\pi}\exp{\left[\frac{a^2b^2}{4}\right]}\,.$$}
we get
\begin{equation}
f(\mu) = f(\mu|x_1, \ldots,f_\circ(z)) =
\frac{1}{\sqrt{2\pi}\sqrt{\sigma_1^2+\sigma_Z^2}}
   \exp{\left[-\frac{(\mu-x_1)^2}{2(\sigma_1^2+\sigma_Z^2)}\right]}\,.
\end{equation}
The result is that $f(\mu)$ is still a gaussian, but with
a larger variance. The global standard uncertainty
is the quadratic combination of
that due to the statistical fluctuation of the data sample
and the uncertainty due to the imperfect knowledge of the
{\it systematic effect}:
\begin{equation}
\sigma_{tot}^2 = \sigma_1^2+\sigma_Z^2\,.
\end{equation}
This result is well known, although there are still
some ``old-fashioned'' recipes which require
different combinations
of the contributions to be performed.

One has to notice that in this framework it makes no sense
 to speak of ``statistical'' and ``systematical'' uncertainties,
 as if they were of a different nature.
 They have the same \underline{probabilistic} nature:
 $\overline{Q}_{n_1}$ is around $\mu$ with a standard deviation
 $\sigma_1$, and $Z$ is around 0 with standard deviation $\sigma_Z$.
 What distinguishes the two components
 is how the knowledge of the uncertainty is gained: in one case
 ($\sigma_1$) from repeated measurements; in the second case ($\sigma_Z$)
 the evaluation was done by somebody else (the constructor
  of the instrument),
 or in a previous experiment, or guessed from the knowledge of the
 detector, or by simulation, etc. This is the reason why the ISO Guide
 \cite{ISO} prefers the generic names {\it Type A} and {\it Type B}
 for the two kinds of contribution to global
  uncertainty. In particular
 the name ``systematic uncertainty'' should be avoided, while
 it is correct to speak about ``uncertainty due to a systematic effect''.

\subsection{Correction for known systematic errors}\label{ss:known_syst}
It is easy to be convinced that if our prior knowledge
about $Z$ was of the kind
\begin{equation}
Z\sim {\cal N}(z_\circ,\sigma_Z)
\end{equation}
the result would have been
\begin{equation}
\mu \sim {\cal N}\left(x_1-z_\circ,
                       \sqrt{\sigma_1^2+\sigma_Z^2}\right)\,,
\end{equation}
i.e. one has first to \underline{correct} the result
\underline{for} the best value of the \underline{systematic error}
 and then include in the \underline{global uncertainty} a term due to
 imperfect knowledge
 about it. This is a well known and practised
 procedure, although there are still people who
 confuse $z_\circ$ with its uncertainty.

\subsection{Measuring two quantities with the same instrument
having an uncertainty of the scale offset}\label{sec:off_err}
Let us take an example which is a bit more complicated (at least from
the mathematical point of view) but conceptually very
simple and also very common in  laboratory practice.
We measure two physical quantities with the same instrument,
assumed
to have an uncertainty on the ``zero'',
modeled with a normal distribution as in the
previous sections. For each of the
quantities we collect a sample of data \underline{under the same
conditions}, which means that the unknown offset error does not
change from one set of measurements to the other.
Calling $\mu_1$ and $\mu_2$ the true
values, $x_1$ and $x_2$ the sample averages, $\sigma_1$ and
$\sigma_2$
 the average's standard deviations,
and $Z$ the true value of the ``zero'',
the initial probability density and the likelihood are
\begin{equation}
f_\circ(\mu_1,\mu_2,z)=f_\circ(\mu_1)f_\circ(\mu_2)f_\circ(z)=
k\frac{1}{\sqrt{2\pi}\sigma_Z}
\exp{\left[-\frac{z^2}{2\sigma_Z^2}\right]}
\end{equation}
and
\begin{eqnarray}
f(x_1,x_2|\mu_1,\mu_2,z) &=&
 \frac{1}{\sqrt{2\pi}\sigma_1}
\exp{\left[-\frac{(x_1-\mu_1-z)^2}{2\sigma_1^2}\right]}
 \frac{1}{\sqrt{2\pi}\sigma_2}
\exp{\left[-\frac{(x_2-\mu_2-z)^2}{2\sigma_2^2}\right]} \nonumber \\
&=& \frac{1}{2\pi\sigma_1\sigma_2}
\exp{\left[-\frac{1}{2}\left(
              \frac{(x_1-\mu_1-z)^2}{\sigma_1^2} +
              \frac{(x_2-\mu_2-z)^2}{\sigma_2^2}
                       \right)
     \right]}\,,
\end{eqnarray}
respectively.
The result of the inference is now the joint probability density
function of $\mu_1$ and $\mu_2$:
\begin{eqnarray}
f(\mu_1,\mu_2|x_1,x_2,\sigma_1,\sigma_2,f_\circ(z))
&=& \frac{\int f(x_1,x_2|\mu_1,\mu_2,z)f_\circ(\mu_1,\mu_2,z)dz}
  {\int\!\!\int\!\!\int
   f(x_1,x_2|\mu_1,\mu_2,z)f_\circ(\mu_1,\mu_2,z)d\mu_1 d\mu_2 dz}\,,\
\end{eqnarray}
where expansion of the functions has been omitted  for the
sake of clarity.
Integrating we get
\begin{eqnarray}
f(\mu_1,\mu_2) &=&
\frac{1}
     {2\pi\sqrt{\sigma_1^2+\sigma_Z^2}
                \sqrt{\sigma_2^2+\sigma_Z^2}\sqrt{1-\rho^2}
     }
\label{eq:bivarm}
 \\
 & & \exp{
     \left\{
     -\frac{1}{2(1-\rho^2)}
      \left[  \frac{(\mu_1-x_1)^2}
                   {\sigma_1^2+\sigma_Z^2}
             -2\rho\frac{(\mu_1-x_1) (\mu_2-x_2)}
                        {\sqrt{\sigma_1^2+\sigma_Z^2}
                         \sqrt{\sigma_2^2+\sigma_Z^2}}
            +\frac{(\mu_2-x_2)^2}
                  {\sigma_2^2+\sigma_Z^2}
      \right]
     \right\}
      }\,. \nonumber
\end{eqnarray}
where
\begin{equation}
\rho = \frac{\sigma_Z^2}{\sqrt{\sigma_1^2+\sigma_Z^2}
                         \sqrt{\sigma_2^2+\sigma_Z^2}}\,.
\label{eq:rho1}
\end{equation}
If $\sigma_Z$ vanishes then (\ref{eq:bivarm}) has the simpler expression
\begin{equation}
f(\mu_1,\mu_2) @>>{\sigma_Z\rightarrow 0}>
\frac{1}{\sqrt{2\pi}\sigma_1}
\exp{\left[-\frac{(\mu_1-x_1)^2}{2\sigma_1^2}\right]}
\frac{1}{\sqrt{2\pi}\sigma_2}
\exp{\left[-\frac{(\mu_2-x_2)^2}{2\sigma_2^2}\right]}\,,
\end{equation}
i.e. if there is no uncertainty on the offset calibration the
joint density function then $f(\mu_1,\mu_2)$ is equal to the
product of two \underline{independent}
 normal functions, i.e. $\mu_1$ and $\mu_2$
are independent.
In the general case we have to conclude that:
\begin{itemize}
\item
the effect of the {\it common uncertainty} $\sigma_Z$ makes the two
values \underline{correlated}, since they are affected by a common
unknown
systematic error; the correlation coefficient is always non negative
($\rho \ge 0$), as intuitively expected from the definition
 of systematic error;
\item
the joint density function is a {\it multinormal distribution}
of parameters
$x_1$, $\sigma_{\mu_1}=\sqrt{\sigma_1^2+\sigma_Z^2}$,
$x_2$, $\sigma_{\mu_2}=\sqrt{\sigma_2^2+\sigma_Z^2}$, and $\rho$
(see example of Fig. ~\ref{fig:bivar});
\item
the marginal distributions are still normal:
\begin{eqnarray}
\mu_1 &\sim& {\cal N}\left(x_1, \sqrt{\sigma_1^2+\sigma_Z^2}\right) \\
\mu_2 &\sim& {\cal N}\left(x_2, \sqrt{\sigma_2^2+\sigma_Z^2}\right)\,;
\end{eqnarray}
\item
the covariance between $\mu_1$ and $\mu_2$ is
\begin{eqnarray}
Cov(\mu_1,\mu_2) &=& \rho\sigma_{\mu_1}\sigma_{\mu_2} \nonumber \\
                 &=& \rho\sqrt{\sigma_1^2+\sigma_Z^2}
                   \sqrt{\sigma_2^2+\sigma_Z^2}
                 = \sigma_Z^2\,.
\label{eq:covm1m2}
\end{eqnarray}
\item
the distribution of any function $g(\mu_1,\mu_2)$ can be calculated
using the standard methods of  probability theory. For example,
one can demonstrate that the sum $S=\mu_1+\mu_2$ and the difference
$D=\mu_1-\mu_2$ are also normally distributed (see also the
introductory discussion to the central limit theorem
and section ~\ref{sec:cov} for the calculation of averages
and standard deviations):
\begin{eqnarray}
S & \sim & {\cal N}\left(x_1+x_2,
                         \sqrt{\sigma_1^2+\sigma_2^2+(2\sigma_Z)^2}\right)\\
D & \sim & {\cal N}\left(x_1-x_2,
                         \sqrt{\sigma_1^2+\sigma_2^2}\right)\,.
\end{eqnarray}
The result can be interpreted in the following way:
  \begin{itemize}
  \item
  the uncertainty on the difference does not depend on the
  common offset uncertainty: whatever the value of the true ``zero'' is,
  it cancels in differences;
  \item
  in the sum, instead, the effect of the common
  uncertainty is somewhat amplified since it enters ``in phase''
  in the global uncertainty of each of the quantities.
  \end{itemize}
\end{itemize}
\subsection{Indirect calibration}
Let us use the result of the previous section to solve
another typical problem of measurements. Suppose that
after (or before, it doesn't matter) we have done the measurements
of $x_1$ and $x_2$ and we have the final result, summarized in
(\ref{eq:bivarm}), we know the ``exact'' value of $\mu_1$
(for example we perform the measurement on a reference).
Let us call it $\mu_1^\circ$.
Will this information provide a better knowledge of $\mu_2$?
In principle yes: the difference between $x_1$ and
$\mu_1^\circ$ defines the systematic error
(the true value of the ``zero'' $Z$). This error can
then be subtracted from $\mu_2$ to get a corrected value.
Also the overall uncertainty of $\mu_2$ should change, intuitively
it ``should'' decrease, since we are adding new information.
But its value doesn't seem to be obvious, since the
logical link between $\mu_1^\circ$ and $\mu_2$ is
$\mu_1^\circ\rightarrow Z \rightarrow \mu_2$.

The problem can be solved exactly using the concept of conditional
probability density function $f(\mu_2|\mu_1^\circ)$
(see (\ref{eq:y_cond}-\ref{eq:y_cond1})). We get
\begin{equation}
\Large{\mu_{2|\mu_1^\circ}
\sim} {\Large \cal N}\left(x_2+\frac{\sigma_Z^2}{\sigma_1^2+\sigma_Z^2}
(\mu_1^\circ-x_1),\  \sqrt{\sigma_2^2+\left(
               \frac{1}{\sigma_1^2}+\frac{1}{\sigma_Z^2}
                                    \right)^{-1}}\right)\,.
\label{eq:sigma2_1}
\end{equation}
The best value of $\mu_2$ is shifted by an amount $\Delta$,
with respect to the measured value $x_2$, which is
not exactly $x_1-\mu_1^\circ$, as
 was na\"\i vely guessed,
and the uncertainty depends on $\sigma_2$, $\sigma_Z$
and $\sigma_1$. It is easy to be convinced that the
exact result is more resonable than the (suggested) first guess.
Let us rewrite $\Delta$ in two different ways:
\begin{eqnarray}
\Delta& = &\frac{\sigma_Z^2}{\sigma_1^2+\sigma_Z^2}(\mu_1^\circ-x_1)
\label{eq:delta1}\\
      & = & \frac{1}{\frac{1}{\sigma_1^2}+\frac{1}{\sigma_Z^2}}
      \left[\frac{1}{\sigma_1^2}\cdot(x_1-\mu_1^\circ)
            + \frac{1}{\sigma_Z^2}\cdot 0
      \right]\,.
\label{eq:delta2}
\end{eqnarray}
\begin{itemize}
\item
Eq. (\ref{eq:delta1}) shows that one has to apply  the
correction $x_1-\mu_1^\circ$ only if $\sigma_1=0$. If instead
$\sigma_Z=0$ there is no correction to be applied, since the
instrument is perfectly calibrated. If $\sigma_1\approx \sigma_Z$
the correction is half of the measured difference between
$x_1$ and $\mu_1^\circ$;
\item
Eq. (\ref{eq:delta2}) shows explicitly what is going on and
why the result is consistent with the way we have modeled the uncertainties.
In fact we have performed two independent calibrations: one
of the offset and one of $\mu_1$. The best estimate of the
true value of the ``zero'' $Z$ is the weighted average of the
two measured offsets;
\item
the new uncertainty of $\mu_2$ (see (\ref{eq:sigma2_1}))
is a combination of $\sigma_2$ and the uncertainty of the
weighted average of the two offsets. Its value is smaller than
what one would have with only one calibration and, obviously,
larger than that due to the sampling fluctuations alone:
\begin{equation}
\sigma_2 \le \sqrt{\sigma_2^2+\frac{\sigma_1^2\sigma_Z^2}
                                   {\sigma_1^2+\sigma_Z^2}}
         \le \sqrt{\sigma_2^2+\sigma_Z^2}\,.
\end{equation}
\end{itemize}
\subsection{Counting measurements in presence of background}
As an example of a different kind of systematic effect, let
us think of counting experiments in the presence of background. For
example we are searching for a new particle, we make some
selection
cuts and count $x$ events. But we also expect
an average number of background events $\lambda_{B_\circ}\pm\sigma_B$,
where $\sigma_B$ is the standard uncertainty of $\lambda_{B_\circ}$,
\underline{not} to be confused with
$\sqrt{\lambda_{B_\circ}}$. What can we say about
$\lambda_S$, the true value of the average number associated to the
signal? First we will treat the case on which the determination of the
expected number
of background
events is well known ($\sigma_B/\lambda_{B_\circ}\ll 1$), and then the
general case:
\begin{description}
\item[\fbox{$\sigma_B/\lambda_{B_\circ}\ll 1$}:] the true value of the
sum of signal and background is $\lambda=\lambda_S+\lambda_{B_\circ}$.
The likelihood is
\begin{equation}
P(x|\lambda) =\frac{e^{-\lambda}\lambda^x}{x!}\,.
\end{equation}
Applying  Bayes's theorem we have
\begin{eqnarray}
f(\lambda_S|x,\lambda_{B_\circ})
&=& \frac{
e^{-(\lambda_{B_\circ}+\lambda_S)}
           (\lambda_{B_\circ}+\lambda_S)^x
f_\circ(\lambda_S)
       }
       { \int_0^\infty
e^{-(\lambda_{B_\circ}+\lambda_S)}
(\lambda_{B_\circ}+\lambda_S)^x
f_\circ(\lambda_S)d\lambda_S
       }\,.
\end{eqnarray}
Choosing again $f_\circ(\lambda_S)$ uniform (in a reasonable interval)
this gets simplified. The integral at the denominator
can be done easily by parts and the final result is:
\begin{eqnarray}
f(\lambda_S|x,\lambda_{B_\circ}) &=&
\frac{e^{-\lambda_S}(\lambda_{B_\circ}+\lambda_S)^x}
     {x!\sum_{n=0}^x\frac{\lambda_{B_\circ}^n}{n!}}
           \label{eq:inv_p_a}\\
F(\lambda_S|x,\lambda_{B_\circ})
&=& 1 -
\frac{e^{-\lambda_S} \sum_{n=0}^x\frac{(\lambda_{B_\circ}+\lambda_S)^n}{n!}}
     {\sum_{n=0}^x\frac{\lambda_{B_\circ}^n}{n!}} \label{eq:inv_p_b}\,.
\end{eqnarray}
{}From (\ref{eq:inv_p_a}-\ref{eq:inv_p_b})
it is possible to calculate in the usual way the best estimate
and the
credibility  intervals of $\lambda_S$.
Two particular cases are of interest:
\begin{itemize}
\item
if $\lambda_{B_\circ}=0$ then formulae
(\!\ref{eq:inv_poiss1}-\!\ref{eq:inv_poiss2}) are recovered.
In such a  case one measured count is enough to claim for a
signal (if somebody is willing to believe that
really $\lambda_{B_\circ}=0$ without any uncertainty$\ldots$);
\item
if $x=0$ then
\begin{equation}
f(\lambda|x,\lambda_{B_\circ}) = e^{-\lambda_S}\,,
\end{equation}
\underline{independently} of $\lambda_{B_\circ}$.
This result is not really obvious.
\end{itemize}
\item[\fbox{Any $g(\lambda_{B_\circ})$}:]
In the general case, the true value of the average
number of background events $\lambda_B$ is unknown.
We only known that it is distributed around
$\lambda_{B_\circ}$ with standard deviation $\sigma_B$
and probability density function $g(\lambda_B)$,
not necessarily a gaussian. What changes with respect to
the previous case is the initial distribution, now a joint
function
of $\lambda_S$ and of $\lambda_B$. Assuming
$\lambda_B$ and $\lambda_S$ independent the prior density function
is
\begin{equation}
f_\circ(\lambda_S,\lambda_B)=f_\circ(\lambda_S)g_\circ(\lambda_B)\,.
\end{equation}
We leave $f_\circ$ in the form of a joint distribution to
indicate that the result we shall get is the most general
for this kind of problem.
The likelihood, on the other hand, remains
the same as in  the previous example.
The inference of $\lambda_S$ is done in the
usual way, applying  Bayes' theorem and marginalizing with respect to
$\lambda_S$:
\begin{equation}
f(\lambda_S|x,g_\circ(\lambda_B))
=
\frac{\int e^{-(\lambda_B+\lambda_S)} (\lambda_B+\lambda_S)^x
      f_\circ(\lambda_S,\lambda_B)d\lambda_B}
{\int\!\!\int e^{-(\lambda_B+\lambda_S)} (\lambda_B+\lambda_S)^x
      f_\circ(\lambda_S,\lambda_B)d\lambda_Sd\lambda_B}\,.
\end{equation}
The previous case (formula (\ref{eq:inv_p_a}))
     is recovered if the only value allowed
for $\lambda_B$ is $\lambda_{B_\circ}$ and $f_\circ(\lambda_S)$
is uniform:
\begin{equation}
f_\circ(\lambda_S,\lambda_B) = k\delta(\lambda_B-\lambda_{B_\circ})\,.
\end{equation}
\end{description}

\section{Approximate methods}
\subsection{Linearization}
We have seen in the above examples how to use the
general formula (\ref{eq:ginf1}) for practical applications.
 Unfortunately,
when the problem becomes  more complicated
one starts facing integration problems.
 For this reason
approximate methods are generally used.
We will derive the approximation rules consistently
with the approach followed in these notes
and then the resulting formulae will
be compared with the ISO recommendations.
To do this, let us neglect for
a while all quantities of influence which could produce
unknown systematic errors. In this case (\ref{eq:ginf1})
can be replaced by (\ref{eq:ginf2}), which can be further simplified
if we remember that correlations between the results
are originated by unknown systematic errors. In absence of these,
the joint distribution of all quantities $\underline{\mu}$
is simply the product of
marginal ones:
\begin{equation}
f_{R_i}(\underline{\mu_i}) =
\prod_i f_{R_i}(\mu_i)\,,
\end{equation}
with
\begin{equation}
f_{R_i}(\mu_i) = f_{R_i}(\mu_i|x_i,\underline{h}_\circ) =
\frac{f(x_i|\mu_i, \underline{h}_\circ)f_\circ(\mu_i)}
     {\int f(x_i|\mu_i, \underline{h}_\circ)
           f_\circ(\mu_i) d\mu_i}\,.
\label{eq:ginf2a}
\end{equation}
The symbol $f_{R_i}(\mu_i)$ indicates that we are dealing with
\underline{\it raw values}\footnote{The choice of the adjective
``raw'' will become clearer in a later on.} evaluated at
$\underline{h}=\underline{h}_\circ$. Since for any variation
of $\underline{h}$ the inferred values of $\mu_i$ will change,
it is convenient, to name with the same subscript $R$ the
quantity obtained for $\underline{h}_\circ$:
\begin{equation}
f_{R_i}(\mu_i) \longrightarrow f_{R_i}(\mu_{R_i})\,.
\end{equation}

Let us indicate with $\widehat{\mu}_{R_i}$
and $\sigma_{R_i}$ the best estimates and the standard uncertainty
of the raw values:
\begin{eqnarray}
\widehat{\mu}_{R_i} &=& E[\mu_{R_i}]\\
\sigma_{R_i}^2 &=& Var(\mu_{R_i})\,.
\end{eqnarray}
 For any possible configuration of conditioning
hypotheses $\underline{h}$, \underline{\it corrected} values $\mu_i$
are obtained:
\begin{equation}
\mu_i=\mu_{R_i} +  g_i(\underline{h})\,.
\label{eq:correction}
\end{equation}
The function which relates the corrected value to the raw value
and to the systematic effects has been denoted by $g_i$ not to
be confused with a probability density function.
Expanding (\ref{eq:correction})
in series around $\underline{h}_\circ$
we finally arrive at the expression
which will allow us to make the approximated
evaluations of uncertainties:
\begin{equation}
\boxed{
\mu_i= \mu_{R_i}
      + \sum_l \frac{\partial g_i}{\partial h_l}
               (h_l-h_{\circ_l}) + \ldots\,
}
\label{eq:linearizzazione}
\end{equation}
(All derivatives are \underline{evaluated at}
$\{\widehat{\mu}_{R_i},\underline{h}_\circ\}$. To simplify
the notation a similar convention
will be used in the following formulae).

Neglecting the terms of the expansion above the first order,
and taking the expected values, we get
\begin{eqnarray}
\widehat{\mu}_i &=&  E[\mu_i] \nonumber \\
&\approx&  \widehat{\mu}_{R_i}\,; \\
\sigma_{\mu_i}^2 & = & E\left[(\mu_i-E[\mu_i])^2\right] \nonumber \\
       &\approx  &
            \sigma_{R_i}^2 +
            \sum_l\left(\frac{\partial g_i}{\partial h_l}\right)^2
            \!\sigma_{h_l}^2  \nonumber \\
       &  &\left\{ +
           2\sum_{l< m} \left(\frac{\partial g_i}{\partial h_l}\right)
                          \left(\frac{\partial g_i}{\partial h_m}\right)
                          \rho_{lm}\sigma_{h_l}\sigma_{h_m}
           \right\} \,; \label{eq:propag1} \\
Cov(\mu_i,\mu_j) &=& E\left[(\mu_i-E[\mu_i])(\mu_j-E[\mu_j])\right]
        \nonumber \\
  &\approx & \sum_l\left(\frac{\partial g_i}{\partial h_l}\right)
             \left(\frac{\partial g_j}{\partial h_l}\right)\sigma_{h_l}^2
              \nonumber \\
        &  & \left\{ +
             2\sum_{l< m} \left(\frac{\partial g_i}{\partial h_l}\right)
                          \left(\frac{\partial g_j}{\partial h_m}\right)
                          \rho_{lm}\sigma_{h_l}\sigma_{h_m}
             \right\}\,. \label{eq:propag2}
\end{eqnarray}
The terms included within $\{\cdot\}$ vanish if the unknown systematic
errors are uncorrelated, and the formulae become simpler.
 Unfortunately, very often this is not the
case, as when several calibration constants
are simultaneously obtained from a fit (for example, in most linear
fits slop and intercept have a correlation coefficient close to $-0.9$).

Sometimes the expansion
(\ref{eq:linearizzazione}) is not performed around the best values
of $\underline{h}$ but around their \underline{nominal values}, in the
sense that the correction for the known value of the systematic errors
has not yet been applied
(see section \ref{ss:known_syst}). In this case (\ref{eq:linearizzazione})
should be replaced by
\begin{equation}
\mu_i=\mu_{R_i}
      + \sum_l \frac{\partial g_i}{\partial h_l}
               (h_l-h_{N_l}) + \ldots\,
\label{eq:linearizzazione1}
\end{equation}
where the subscript $N$ stands for {\it nominal}. The best value of $\mu_i$
is then
\begin{eqnarray}
\widehat{\mu}_i &=&  E[\mu_i] \nonumber \\
&\approx&  \widehat{\mu}_{R_i}
  + E\left[\sum_l \frac{\partial g_i}{\partial h_l}(h_l-h_{N_l})\right]
  \nonumber \\
&=&  \widehat{\mu}_{R_i} + \sum_l\delta \mu_{i_l}
\label{eq:syst_corr}\,.
\end{eqnarray}
(\ref{eq:propag1}) and (\ref{eq:propag2}) instead remain valid,
with the condition that the derivative is calculated at
$\underline{h}_N$.
If \underline{$\rho_{lm}=0$} it is possible to
rewrite (\ref{eq:propag1}) and (\ref{eq:propag2})
in the following way, which is very convenient for practical applications:
\begin{eqnarray}
\sigma_{\mu_i}^2 &\approx  &
            \sigma_{R_i}^2 +
            \sum_l\left(\frac{\partial g_i}{\partial h_l}\right)^2
            \!\sigma_{h_l}^2 \\
            &=& \sigma_{R_i}^2 + \sum_l u_{i_l}^2
            \,; \label{eq:propag1a} \\
Cov(\mu_i,\mu_j)
  &\approx & \sum_l\left(\frac{\partial g_i}{\partial h_l}\right)
             \left(\frac{\partial g_j}{\partial h_l}\right)\sigma_{h_l}^2
              \\
  &=&         \sum_l s_{ij_{l}}
              \left|\frac{\partial g_i}{\partial h_l}\right|\sigma_{h_l}
              \left|\frac{\partial g_j}{\partial h_l}\right|\sigma_{h_l}
              \label{eq:propag2a} \\
  &=& \sum_l s_{ij_{l}}  u_{i_l}u_{j_l} \\
  &=&  \sum_l Cov_l(\mu_i,\mu_j)
\,. \label{eq:propag2b}
\end{eqnarray}
$u_{i_l}$ is the component to the standard uncertainty due to effect $h_l$.
$s_{ij_{l}}$ is equal to the product of signs of the derivatives,
which takes
 into account whether the uncertainties are positively or negatively
 correlated.

To summarize, when systematic effects are not correlated
with each other,
the following quantities are needed
 to evaluate the corrected result, the
combined uncertainties and the correlations:
\begin{itemize}
\item
the raw $\widehat{\mu}_{R_i}$ and $\sigma_{R_i}$;
\item
the best estimates of the corrections $\delta \mu_{i_l}$ for each
systematic effect $h_l$;
\item
the best estimate of the standard deviation $u_{i_l}$ due to the
imperfect knowledge of the systematic effect;
\item
for any pair $\{\mu_i,\mu_j\}$ the sign of the correlation
$s_{ij_{l}}$ due to the effect $h_l$.
\end{itemize}

In High Energy Physics applications it is frequently the case that
the derivatives appearing in
(\ref{eq:syst_corr}-\ref{eq:propag2a}) cannot be calculated directly,
as for example when $h_l$ are parameters of a simulation program,
or acceptance cuts. Then variations of $\underline{\mu}_i$ are
usually studied varying a particular $h_l$ within
a {\it reasonable} interval, holding the other influence
quantities at the nominal value.
$\delta \mu_{i_l}$ and $u_{i_l}$ are calculated from
the interval $\pm\Delta_i^\pm$ of variation of the true value
for a given variation $\pm\Delta_{h_l}^\pm$ of $h_l$
and from the probabilistic meaning of the  intervals (i.e.
from the assumed distribution of the true value).
This empirical procedure for determining
 $\delta \mu_{i_l}$ and $u_{i_l}$ has the advantage that it
 can take into account non linear effects, since it
 directly measures the difference
  $\widehat{\mu}_i -  \widehat{\mu}_{R_i}$ for a given difference
 $h_l-h_{N_l}$.

Some examples are given
in section \ref{ss:examples},
and two typical experimental applications
will be discussed in more detail
in section \ref{sec:cov}.
\subsection{BIPM and ISO recommendations}
In this section we compare the results obtained in the previous
section with the recommendations of the Bureau International des Poids
et M\'esures (BIPM) and the International Organization for Standardization
(ISO) on {\sl ``the expression of experimental uncertainty''}.
\begin{enumerate}
\item
\begin{quote}
{\sl\small The uncertainty in the result of a measurement generally consists
of several components which may be grouped into two categories according
to the way in which their numerical value is estimated:
\begin{description}
\item[A:] those which are evaluated by statistical methods;
\item[B:] those which are evaluated by other means.
\end{description}

There is not always a simple correspondence between the classification into
categories A or B and the previously used classification into
``random'' and ``systematic'' uncertainties.
The term ``systematic uncertainty'' can be misleading
and should be avoided.

The detailed report of the uncertainty should
consist of a complete list of the components, specifying for each the
method used to obtain its numerical result.
}
\end{quote}
 Essentially
the first recommendation states  that all  uncertainties
can be treated probabilistically. The distinction between types A and B
is subtle and can be misleading if one thinks of ``statistical
methods'' as synonymous with ``probabilistic methods'' - as currently
done in High Energy Physics.
Here ``statistical'' has the classical meaning of
repeated measurements.

\item
\begin{quote}
{\sl \small
The components in category A are characterized by the estimated
 variances $s_i^2$ (or the estimated ``standard deviations'' $s_i$)
 and the number of degrees of freedom $\nu_i$. Where appropriate,
 the covariances should be given.
}
\end{quote}
The estimated variances correspond to $\sigma_{R_i}^2$ of the
previous section. The degrees of freedom have are related to small
samples and to the {\it Student t} distribution. The problem of
small samples is not discussed in these notes,
but clearly this recommendation
is a relic of frequentistic methods. With the approach followed in
theses notes
there is no need to talk about degrees of freedom,
since the Bayesian inference defines the final
probability function $f(\mu)$ completely.

\item
\begin{quote}
{\sl\small The components in category B should be characterized
by quantities $u_j^2$, which may be considered as approximations
to the corresponding variances, the existence of which is assumed. The
quantities $u_j^2$ may the treated like variances and the quantities
$u_j$ like standard deviations. Where appropriate,
the covariances should be treated in a similar way.
}
\end{quote}
Clearly, this
recommendation is meaningful only in a Bayesian framework.

\item
\begin{quote}
{\sl \small
 The combined uncertainty should be characterized by the numerical
  value obtained by applying the usual method for the combination
  of variances. The combined uncertainty and its components should
  be expressed in the form of ``standard deviations''.
}
\end{quote}
This is
what we have found in (\ref{eq:propag1}-\ref{eq:propag2}).
\item
\begin{quote}
{\sl \small
If, for particular applications, it is necessary to multiply
the combined uncertainty by a factor to obtain
an overall uncertainty, the multiplying factor used must always
stated.
}
\end{quote}
This last recommendation states once more that the uncertainty is ``by
default'' the standard deviation
of the true value distribution. Any other quantity
calculated
to obtain a credibility interval with a certain
probability level should be clearly stated.
\end{enumerate}
To summarize, these are the basic ingredients of
the BIPM/ISO recommendations:
\begin{description}
\item[subjective definition of probability:] it allows variances to
be assigned conceptually to any physical quantity
which has an uncertain value;
\item[uncertainty as standard deviation]\
\begin{itemize}
\item
it is ``standard'';
\item
the rule of combination
 (\ref{eq:linc2}-\ref{eq:linc5}) applies to standard deviations and not
to confidence intervals;
\end{itemize}
\item[combined standard uncertainty:] it is obtained by the usual formula
of ``error propagation'' and it makes use on variances,
covariances and first derivatives;
\item[central limit theorem:] it makes, under proper conditions,
 the true value
normally distributed if one has several sources of uncertainty.
\end{description}

Consultation of the {\it Guide}\cite{ISO}
 is recommended
for further explanations about the justification of the norms,
for the
description of evaluation procedures, as well for as examples.
 I would just like to
end this section with some examples of the evaluation of
type B uncertainties and with some words of caution
concerning the use of approximations and of linearization.
\subsection{Evaluation of type B uncertainties}
The ISO {\it Guide} states that
\begin{quote}
{\sl \small
For estimate $x_i$ of an input quantity\footnote{
By ``input quantity'' the ISO {\it Guide} mean
any of the contributions $h_l$
or $\mu_{R_i}$ which enter into (\ref{eq:propag1}-\ref{eq:propag2}).}
$X_i$ that has not been
obtained from repeated observations, the $\ldots$ standard
uncertainty $u_i$ is evaluated by scientific judgement based on all the
available information on the possible variability of $X_i$. The pool
of information may include
\begin{itemize}
\item
previous measurement data;
\item
experience with or general knowledge of the behaviour and properties of
relevant materials and instruments;
\item
manufacturer's specifications;
\item
data provided in calibration and other certificates;
\item
uncertainties assigned to reference data taken from handbooks.
\end{itemize}
}
\end{quote}
\subsection{Examples of type B uncertainties}\label{ss:examples}
\begin{enumerate}
\item
A manufacturer's calibration certificate states that the uncertainty,
defined as  \underline{$k$ standard deviations},
is ``$\pm\Delta$'':
$$u=\frac{\Delta}{k}\,.$$
\item
A result
is reported
in a publication
as $\overline{x}\pm \Delta$,
stating that the average has been performed on 4 measurements
and the uncertainty is a $95\,\%$ confidence interval.
One has to conclude that the confidence interval has been calculated
using the \underline{{\it Student} $t$}:
$$u=\frac{\Delta}{3.18}\,.$$
\item
a manufacturer's specification states that the
error on a quantity should not exceed $\Delta$. With this
limited information one has to assume a \underline{uniform distribution}:
$$u=\frac{2\Delta}{\sqrt{12}}=\frac{\Delta}{\sqrt{3}}\,;$$
\item
A physical parameter of a Monte Carlo is believed to lie in the
interval of $\pm \Delta$ around its best value,
but not with uniform distribution:
the probability that the parameter is
at center is higher than than that it is at the edges of the
interval. With this information a \underline{triangular distribution}
can be reasonably assumed:
$$u=\frac{\Delta}{\sqrt{6}}\,.$$
{\bf Note} that the coefficient in front of $\Delta$
changes from the $0.58$ of the
previous example to the $0.41$ of this. If the interval
$\pm\Delta$ were a $3\sigma$ interval then the coefficient
would have been
equal to $0.33$. These variations - to
be considered extreme - are smaller than
the statistical fluctuations of empirical standard
deviations estimated from $\approx 10$ measurements.
This shows that one should not be worried that the type B
uncertainties are less accurate than
 type A, especially if one tries
to model
 the distribution of the physical quantity
{\it honestly}.
\item
The absolute energy calibration of  an electromagnetic
calorimeter module is not
exactly known and it is estimated to be between the nominal one
and $+10\,\%$. The ``statistical'' resolution is known by test beam
 measurements to be $18\%/\sqrt{E/\mbox{GeV}}$. What is the uncertainty
 on the energy measurement of an electron which has apparently released
 30 GeV?
 \begin{itemize}
 \item
 The energy has to be \underline{corrected} for the best estimate
 of the calibration constant: $+5\,\%$:
 $$E_R=31.5\pm 1.0\,\mbox{GeV}\,.$$
 \begin{itemize}
 \item
 assuming a \underline{uniform}
 distribution of the true calibration constant:
 $u=31.5 \times 0.1/\sqrt{12} = 0.9\, \mbox{GeV}$:
 $$E=31.5\pm 1.3\, \mbox{GeV}\,;$$
 \item
 assuming a \underline{triangular} distribution: $u=1.3\, \mbox{GeV}$:
 $$E=31.5\pm 1.6\, \mbox{GeV}\,;$$
 \end{itemize}
 \item
 interpreting the maximum deviation from the nominal calibration
 as uncertainty
 (see comment at the end of section \ref{ss:known_syst}):
 $$E=30.0\pm 1.0\pm 3.0 \, \mbox{GeV} \rightarrow E=30.0\pm 3.2
 \, \mbox{GeV} \,;$$
 As already remarked earlier in these notes,
 while reasonable assumptions (in this case
 the first two) give consistent results, this is not true if one
 makes inconsistent use of the information just for the sake
 of giving ``safe'' uncertainties.
 \end{itemize}
\item
As a more realistic and slightly more complicated example, let us
take the case of a measurement of two physical quantities performed
with the same apparatus. The result, before the correction
 for systematic
effects and only with type {\it A} uncertainties is
$\mu_{R_1}=1.50\pm 0.07$ and
$\mu_{R_2}=1.80\pm 0.08$ (arbitrary units).
Let us assume that the measurements
depend on eight influence quantities $h_l$, and that most of them
influence both physical quantities. For simplicity we consider
$h_l$ to be independent from each other.
{ \footnotesize
\begin{table}
\begin{center}
\begin{tabular}{|cc|ccc|ccc|c|} \hline
\multicolumn{2}{|c|}{$h_l$} &
\multicolumn{3}{|c|}{$\mu_{R_1}=1.50\pm 0.07$} &
\multicolumn{3}{|c|}{$\mu_{R_2}=1.80\pm 0.08$} &
correlation  \\ \hline
$l$   & model &
$\Delta_{1_l}^\pm$ & $\delta \mu_{1_l}$ & $u_{1_l}$ &
$\Delta_{2_l}^\pm$ & $\delta \mu_{2_l}$ & $u_{2_l}$ &
$Cov_l(\mu_1,\mu_2)$ \\ \hline
1  & normal    &
   $\pm 0.05$  &   0    &  0.05  &
       0       &   0    &   0    &    0 \\
2  & normal    &
   $ 0 $       &   0    &  0  &
   $\pm 0.08$  &   0.00 &  0.08 &    0 \\
3  & normal    &
   $ \left\{^{+0.10}_{-0.04}\right. $  & $+0.03$ &  0.07  &
   $ \left\{^{+0.12}_{-0.05}\right. $  & $+0.035$&  0.085    & $+0.0060$ \\
4  & uniform   &
   $ \left\{^{+0.00}_{-0.15}\right. $  & $-0.075$    &  0.04  &
   $ \left\{^{+0.07}_{-0.00}\right. $  & $+0.035$    &  0.02    & $-0.0008$ \\
5  & triangular &
   $\pm 0.10$  &  0.00  &  0.04  &
   $\pm 0.02$  &  0.00  &  0.008 &  $+0.0003$ \\
6  & triangular&
   $ \left\{^{+0.02}_{-0.08}\right. $   &  $-0.03$    &  0.02  &
   $ \left\{^{+0.01}_{-0.03}\right. $   &  $-0.010$   &  0.008 & $+0.0016$ \\
7  & uniform   &
   $ \left\{^{+0.10}_{-0.06}\right. $   &  $+0.02$    &  0.05  &
   $ \left\{^{+0.14}_{-0.06}\right. $   &  $+0.04$    &  0.06  &   $+0.0030$ \\
8  & triangular&
   $ \left\{^{+0.03}_{-0.02}\right. $   &  $+0.005$    & 0.010 &
   $\pm 0.03$             &   0.000     & 0.012    &  $+0.00012$ \\ \hline
``$\sum_{h_l}$'' & normal   &
   & $-0.05$ & 0.12 & & $+0.10$ & 0.13 & +0.010 \\ \hline
\multicolumn{2}{|c|}{}  &
\multicolumn{3}{|c|}{$\mu_{1}=1.45\pm 0.14$} &
\multicolumn{3}{|c|}{$\mu_{2}=1.90\pm 0.15$} &
+0.010  \\
\multicolumn{2}{|c|}{}  &
\multicolumn{3}{|c|}{$(\mu_{1}=1.45\pm 0.10 \pm 0.10$)} &
\multicolumn{3}{|c|}{$(\mu_{2}=1.90\pm 0.11 \pm 0.10$)} &
($\rho = +0.49$)  \\ \hline
\multicolumn{2}{|c|}{}   &
\multicolumn{6}{|c|}{$\mu_2+\mu_1 = 3.35\pm 0.25$} & \\ \hline
\multicolumn{2}{|c|}{}   &
\multicolumn{6}{|c|}{$\mu_2-\mu_1 = 0.45\pm 0.15$} & \\ \hline
\multicolumn{2}{|c|}{}   &
\multicolumn{6}{|c|}{$\overline{\mu} = 1.65\pm 0.12$} & \\ \hline
\end{tabular}
\end{center}
\caption{\sf Example of the result of two physical quantities
corrected by several systematic effects (arbitrary units).}
\label{tab:syst}
\end{table}
} 
Tab. \ref{tab:syst} gives
the details of correction for the systematic effects and
of the uncertainty evaluation,
performed using (\ref{eq:syst_corr}), (\ref{eq:propag1a})
and (\ref{eq:propag2a}).
 To see the importance
of the correlations, the result of the sum and of the difference
of $\mu_1$ and $\mu_2$ is also reported.

In order to split  the final result into ``individual'' and ``common''
uncertainty (between parenthesis in Tab. \ref{tab:syst})
we have to remember that,
if the error is additive, the covariance between $\mu_1$
and $\mu_2$ is
given by the variance
of the unknown systematic error (see \ref{eq:covm1m2}).

The average $\overline{\mu}$ between $\mu_1$ and $\mu_2$,
assuming it
has a physical meaning, can be evaluated either using
the results of section \ref{sec:cov}, or simply calculating the
average weighted with the inverse of the
variances due to the individual uncertainties,
and then adding quadratically
the common uncertainty at the end. Also $\overline{\mu}$ is reported in
Tab. \ref{tab:syst}.
\end{enumerate}
\subsection{Caveat concerning a blind use of approximate methods}
The mathematical apparatus of variances and covariances
of (\ref{eq:propag1}-\ref{eq:propag2})
is often seen as the most complete description of uncertainty
and in most cases used blindly in further uncertainty calculations.
It must be clear, however,  that
this is just an approximation based on linearization. If the
function which relates the corrected value to the raw value and the
systematic effects is not linear then the linearization may cause
trouble.
An interesting case is discussed in section \ref{sec:cov}.

There is another problem which may arise from the simultaneous use
of Bayesian estimators \underline{and} approximate methods.
Let us introduce the problem with an example.
\begin{description}
\item[Example 1:] 1000 independent
measurements of the efficiency of a detector have been
performed (or 1000 measurements of branching ratio, if you
prefer).
Each measurement was
carried out on a base of 100 events and each time 10 favorable events
were observed (this obviously strange - though not impossible -
but it simplifies the calculations). The result of each
measurement will be (see (\ref{eq:infbinom1}-\ref{eq:infbinom2})):
\begin{eqnarray}
\widehat{\epsilon}_i &=& \frac{10+1}{100+2} = 0.1078 \\
\sigma(\epsilon_i) & = & \sqrt{\frac{11\cdot 91}{103\cdot 102^2}} = 0.031\,;
\end{eqnarray}
Combining the 1000 results using
the standard weighted average
procedure gives
\begin{equation}
\epsilon = 0.1078 \pm 0.0010\,.
\end{equation}
Alternatively, taking the complete set of results to be equivalent to
100000 trials with 10000 favorable events, the combined result is
\begin{equation}
\epsilon^\prime = 0.10001\pm 0.0009\,.
\end{equation}
(the same as if one had used
the Bayes theorem
iteratively
to infer $f(\epsilon)$ from the the partial 1000 results.)
The conclusions are in disagreement and the
first result is clearly mistaken.
\end{description}
The same problem arises  in the case of inference of the
 Poisson distribution
parameter $\lambda$ and, in general, whenever $f(\mu)$ is not symmetrical
around $E[\mu]$.
\begin{description}
\item[Example 2:] Imagine an experiment running
continuously for one year,
searching for monopoles and identifying none.
The consistency with zero can be stated either
quoting $E[\lambda]=1$ and $\sigma_\lambda=2$, or
a $95\,\%$
upper limit $\lambda < 3$. In terms of rate (number of monopoles
per day) the result would be either $E[r]=2.7\cdot 10^{-3}$,
$\sigma(r)=5.5\cdot 10^{-3}$, or an  upper limit $r<8.2\cdot 10^{-3}$.
It easy to show that, if we take the 365 results for
each of the running days  and combine them
using
the standard weighted average, we get
 $r=1.0\pm 0.1$ monopoles/day! This absurdity is not
 caused by
 the Bayesian method, but by the standard rules for combining the
 results (the weighted average formulae
 (\ref{eq:waver1}) and (\ref{eq:waver2})
 are derived from the normal distribution hypothesis).
 Using Bayesian inference  would have led to
 a consistent and reasonable result no matter how the 365 days of running
 had  been subdivided for partial analysis.
\end{description}
This suggests that in some cases it could be preferable to
give the  result in terms
of the value of $\mu$
which maximizes $f(\mu)$ ($p_m$ and $\lambda_m$ of
sections \ref{ss:binom} and \ref{ss:poisson}). This way of presenting
the results is similar to that suggested by the maximum likelihood
approach, with the difference that for $f(\mu)$ one should take
the final probability density function and not simply the likelihood.
Since it is practically impossible to summarize
the outcome of an inference
in only two
numbers (best value and uncertainty),
a description of the method
used to evaluate them should be provided, except when
$f(\mu)$ is approximately normally distributed
(fortunately this happens most of the time).

\section{Indirect measurements}
Conceptually this is a very simple task in the Bayesian framework,
whereas
the frequentistic one requires a lot of gymnastics
going back and forward from the logical level of true values
to the logical level of estimators. If one accepts that
the true values are just random variables\footnote{
To make the formalism lighter, let us call
both the
random variable associated to the quantity and the quantity itself
by the same name  $X_i$ (instead of $\mu_{x_i}$).},
 then,
calling $Y$ a function of other quantities $X$,
each
having a probability density function $f(x)$,
the probability density function
of $Y$ $f(y)$ can be calculated with the
 standard formulae
which follow from the rules
probability.
Note that in the approach presented in these notes
uncertainties due to systematic effects
are treated in the same way as indirect measurements.
It is worth repeating that
there is no conceptual
distinction between various components
of the measurement uncertainty.
When approximations are sufficient,
formulae (\ref{eq:propag1}) and (\ref{eq:propag2}) can be used.

Let us take an example for which the linearization does not give
the right result:
\begin{description}
\item[Example:] The speed of a proton is measured with a time-of-flight
system. Find the $68$, $95$ and $99\,\%$ probability intervals
for the energy, knowing that $\beta=v/c=0.9971$,
and that distance and time have been measured with a $0.2\,\%$ accuracy.

The relation
$$E=\frac{mc^2}{\sqrt{1-\beta^2}}$$
is strongly non linear. The results given by the approximated method
and the correct one are, respectively:\\
\begin{center}
\begin{tabular}{|c|c|c|}\hline
C.L. & linearization & correct \\
(\%) &   $E$ (GeV)   & $E$ (GeV)        \\ \hline
68 & $6.4 \le E \le 18$  & $8.8 \le E \le 64$ \\
95   & $0.7 \le E \le 24$  & $7.2 \le E < \infty$ \\
99   &  $0. \le E \le 28$  & $6.6 \le E < \infty$ \\ \hline
\end{tabular}
\end{center}
\end{description}

\section{Covariance matrix of
experimental results}\label{sec:cov}
\subsection{Building the covariance matrix of experimental data}
In physics applications,
it is rarely the case that the covariance
between the best estimates
of two physical quantities\footnote{
In this section\cite{syst}
  the symbol $X_i$ will indicate
the variable associated to the $i$-th physical quantity
and $X_{ik}$ its $k$-th direct measurement; $x_i$
the best estimate of
its value, obtained by an average over many direct measurements or
indirect measurements, $\sigma_i$ the
standard deviation, and $y_i$ the value corrected for the calibration
constants. The weighted average of several $x_i$ will be
denoted by $\overline{x}$.},
each given by the
arithmetic average of direct measurements
($x_i = \overline{X_i} = \frac{1}{n}\sum_{k=1}^n X_{ik}$),
can be
evaluated from the sample covariance of the two averages
\begin{equation}
Cov(x_i,x_j)
=\frac{1}{n(n-1)}\sum_{k=1}^{n}(X_{ik}-\overline{X}_i)(X_{jk}-\overline{X}_j)\
{}.
\label{eq:cov1}
\end{equation}

More frequent is the well understood
case in which the physical
quantities are obtained as a result of a $\chi^2$
minimization, and the terms of the inverse of the
covariance matrix are related to the curvature of $\chi^2$
at its minimum:
\begin{equation}
\left(V^{-1}\right)_{ij} = \frac{1}{2}\left.
                \frac{\partial^2\chi^2}{\partial X_i\partial X_j}
                \right|_{x_i,x_j}\, .
\label{eq:cov2}
\end{equation}

In most cases one determines independent values
of physical quantities
with the same
detector, and the correlation between them originates from
the detector calibration uncertainties.
Frequentistically, the use of (\ref{eq:cov1}) in this case
would correspond to having a ``sample
of detectors'', with each of which a
 measurement of all the physical quantities is performed.

A way of building the covariance matrix from the direct measurements
is to consider the original measurements and the calibration
constants as a common set of independent and uncorrelated
measurements, and then to calculate corrected values that take into
account the calibration constants.
The variance/covariance propagation will automatically provide the full
covariance matrix of the set of results.
Let us derive it for two cases that happen frequently, and then
proceed to the general case.
\subsubsection{Offset uncertainty}
Let
 $x_i\pm\sigma_i$ be the $i=1\ldots n$ results
of independent measurements
and ${\bf V}_X$ the (diagonal) covariance matrix.
Let us assume that they are all affected by the same calibration
constant $c$, having a standard uncertainty $\sigma_c$.
The corrected results are then $y_i = x_i + c$.
We can  assume, for
simplicity, that the most probable value of $c$ is 0, i.e.
the detector is well calibrated.
One has to
consider the calibration constant as
the physical quantity $X_{n+1}$, whose best estimate is
$x_{n+1} = 0$.
A term $V_{X_{n+1,n+1}} = \sigma^2_c$ must be added to the
 covariance matrix.

The covariance matrix of the corrected results is given by the
transformation:
\begin{equation}
{\bf V}_Y = {\bf M}{\bf V}_X{\bf M}^T\,,
\end{equation}
where $M_{ij}= \left.\frac{\partial Y_i}{\partial X_j}
               \right|_{x_j}$.
The elements of ${\bf V}_Y$ are given by
\begin{equation}
     V_{Y_{kl}} = \sum_{ij}
                    \left.
                    \frac{\partial Y_k}{\partial X_i}
                    \right|_{x_i}
                    \left.
                    \frac{\partial Y_l}{\partial X_j}
                    \right|_{x_j}
                    V_{X_{ij}}\, .
\end{equation}
In this case we get:
\begin{eqnarray}
\sigma^2(Y_i) & = & \sigma_i^2+\sigma_c^2 \\
Cov(Y_i,Y_j) & = & \sigma_c^2 \hspace{1.3 cm} (i\ne j)\\
\rho_{ij} & = & \frac{\sigma_c^2}
                     {\sqrt{\sigma_i^2+\sigma_c^2}
                     \sqrt{\sigma_j^2+\sigma_c^2}} \\
          &=&     \frac{1}
                     {\sqrt{1+\left(\frac{\sigma_i}{\sigma_c}\right)^2}
                      \sqrt{1+\left(\frac{\sigma_j}{\sigma_c}\right)^2}}\, ,
\end{eqnarray}
reobtaining the results of section \ref{sec:off_err}.
The total uncertainty on the single measurement is given by the
combination in quadrature of the individual and the common
standard uncertainties, and all the covariances are equal to $\sigma^2_c$.
To verify, in a simple case, that the result is reasonable,
let us consider only two independent quantities $X_1$ and $X_2$,
and a calibration constant $X_3  = c$, having
an expected value equal to zero. From these we can calculate
the correlated quantities $Y_1$ and $Y_2$ and finally their
sum ($S\equiv Z_1$) and difference ($D\equiv Z_2$). The results are:
\begin{eqnarray}
{\bf V}_Y &=&
\left( \begin{array}{cc}
                 \sigma_1^2+\sigma_c^2 & \sigma_c^2 \\
                 \sigma_c^2 & \sigma_2^2+\sigma_c^2
                 \end{array}
           \right)\\
& & \\
& & \\
{\bf V}_Z &=&
\left( \begin{array}{cc}
                 \sigma_1^2 + \sigma_2^2+
                 4\cdot\sigma_c^2 &\ \sigma_1^2-\sigma_2^2 \\
                 \sigma_1^2-\sigma_2^2 & \ \sigma_1^2 + \sigma_2^2
                 \end{array}
           \right) \, .
\end{eqnarray}
It follows that
\begin{eqnarray}
\sigma^2(S) & = & \sigma_1^2 +\sigma_2^2 +(2\cdot\sigma_c)^2 \\
\sigma^2(D) & = & \sigma_1^2 + \sigma_2^2\, ,
\end{eqnarray}
as intuitively expected.
\subsubsection{Normalization uncertainty}
Let us consider now the case where the calibration constant
is the scale factor $f$, known with a standard uncertainty $\sigma_f$.
Also in this case, for simplicity and without losing generality,
let us suppose that the most probable value of $f$ is 1.
Then
$ X_{n+1}  =  f$, i.e.
$ x_{n+1} = 1$, and
$V_{X_{n+1,n+1}} = \sigma^2_f$.
Then
\begin{eqnarray}
\sigma^2(Y_i) & = & \sigma_i^2 + \sigma_f^2  x_i^2 \\
Cov(Y_i,Y_j) & = & \sigma_f^2  x_i x_j
             \hspace{1.3 cm}(i\ne j) \\
\rho_{ij} & = & \frac{x_i x_j}
                     {\sqrt{x_i^2+\frac{\sigma_i^2}{\sigma_f^2}}
                      \sqrt{x_j^2+\frac{\sigma_j^2}{\sigma_f^2}}} \\
|\rho_{ij}| & = & \frac{1}
                     {\sqrt{1+\left(\frac{\sigma_i}{\sigma_f x_i}\right)^2}
                      \sqrt{1+\left(\frac{\sigma_j}{\sigma_f x_j}\right)^2}
                     } \\
\, .
\end{eqnarray}
To verify the results let us consider two independent measurements
$X_1$ and $X_2$, let us
calculate the correlated quantities $Y_1$ and $Y_2$, and finally their
product ($P\equiv Z_1$) and their ratio ($R\equiv Z_2$):
\begin{eqnarray}
{\bf V}_Y & = &
\left( \begin{array}{cc}
                 \sigma_1^2+\sigma_f^2\cdot x_1^2
                 & \sigma_f^2\cdot x_1\cdot x_2 \\
                 & \\
                 \sigma_f^2\cdot x_1\cdot x_2
                 & \sigma_2^2+\sigma_f^2\cdot x_2
                 \end{array}
           \right)\\
& & \\
& & \\
{\bf V}_Z & = &
\left( \begin{array}{cc}
                 \sigma_1^2\cdot x_2^2 +
                 \sigma_2^2\cdot x_1^2 +
                 4\cdot\sigma_f^2\cdot x_1^2\cdot x_2^2 &
                 \ \sigma_1^2-\sigma_2^2\cdot\frac{x_1^2}{x_2^2} \\
                 & \\
                 \sigma_1^2 -
                 \sigma_2^2\cdot\frac{x_1^2}{x_2^2}
                   &
                 \ \frac{\sigma_1^2}{x_2^2} +
                  \sigma_2^2\cdot\frac{x_1^2}{x_2^4}
                 \end{array}
           \right)\, .
\end{eqnarray}
It follows that:
\begin{eqnarray}
\sigma^2(P) & = & \sigma_1^2\cdot x_2^2 +
                  \sigma_2^2\cdot x_1^2 +
                  (2\cdot\sigma_f\cdot x_1\cdot x_2)^2 \\
\sigma^2(R) & = & \frac{\sigma_1^2}{x_2^2} +
                  \sigma_2^2\cdot\frac{x_1^2}{x_2^4} \, .
\end{eqnarray}
Just as an unknown common offset error cancels in differences
and is enhanced in sums, an unknown normalization error has
a similar effect
on the ratio and the product. It is also interesting to calculate
the standard uncertainty of a difference in case of a normalization error:
\begin{eqnarray}
\sigma^2(D) & = & \sigma_1^2+\sigma_2^2
                  +\sigma_f^2\cdot(x_1-x_2)^2\, .
\end{eqnarray}
The contribution from an unknown
 normalization error vanishes if the two
values are equal.
\subsubsection{General case}
Let us assume there are $n$ independently
measured values $x_i$ and
$m$
 {\it calibration constants} $c_j$
with their covariance matrix
${\bf V}_c$. The latter
can also be theoretical parameters influencing the data, and
moreover they may be
correlated, as usually
happens if, for example, they are parameters of a calibration fit.
We can then include the $c_j$ in the vector that contains the
measurements and ${\bf V}_c$ in the covariance matrix ${\bf V}_X$:
\begin{equation}
\underline{x}  =  \left( \begin{array}{c}
                             x_1 \\
                             \vdots \\
                             x_n \\
                             c_1 \\
                             \vdots \\
                             c_m
                             \end{array}
                       \right)\, , \ \ \ \ \ \ \
{\bf V}_X  =  \left( \begin{array}{cccc|c}
                      \sigma_1^2 & 0 & \cdots & 0 & \\
                      0 & \sigma_2^2  & \cdots &  0 & \\
                      \cdots & \cdots  & \cdots &  \cdots & {\bf 0} \\
                      0 & 0 & \cdots & \sigma_n^2 & \\
                      \hline
                      & & {\bf 0} & & {\bf V}_c
                      \end{array}
               \right)\, .
\end{equation}
The corrected  quantities are obtained from the most general
function
\begin{equation}
Y_i = Y_i(X_i,\underline{c}) \hspace{2.0 cm} (i=1,2, \ldots,
n)\, ,
\end{equation}
and the covariance matrix
 ${\bf V}_Y$ from the covariance propagation
${\bf V}_Y = {\bf M}{\bf V}_X{\bf M}^T$.

As a frequently encountered example, we can think of several
normalization constants, each affecting a subsample of the data -
as is
the case where each of several detectors
measures a set of physical quantities.
Let us consider consider just three quantities
 ($X_i$) and three
uncorrelated
normalization standard uncertainties ($\sigma_{f_j}$),
the first one common to
 $X_1$ and $X_2$, the second to
$X_2$ and $X_3$ and the third to all three.
We get the following covariance matrix:
$$
\left( \begin{array}{ccc}
                 \sigma_1^2 +
                 \left(\sigma_{f_1}^2 +
                 \sigma_{f_3}^2 \right)\cdot x_1^2
           &     \left(\sigma_{f_1}^2  +
                 \sigma_{f_3}^2 \right)\cdot x_1\cdot x_2
           &     \sigma_{f_3}^2\cdot x_1\cdot x_3   \\
           & & \\
                 \left(\sigma_{f_1}^2 +
                 \sigma_{f_3}^2\right) \cdot x_1\cdot x_2
           &     \sigma_2^2 +
                 \left(\sigma_{f_1}^2 + \sigma_{f_2}^2 +
                 \sigma_{f_3}^2 \right)\cdot x_2^2
           &     \left( \sigma_{f_2}^2 +
                 \sigma_{f_3}^2 \right)\cdot x_2\cdot x_3   \\
           & & \\
                 \sigma_{f_3}^2 \cdot x_1 \cdot x_3
           &     \left( \sigma_{f_2}^2 +
                 \sigma_{f_3}^2\right) \cdot x_2\cdot x_3
           &     \sigma_3^2 +
                 \left( \sigma_{f_2}^2 +
                 \sigma_{f_3}^2\right) \cdot x_3^2
                 \end{array}
           \right)\, .
$$
\subsection{Use and misuse of the covariance matrix to fit correlated data}
\subsubsection{Best estimate of the true value from two correlated
values.}
Once the covariance matrix is built
 one uses it in a $\chi^2$ fit to get the
 parameters of a function.
The quantity to be minimized is
$\chi^2$, defined as
\begin{equation}
 \chi^2 = \underline{\Delta}^T {\bf V}^{-1}\underline{\Delta}\, ,
 \end{equation}
 where $\underline{\Delta}$ is the vector of the differences
 between the theoretical and the experimental values.
Let us
consider the simple case in which two results of the same physical quantity
are available, and the individual and the common
standard uncertainty are known.
The best estimate of the true value of the physical quantity
is then obtained by fitting the constant
$Y=k$ through the data points. In this simple case
the $\chi^2$ minimization can be performed easily.
We will consider
the two cases of offset and normalization uncertainty. As before,
we assume that the detector is well calibrated, i.e. the most
probable value of the calibration constant is, respectively
for the two cases, 0 and 1, and hence $y_i=x_i$
\subsubsection{Offset uncertainty}
Let $x_1\pm\sigma_1$ and $x_2\pm\sigma_2$ be the two measured values,
and $\sigma_c$ the common standard uncertainty.
The $\chi^2$ is
\begin{eqnarray}
\chi^2 & = &  \frac{1}{D} \left[
              (x_1-k)^2\cdot (\sigma_2^2+\sigma_c^2)
              +(x_2-k)^2\cdot (\sigma_1^2+\sigma_c^2)\right. \\
       & &    \hspace{0.7 cm} \left. -2\cdot (x_1-k)\cdot
              (x_2-k)\cdot\sigma_c^2
              \right]\, ,
\end{eqnarray}
where
$D=\sigma_1^2\cdot\sigma_2^2+ (\sigma_1^2+\sigma_2^2)\cdot\sigma_c^2$
 is the determinant of the covariance matrix.

Minimizing $\chi^2$ and using the second derivative
calculated at the minimum we obtain the best value of $k$ and its
standard deviation:
\begin{eqnarray}
\widehat{k} &=& \frac{x_1\cdot\sigma_2^2+x_2\cdot\sigma_1^2}
                     {\sigma_1^2+\sigma_2^2}
                     \hspace{0.3 cm}(= \overline{x}) \\
& & \\
\sigma^2(\widehat{k})
            &=& \frac{\sigma_1^2\cdot\sigma_2^2}
                       {\sigma_1^2+\sigma_2^2} + \sigma_c^2\, .
\end{eqnarray}
The most probable value of
the physical quantity is exactly what one obtains
from the
average $\overline{x}$
weighted with the inverse of the individual variances.
Its overall uncertainty is the quadratic sum of the standard
deviation of the weighted
average and the common one. The result coincides with the simple
expectation.
\subsubsection{Normalization  uncertainty}
\begin{figure}
\centering\epsfig{file=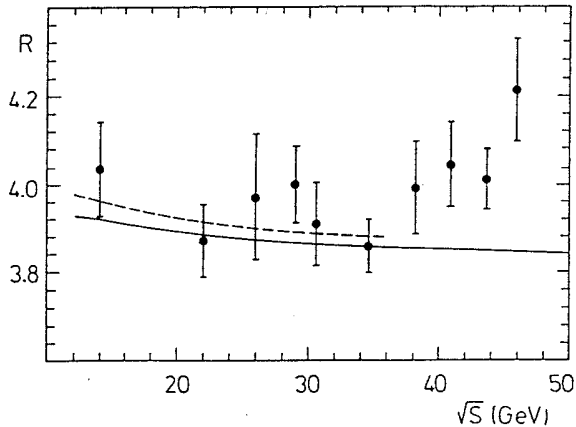,width=9cm,clip=}
\caption{\sf {\it R} measurements
from PETRA and PEP experiments
 with the best fits
 of QED+QCD to all the data (full line) and only below
 $36\,$GeV (dashed line). All data points are correlated (see text).}
\label{fig:cello}
 \end{figure}

Let $x_1\pm\sigma_1$ and $x_2\pm\sigma_2$ be the two measured values,
and $\sigma_f$ the common standard uncertainty on the scale.
The $\chi^2$ is
\begin{eqnarray}
\chi^2 & = &  \frac{1}{D} \left[
              (x_1-k)^2\cdot (\sigma_2^2+x_2^2\cdot\sigma_f^2)
              +(x_2-k)^2\cdot (\sigma_1^2+x_1^2\cdot\sigma_f^2)\right. \\
       & &    \hspace{0.7 cm} \left. -2\cdot (x_1-k)\cdot
              (x_2-k)\cdot x_1\cdot x_2\cdot\sigma_f^2
              \right]\, ,
\end{eqnarray}
where $D=\sigma_1^2\cdot\sigma_2^2 +
(x_1^2\cdot\sigma_2^2 +x_2^2\cdot\sigma_1^2)\cdot\sigma_f^2\,$\,.
We obtain in this case the following result:
\begin{eqnarray}
\widehat{k} &=& \frac{x_1\cdot\sigma_2^2+x_2\cdot\sigma_1^2}
                {\sigma_1^2+\sigma_2^2+(x_1-x_2)^2\cdot\sigma_f^2} \\
& & \\
\sigma^2(\widehat{k})
            &=& \frac{\sigma_1^2\cdot\sigma_2^2+
                 (x_1^2\cdot\sigma_2^2+x_2^2\cdot\sigma_1^2)\cdot\sigma_f^2}
                 {\sigma_1^2+\sigma_2^2 + (x_1-x_2)^2\cdot\sigma_f^2}\, .
\end{eqnarray}
With respect to the previous case, $\widehat{k}$
has a new term $(x_1-x_2)^2\cdot\sigma_f^2$ in the denominator. As long as
this is negligible with respect to the individual variances
we still get the  weighted average $\overline{x}$,
otherwise a smaller value is obtained.
Calling $r$ the ratio between $\widehat{k}$ and
$\overline{x}$, we obtain
\begin{equation}
 r = \frac{\widehat{k}}{\overline{x}} = \frac{1}
                                          {1+\frac{(x_1-x_2)^2}
                    {\sigma_1^2+\sigma_2^2}\cdot\sigma_f^2 }\, .
\end{equation}
Written in this way, one can see that the deviation from the
simple average value depends on the compatibility of the two values
and on  the normalization uncertainty.
This can be understood in the following way:
as soon as the two values are in some disagreement, the fit
starts to vary the normalization factor
- in a hidden way -
and to squeeze the scale
by an amount allowed by $\sigma_f$, in order to minimize the
$\chi^2$.
The reason the fit prefers,
 normalization factors smaller than 1
under these conditions
lies in the standard formalism of the covariance propagation,
where only first derivatives are considered. This
implies that the individual
standard deviations are not rescaled by lowering the normalization
factor, but
the points get closer.
\begin{description}
\item[Example 1.] Consider
the results of two measurements,
$8.0\cdot (1\pm 2\,\%)$ and $8.5\cdot(1\pm 2\,\%)$, having
a $10\,\%$ common normalization error.
Assuming that the two measurements
refer to the same physical quantity,
 the best estimate
of its true value can be obtained
 by fitting the points to a constant.
Minimizing
$\chi^2$ with
${\bf V}$ estimated empirically by the data, as explained
in the previous section, one obtains a value of
$7.87\pm0.81$, which is surprising to say the least,
since the most
probable result is outside the interval determined by the two
measured values.
\item[Example 2.] A real life
case of this strange effect which occurred during
the global analysis of the $R$ ratio in $e^+e^-$
performed by the CELLO colla\-bo\-ration\cite{CELLO},
is shown in
Fig. ~\ref{fig:cello}.
The data points represent the averages in
energy bins of the results of the PETRA and PEP experiments. They
are all correlated and
the error bars show the
total error
(see \cite{CELLO} for details).  In particular, at the
intermediate stage of the analysis shown in the figure, an
overall $1\,\%$ systematic error due theoretical uncertainties
was included in the covariance matrix.
The $R$ values above $36\,$GeV
show the first hint of the rise of the $e^+e^-$ cross section
due to the $Z^\circ$ pole.
At that time it was
 very interesting to prove that the observation was not
just a statistical fluctuation.
In order to test this,  the
$R$ measurements were fitted with
a theoretical function having \underline{no}
$Z^\circ$ contributions,
using only data below a certain energy.
It was expected that a fast increase of
$\chi^2$ per number of degrees of freedom $\nu$
would be observed
above $36\,$GeV,
indicating that a theoretical prediction without
$Z^\circ$ would be inadequate for describing the high energy data.
 The surprising result
was a ``repulsion''
(see Fig. ~\ref{fig:cello})
between the experimental data and the fit:
including the high energy points with larger $R$ a lower
curve was obtained,
while $\chi^2/\nu$ remained almost constant.
\end{description}
To see the source
of this effect more explicitly let us consider an alternative way
often used to take
the normalization uncertainty
into account.
A scale factor $f$, by which all
data points are multiplied, is introduced to the
expression of the $\chi^2$:
\begin{equation}
\chi^2_A = \frac{(f\cdot x_1 - k)^2}{(f\cdot\sigma_1)^2} +
\frac{(f\cdot x_2 - k)^2}{(f\cdot\sigma_2)^2} +
\frac{(f-1)^2}{\sigma_f^2}\, .
\label{eq:chi2_a}
\end{equation}
Let us also consider the same expression when the individual
standard deviations
are not rescaled:
\begin{equation}
\chi^2_B = \frac{(f\cdot x_1 - k)^2}{\sigma_1^2} +
\frac{(f\cdot x_2 - k)^2}{\sigma_2^2} +
\frac{(f-1)^2}{\sigma_f^2}\, .
\label{eq:chi2_b}
\end{equation}
The use of $\chi^2_A$ always gives the result $\widehat{k} = \overline{x}$,
because the term
$(f-1)^2/\sigma_f^2$ is harmless\footnote{
This can be seen
rewriting (\ref{eq:chi2_a}) as
\begin{equation}
\frac{(x_1 - k/f)^2}{\sigma_1^2} +
\frac{(x_2 - k/f)^2}{\sigma_2^2} +
\frac{(f-1)^2}{\sigma_f^2}\, .
\end{equation}
For any $f$,
the first two terms determine the value of $k$, and the
third one binds $f$ to 1.}
 as far as the value of the minimum $\chi^2$
and the determination on $\widehat{k}$ are concerned.
Its only influence is
on $\sigma(\widehat{k})$, which turns out to be equal to
quadratic combination of the
weighted average standard deviation with
$\sigma_f\cdot\overline{x}$,
the normalization uncertainty on the
average.
This result corresponds to the usual one
when the normalization factor
in the definition of $\chi^2$ is not included,
and the overall uncertainty is added
at the end.

Instead,
the use of $\chi^2_B$ is equivalent to the
covariance matrix: the same values of the minimum $\chi^2$,
of $\widehat{k}$ and of $\sigma(\widehat{k})$ are obtained, and
$\widehat{f}$ at the minimum turns out to be exactly the $r$ ratio
defined above. This demonstrates that the effect happens
when the data values are rescaled independently of their
standard uncertainties. The effect can become huge if the
data show mutual disagreement.
The equality of the results obtained with $\chi^2_B$
with those obtained with the covariance matrix allows us to
study, in a simpler way,
the behaviour of $r$ (= $\widehat{f}$)
 when an arbitrary amount of data points are analysed.
The fitted value of the normalization factor is
\begin{equation}
\widehat{f} = \frac{1}
    {1+\sum_{i=1}^n\frac{(x_i-\overline{x})^2}{\sigma_i^2}\cdot\sigma_f^2}\,.
\end{equation}
If the values of $x_i$ are consistent with
 a common
true value it can be shown
 that the expected value of  $f$ is
\begin{equation}
   <f> = \frac{1}{1+(n-1)\cdot\sigma_f^2}\,.
\end{equation}
Hence, there is a bias on the result when for a non-vanishing
$\sigma_f$ a large number of data points are fitted. In particular,
the fit on average produces a bias larger than the
normalization uncertainty itself if $\sigma_f > 1/(n-1)$.
One can also see that $\sigma^2(\widehat{k})$ and the
minimum of $\chi^2$
obtained with the
covariance matrix or with $\chi^2_B$ are smaller by the same factor
$r$
than those obtained with $\chi^2_A$.

\subsubsection{Peelle's Pertinent Puzzle}
To summarize, where
there is an overall uncertainty
due to an unknown systematic error and
the covariance matrix is used to define
$\chi^2$, the behaviour of the fit
depends on whether
the uncertainty is on the offset or on the scale. In the first case
the best estimates of the function parameters are exactly
those obtained without overall uncertainty,
and only the parameters' standard deviations are affected.
 In the case of unknown \underline{normalization}
errors, biased results  can be obtained. The
size of the bias depends on the
fitted function, on the
magnitude of the overall uncertainty and on the number of data points.

 It has also been shown that
this bias comes from the linearization performed in the
usual covariance propagation. This means that, even though the use
of the covariance matrix can be very useful in analyzing the
data in a compact way using available computer algorithms,
care is required
if there is one large normalization uncertainty
which affects all the data.

The effect discussed above has also been observed independently
by R.W. Peelle
and reported the year after the analysis
of the CELLO data\cite{CELLO}. The problem has been
extensively discussed among the community of
nuclear physicists, where it is presently
known as  ``Peelle's
Pertinent Puzzle''\cite{CS}.

A recent case in High Energy Physics in which this effect has
been found to have biased the result is discussed in \cite{Morris}.
\section{Multi-effect multi-cause inference: unfolding}
\subsection{Problem and typical solutions}
In any experiment the distribution of the measured observables
differs from that of the corresponding
{\it true} physical quantities due to
physics and detector effects. For example,  one may be interested
in measuring in
the variables  $x$ and $Q^2$
 Deep Inelastic Scattering events.
In such a case one is able to build
statistical estimators
which
in principle have a physical
meaning similar to the true quantities,
but which have a non-vanishing variance and  are also distorted
due to QED and QCD radiative corrections, parton fragmentation,
particle decay and limited detector performances.
The aim of the experimentalist
is to {\it unfold}
 the observed distribution from all these distortions
so as
to extract the true distribution
(see also \cite{Blobel} and \cite{Zech}).
This requires
a satisfactory knowledge of the overall effect of the
distortions on the true physical quantity.

When dealing with only one physical variable the usual method
for handling this problem is the so called
{\it bin-to-bin} correction: one evaluates
a {\it generalized efficiency} (it
may even be larger than unity)
calculating the ratio between the number of events falling in a certain
bin of the reconstructed variable and the number of events  in the
\underline{same} bin of the true variable
with a Monte Carlo simulation.
 This efficiency is then used
to estimate the number of true events from the
number of events observed in that bin. Clearly this method requires the
same subdivision in bins of the true and the experimental variable
and hence it cannot take into account
large migrations of events from one
bin to the others.
Moreover it neglects the
unavoidable correlations between adjacent bins. This approximation is valid
only if the amount of migration is negligible
and if the standard deviation of the
smearing is smaller than the bin size.

An attempt to solve the problem
of migrations is sometimes
made building a matrix which connects the
number of events generated in one bin to
 the number of events observed
in the other bins. This matrix is then inverted and applied to the measured
distribution. This immediately produces inversion problems
if the matrix is singular. On the other hand, there is no reason
from a probabilistic point of view why the inverse matrix
should exist. This  as  can easily be seen
taking the  example of two bins of the
true quantity both of which have
the same probability of being observed
in each of the bins of the measured quantity.
It follows that treating probability distributions
as vectors in space is  not correct, even
in principle. Moreover
the method is not able to handle large statistical fluctuations
even if the matrix can be inverted (if we have, for example,
a very large number of events with which
to estimate its elements and we choose
the binning in such a way as to make the matrix not singular).
The easiest way to see this is to think of the unavoidable negative
terms of the inverse of the matrix which in some extreme cases
may yield negative numbers of unfolded events.
Quite apart from these
theoretical reservations,
the actual experience of those who have used this method is
rather discouraging, the results being highly unstable.
%
%
\subsection{Bayes' theorem stated in terms of causes and effects}
Let us
 state Bayes' theorem in terms of several independent
{\it causes} ($C_i,\ i=1, 2, \ldots, n_C$) which
can produce one {\it effect} ($E$).
For example, if we consider Deep Inelastic Scattering events,
 the effect $E$ can be the observation of
an event in a cell of the measured quantities
$\{\Delta Q^2_{meas}, \Delta x_{meas}\}$.
The causes $C_i$ are then all the possible cells of the true values
$\{\Delta Q^2_{true}, \Delta x_{true}\}_i$.
Let us assume we know the
{\it initial probability} of the causes $P(C_i)$ and the conditional
probability that the $i$-th cause will produce the effect $P(E|C_i)$.
The Bayes formula is then
\begin{equation}
P(C_i|E) = \frac{P(E|C_i)\cdot P(C_i)}
            {\sum_{l=1}^{n_C} P(E|C_l)\cdot P(C_l)}\, .
\label{eq:bayes_unf}
\end{equation}
$P(C_i|E)$ depends on the initial
probability of the causes.
If one has no better prejudice
concerning $P(C_i)$
the process of inference can be started
from a uniform distribution.

The final distribution depends also on $P(E|C_i)$. These probabilities must
be calculated or estimated with Monte Carlo methods. One
has to keep in mind
that, in contrast to $P(C_i)$, these probabilities are not updated
by the observations. So if there are ambiguities
concerning the choice of $P(E|C_i)$ one has to try
them all in order to evaluate
their {\it systematic effects} on the results.

\subsection{Unfolding an experimental distribution}
If one observes $n(E)$ events with effect $E$, the expected
number of events assignable to each of the causes is
\begin{eqnarray}
\widehat{n}(C_i) = n(E)\cdot P(C_i|E)\,.
\label{eq:nc}
\end{eqnarray}
As the outcome of a measurement one has several possible effects
 $E_j$ ($j=1, 2, \ldots, n_E$) for a given cause $C_i$.
 For each of them the
Bayes formula
(\ref{eq:bayes_unf})
 holds, and
$P(C_i|E_j)$ can be evaluated.
Let us write (\ref{eq:bayes_unf})
again in the case of $n_E$ possible effects\footnote{The
broadening of the distribution due to the smearing suggests
a choice of $n_E$ larger then $n_C$. It is worth remarking
that there is no need to reject events where a measured quantity
has a value outside the range allowed for the physical quantity.
For example, in  the case of
Deep Inelastic Scattering events, cells with
$x_{meas} > 1$ or $Q_{meas}^2 < 0$ give information
about the true distribution too.},
indicating the initial probability of the causes with
$P_\circ (C_i)$:
\begin{equation}
P(C_i|E_j) = \frac{P(E_j|C_i)\cdot P_\circ (C_i)}
            {\sum_{l=1}^{n_C} P(E_j|C_l)\cdot P_\circ (C_l)}\, .
\label{eq:bays_unf}
\end{equation}
 One should note that:
\begin{itemize}
\item
$\sum_{i=1}^{n_C} P_\circ (C_i) = 1$, as usual.
Notice that if the probability of a cause
is initially set to zero it can never change, i.e. if a cause
does not exist it cannot be invented;
\item
$\sum_{i=1}^{n_C} P(C_i|E_j) = 1$\, :
 this normalization condition, mathematically
trivial since it comes directly from (\ref{eq:bays_unf}),
indicates that
each effect must come from one or more of the
causes under examination. This means that if the
observables also contain
a non negligible amount of background, this needs to be included
among the causes;
\item
$0 \le \epsilon_i \equiv \sum_{j=1}^{n_E} P(E_j|C_i) \le 1$\,:
 there is no need for
each cause to produce at least
one of the effects.
$\epsilon_i$ gives the {\it efficiency} of finding the
cause $C_i$ in any of the possible effects.
\end{itemize}

\begin{figure}
\centering\epsfig{file=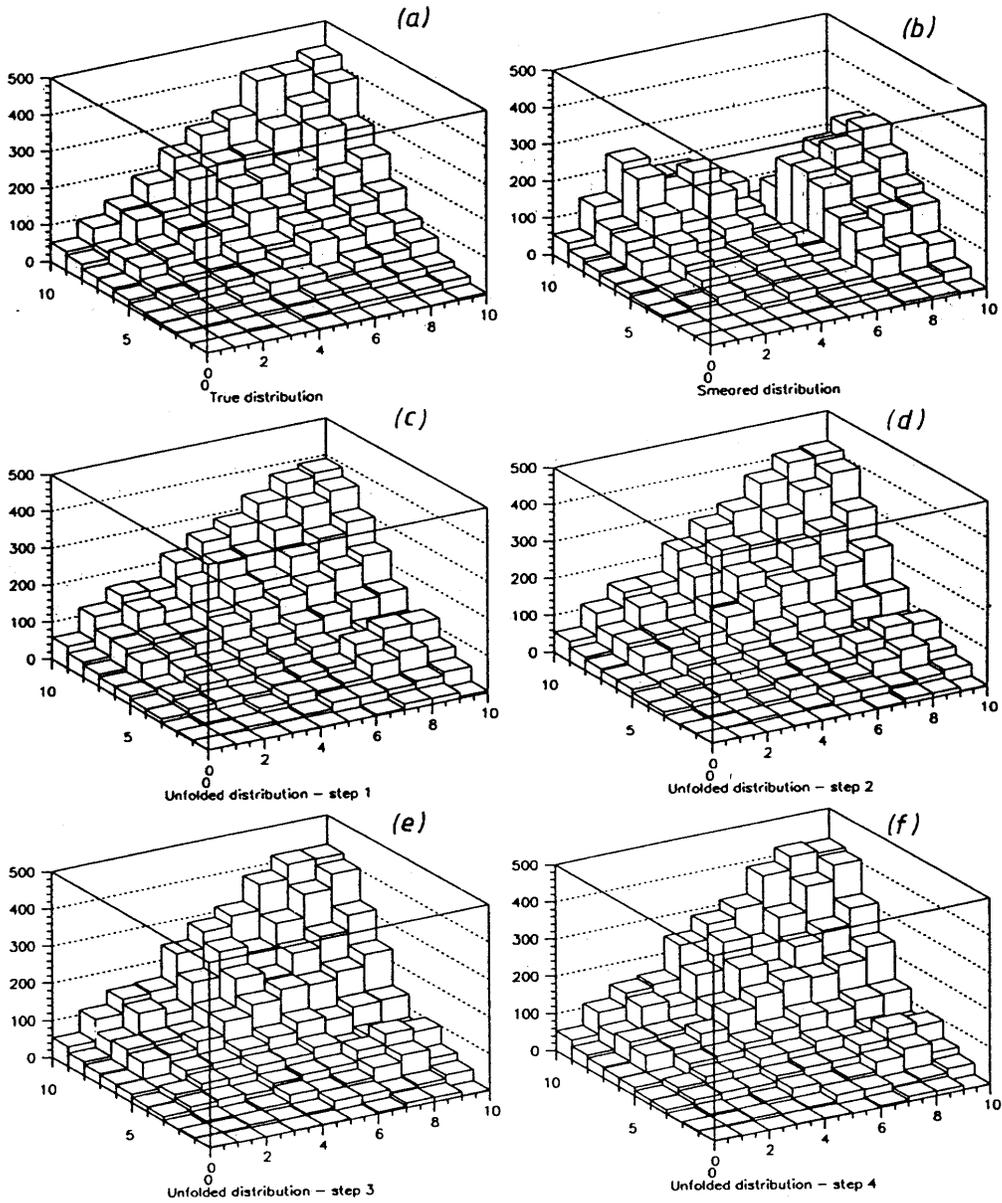,width=\linewidth,clip=}
\caption{\sf Example of a two dimensional unfolding: true distribution
(a), smeared distribution (b)
and results after the first 4 steps ((c) to (f)).}
\label{fig:unf}
 \end{figure}

After $N_{obs}$ experimental observations one obtains
a distribution of frequencies
$\underline{n}(E) \equiv \{n(E_1), n(E_2), \ldots , n(E_{n_E})\} $.
The expected number of events
to be assigned
to each of the causes (taking into account only to the observed events)
can be calculated applying (\ref{eq:nc}) to each effect:
\begin{eqnarray}
\left.\widehat{n}(C_i)\right|_{obs} & =& \sum_{j=1}^{n_E}n(E_j)\cdot P(C_i|E_j)
                     \,.
\end{eqnarray}
When inefficiency\footnote{If $\epsilon_i=0$ then
$\widehat{n}(C_i)$ will
be set to zero, since the experiment is not sensitive to the cause $C_i$.}
is also brought into the picture,
the best estimate of the true
number of events becomes
\begin{eqnarray}
\widehat{n}(C_i)& =& \frac{1}{\epsilon_i}
                   \sum_{j=1}^{n_E}n(E_j)\cdot P(C_i|E_j)
                   \hspace{1. cm}\epsilon_i \ne 0\,.
\end{eqnarray}
{}From these unfolded events we can estimate
the true total number of events,
the final probabilities of the causes and the overall efficiency:
\begin{eqnarray}
\widehat{N}_{true}& =& \sum_{i=1}^{n_C} \widehat{n}(C_i)\nonumber \\
\widehat{P}(C_i) \equiv P(C_i|\underline{n}(E))
 &=&\frac{\widehat{n}(C_i)}{\widehat{N}_{true}}  \nonumber \\
\widehat{\epsilon} &=& \frac{N_{obs}}{\widehat{N}_{true}}\,. \nonumber
\end{eqnarray}
If the initial distribution $\underline{P_\circ} (C)$
is not consistent with the
data, it will not agree with the final distribution
$\underline{\widehat{P}}(C)$.
The closer the initial distribution is to
the true distribution, the better
the agreement is.
For simulated data one can
easily verify for simulated data
that the distribution $\underline{\widehat{P}}(C)$
 lies between $\underline{P_\circ} (C)$
and the true one. This suggests  proceeding iteratively.
Fig. ~\ref{fig:unf} shows an example of a bidimensional distribution
unfolding.

More details about iteration strategy, evaluation of uncertainty,
etc. can be found in \cite{unfolding}.
I would just like to comment on an obvious criticism that may be made:
``{\it the iterative procedure is against the Bayesian spirit}, since
the same data are used many times
for the same inference''. In principle
the objection is valid, but in practice this
technique is  a ``trick'' to give
to the experimental data a weight (an importance) larger than
 that of the priors. A more rigorous procedure which took into
 account uncertainties and correlations of the initial distribution
 would have been much more complicated.
An attempt of this kind can be found in \cite{Weise3}.
Examples of unfolding procedures performed with
non-Bayesian methods are described
in \cite{Blobel} and \cite{Zech}.

\newpage
\section{Conclusions}
These notes have shown how it is possible to build a
powerful theory of measurement uncertainty starting from
a definition  of probability which seems
of little use at  first sight,
and from a formula that - some say -
looks  too trivial to be called a theorem.

The main advantages  the Bayesian approach has
over the others are
(further to the non negligible fact that it is
 able to treat problems on
which the others fail):
\begin{itemize}
\item
the recovery of the intuitive idea of probability as a valid
concept for treating scientific problems;
\item
the simplicity and naturalness of the basic tool;
\item
the capability of combining prior prejudices and experimental
facts;
\item
the automatic updating property as soon as new
facts are observed;
\item
the transparency of the method which allows the
different assumptions on which the inference may depend
to be checked and changed;
\item
the high degree of awareness that it gives to its user.
\end{itemize}

When employed on the problem of measurement errors,
as a special application of conditional probabilities,
it allows all possible sources of uncertainties
to be treated in the most general way.

When the problems get complicated and the general method becomes
too heavy to handle, it is possible to use
approximate methods based on the linearization of
the final probability density function to calculate
the first and second moments of the distribution. Although the
formulae are exactly those of the standard ``error propagation'',
the interpretation of the true value as a random variable simplifies
the picture   and allows easy inclusion of
 uncertainties due to systematic effects.

Nevertheless there are some cases in which the linearization
may  cause severe problems, as shown in Section 14. In such
cases one needs to go back to the general method or to apply
other kinds of approximations which are not just  blind use
of the covariance matrix.

The problem of unfolding dealt with in the last section should
to be considered a side remark with respect to the
mainstream  of the notes. It is in fact a mixture of genuine Bayesian
method (the basic formula), approximations (covariance matrix evaluation)
and  {\it ad hoc} prescriptions of iteration and smoothing,
used to sidestep the formidable problem
of seeking the general solution.

\newpage
 I would like to conclude with three quotations:
\begin{itemize}
\item
\begin{quote}
{\sl \small
``($\ldots$) the best evaluation of the uncertainty ($\ldots$) must be given
($\ldots$) \\
The method stands, therefore, in contrast to certain older methods
that have the following two ideas in common:
\begin{itemize}
\item
The first idea is that the uncertainty reported should be 'safe'
or 'conservative' ($\ldots$) In fact, because the evaluation
of the uncertainty of a measurement result is problematic,
it was often made deliberately large.
\item
The second idea is that the influences that give rise to
uncertainty were always recognizable as either 'random'
or 'systematic' with the two being of different nature; ($\ldots$)''
\end{itemize}
}(ISO {\it Guide}\cite{ISO})

\end{quote}
\item
\begin{quote}
{\sl \small
``Well, QED is very nice and impressive, but when everything
is so neatly wrapped up in blue bows, with all
experiments in exact agreement with each other and with
the theory - that is when one is learning
{\bf absolutely nothing}.''

``On the other hand, when experiments are in hopeless conflict
- or when the observations do not make sense according to
conventional ideas, or when none of the new models seems
to work, in short when the situation is an unholy mess -
{\bf that} is when one is really making hidden progress
and a breakthrough is just around the corner!''}\\
(R. Feynman, 1973 Hawaii Summer Institute,
cited by D. Perkins at 1995 EPS Conference, Brussels)
\end{quote}
\item
\begin{quote}
{\sl \small
``Although this {\it Guide} provides a framework for assessing
uncertainty, it cannot substitute for critical
thinking, intellectual honesty, and professional skill.  The evaluation
of uncertainty is neither a routine task nor a
purely mathematical one; it depends on detailed knowledge
of the nature of the measurand and of the measurement.
The quality and utility of the uncertainty quoted for the result of a
 measurement therefore ultimately depend on the understanding,
 critical analysis, and integrity of those who contribute to
 the assignment of its value''.
} (ISO {\it Guide}\cite{ISO})
\end{quote}
\end{itemize}
\newpage
\section*{Acknowledgements}
It was a great pleasure to give the lectures on which these
notes are based. I thank all the students for the active interests
shown and for questions and comments. Any further criticism
on the text is welcome.

I have benefitted a lot from discussions
on this subject
with my Rome and DESY colleagues,
expecially those of the ZEUS Collaboration. Special
 acknowledgements go to
Dr. Giovanna Jona-Lasinio and Prof. Romano Scozzafava of
`'La Sapienza'', Dr. Fritz Fr\"ohner of FZK Karlsruhe (Germany),
Prof. Klaus Weise of the PTB Braunschweig  (Germany),
and Prof. G\"unter Zech of Siegen University (Germany)
for clarifications on several aspects of probability
and metrology. Finally, I would like to thank
Dr. Fritz Fr\"ohner,
Dr. Gerd Hartner of Toronto University (Canada),
Dr. Jos\`e Repond of Argonne National Laboratory (USA)
and Prof. Albrecht Wagner of DESY (Germany) for
critical comments
on the manuscript.

\section*{Bibliographic note}
The state of the art of Bayesian theory is summarized
in \cite{Bernardo}, where many references can be found.
A concise presentation of the basic principles can instead
been found in \cite{nature}.
Text books that I have consulted are \cite{Winkler}
and \cite{Press}. They contain many references too.
As an introduction to subjective probability
de Finetti's {\it ``Theory of
probability''}\cite{Definetti3}
is
a {\it must}. I have found the reading of \cite{Definetti1}
particular stimulating and that of \cite{Scozzafava}
very convincing (thanks expecially to the many examples and exercises).
Unfortunately these two books are only available
 in Italian for the moment.

Sections \ref{sec:variables} and \ref{sec:clim} can be reviewed
in standard text books. I recommend those familiar to you.

The applied part of these notes, i.e. after section \ref{sec:inference},
is, in a sense, ``original'', as it has been derived
autonomously and, in many cases, without the knowledge that the
results have been  known to experts for two centuries. Some of
the examples of section \ref{sec:unknown} were worked out for these
lectures.
The references in the applied part are given at the appropriate
place in the text - only those actually used have been indicated.
Of particular interest is the Weise and W\"oger theory
of uncertainty\cite{Weise2}, which differs from that of these
notes because of the additional use of the Maximum Entropy Principle.

A consultation of the ISO {\it Guide}\cite{ISO} is advised.
Presently the BIPM recommendations are  also followed
by the american National Institute of Standards and Technology
(NIST), whose {\it Guidelines}\cite{nist} have the advantage,
with respect to the ISO {\it Guide}, of being available on www too.


\begin{thebibliography}{ref99}
{\small
\bibitem{DIN}
DIN Deutsches Institut f\"ur Normung,
{\it ``Grunbegriffe der Messtechnick - Behandlung
von Unsicherheiten bei der Auswertung von Messungen''}
(DIN 1319 Teile 1-4),
Beuth Verlag GmbH, Berlin, Germany, 1985.
Only parts 1-3
are published in English. An English translation
of part 4 can be requested from the authors of \cite{Weise2}.
Part 3 is going to be rewritten in order to be made in agreement
with \cite{ISO} (private communication from K. Weise).

\bibitem{ISO}
International Organization for Standardization (ISO),
{\it ``Guide to the expression of uncertainty in measurement''},
Geneva, Switzerland, 1993.

\bibitem{Jeffreys}
H. Jeffreys, {\it ``Theory of probability''}, Oxford
University Press, 1961.

\bibitem{Winkler}
R. L. Winkler,
``{\it An introduction to Bayesian inference and decision}'',
Holt, Rinehart and Winston, Inc., 1972.

\bibitem{Definetti3}
B. de Finetti, ``{\it Theory of probability}'',
J. Wiley \& Sons, 1974.

\bibitem{Press}
S. J. Press,
``{\it Bayesian statistics: principles, models, and applications}'',
John Wiley \& Sons Ltd, 1989.

\bibitem{Bernardo}
J.M. Bernardo, A.F.M. Smith,
``{\it Bayesian theory}'', John Wiley \& Sons Ltd, Chichester, 1994.

\bibitem{PDG}
Particle Data Group,
{\it ``Review of particle properties''},
{\it Phys. Rev. D}, {\bf 50} (1994) 1173.

\bibitem{gn}
New Scientist, April 28 1995, pag. 18 {\it (``Gravitational
constant is up in the air'')}. The data
of Tab. \ref{tab:Gn} are from H. Meyer's DESY seminar,
June 28 1995.

\bibitem{end}
P.L.  Gaison, {\it ``How experiments end''}, The University of Chicago Press,
1987.

\bibitem{comp}
G. D'Agostini, {\it ``Limits on electron compositeness from the
Bhabha scattering at PEP and PETRA''},
Proceedings of the
$XXV${\it th}
Rencontre de Moriond on ``Z$^\circ$ Physics'',
Les Arcs (France), March 4-11, 1990, pag. 229 (also DESY-90-093).

\bibitem{wroblewski}
A.K. Wr\'oblewski, {\it ''Arbitrariness in the development of physics''},
after-dinner talk at the International Workshop on {\it
Deep Inelastic Scattering and Related Subjects}, Eilat, Israel,
 6-11 February 1994, Ed. A. Levy (World Scientific, 1994), pag. 478.

\bibitem{Jaynes}
E.T. Jaynes, {\it ``Information theory and statistical mechanics''},
{\it Phys. Rev.} {\bf 106} (1957) 620.

\bibitem{Weise2}
K. Weise, W. W\"oger,
{\it ``A Bayesian theory of measurement uncertainty''},
{\it Meas. Sci. Technol.}, {\bf 4}  (1993) 1.

\bibitem{ISOD}
International Organization for Standardization  (ISO),
{\it ``International vocabulary of basic and general terms in
metrology''},
Geneva, Switzerland, 1993.

\bibitem{syst}
G. D'Agostini, {\it ``On the use of the covariance matrix to fit correlated
data''}, Nucl. Instr. Meth. {\bf  A346} (1994) 306.

\bibitem{CELLO}
CELLO Collaboration, H.J. Behrend et al.,
{\it ``Determination of $\alpha_s$ and $\sin^2\theta_w$ from measurements
of total hadronic cross section in $e^+e^-$ annihilation''},
{\it Phys. Lett.}
 {\bf 183B} (1987) 400.

\bibitem{CS}
S. Chiba and D.L. Smith, {\it ''Impacts of data transformations
on least-square solutions and their significance in data analysis and
evaluation''},
{\it J. Nucl. Sc. Tech.} {\bf 31} (1994) 770.

\bibitem{Morris}
M. L. Swartz, {\it ``Reevaluation of the hadronic contribution
to $\alpha(M_Z^2)$''}, SLAC-PUB-95-7001, September 1995, submitted to
{\it Phys. Rev. D}.

\bibitem{Blobel}
V. Blobel, {\it Proceedings of the ``1984 CERN School of Computing'',
Aiguablava, Catalonia}, Spain, 9-12 September 1984,
Published by CERN, July 1985, pag. 88-127.

\bibitem{Zech}
G. Zech, {\it ``Comparing statistical data to Monte Carlo simulation -
parameter fitting and unfolding''}, DESY 95-113, June 1995.

\bibitem{Weise3}
K. Weise, {\it ``Matematical foundation of an analytical
approach to bayesian Monte Carlo spectrum unfolding''},
Physicalish Technische Bundesanstalt, Braunschweig, BTB-N-24,
July 1995.

\bibitem{unfolding}
G. D'Agostini, {\it ``A multidimensional unfolding method based on
Bayes' theorem''}, Nucl. Instr. Meth. {\bf A362} (1995) 487.

\bibitem{nature}
C. Howson and P. Urbach, {\it ``Bayesian reasoning in science''},
Nature, Vol. 350, 4 April 1991, pag. 371.

\bibitem{Definetti1}
B. de Finetti, {\it ``Filosofia della probabilit\`a''},
il Saggiatore, 1995.

\bibitem{Scozzafava}
R. Scozzafava, {\it ``La probabilit\`a soggettiva e le sue
applicazioni''}, Masson, editoriale Veschi, Roma, 1993.

\bibitem{nist}
B.N. Taylor and C.E. Kuyatt,
{\it ``Guidelines for evaluating and expressing
uncertainty of NIST measurement results''}, NIST Technical Note 1297,
September 1994;\\
(www: http://physics.nist.gov/Pubs/guidelines/outline.html).

}
\end{thebibliography}
\end{document}